\documentclass[preprint,prd,showpacs,superscriptaddress,preprintnumbers,nofootinbib]{revtex4}

\def\pslash{\rlap{\hspace{0.02cm}/}{p}}

\def\qslash{\rlap{\hspace{0.01cm}/}{q}}

\def\Dslash{\rlap{\hspace{0.07cm}/}{D}}

\def\Tr{\mathop{\rm Tr}}

\newcommand{\LB}{L\!B}

\usepackage[fleqn]{amsmath}
\usepackage{amssymb}
\usepackage{enumerate}
\usepackage{bm}
\usepackage{dcolumn}
\usepackage{graphicx}
\usepackage{subfigure}

\graphicspath{{fig/}}

\begin{document}

\date{\today}

\preprint{RIKEN-TH-203}

\title{%
  Proper Eighth-Order Vacuum-Polarization Function
  and its Contribution to the Tenth-Order Lepton $g\!-\!2$
}


\author{T.~Aoyama}
\affiliation{Kobayashi-Maskawa Institute for the Origin of Particles and the Universe (KMI), Nagoya University, Nagoya,464-8602, Japan}
\affiliation{Theoretical Physics Laboratory, Nishina Center, RIKEN, Wako, 351-0198, Japan }

\author{M.~Hayakawa}
\affiliation{Theoretical Physics Laboratory, Nishina Center, RIKEN, Wako, 351-0198, Japan }
\affiliation{Department of Physics, Nagoya University, Nagoya, 464-8602, Japan }

\author{T.~Kinoshita}
\affiliation{Theoretical Physics Laboratory, Nishina Center, RIKEN, Wako, 351-0198, Japan }
\affiliation{Laboratory for Elementary Particle Physics, Cornell University, Ithaca, New York, 14853, U.S.A }

\author{M.~Nio}
\affiliation{Theoretical Physics Laboratory, Nishina Center, RIKEN, Wako, 351-0198, Japan }

\begin{abstract}
This paper reports the Feynman-parametric representation
of the vacuum-polarization function consisting of 
105 Feynman diagrams of the eighth order, 
and its contribution to the 
gauge-invariant set called Set I(i) of the tenth-order 
lepton anomalous magnetic moment.
Numerical evaluation of this set is carried out using FORTRAN codes 
generated by an automatic code generation system {\sc gencodevp{\it N}}
developed specifically for this purpose.
The contribution of diagrams containing electron loop
to the electron $g\!-\!2$ is 
$0.017~47~(11)~ (\alpha/\pi)^5$.
The contribution of diagrams containing muon loop is 
$0.000~001~67~(3)~ (\alpha/\pi)^5$.
The contribution of tau-lepton loop is negligible at present.
The sum of all these terms is
$0.017~47~(11)~ (\alpha/\pi)^5$.
The contribution 
of diagrams containing electron loop to the muon $g\!-\!2$ is 
$0.087~1~(59)~ (\alpha/\pi)^5$.
This is to be compared with the unpublished asymptotic analytic result
$(0.252~37+\mathrm{O}(m_e/m_\mu)) (\alpha/\pi)^5$.
The contribution  of tau-lepton loop to $a_\mu$ is 
$0.000~237~(1)~ (\alpha/\pi)^5$.
The total contribution to $a_\mu$, the sum of these terms and 
the mass-independent term, is 
$0.104~8~(59)~ (\alpha/\pi)^5$.
\end{abstract}

%
\pacs{ 13.40.Em, 14.60.Ef, 12.39.Fe, 12.40.Vv }

\maketitle

\section{Introduction}
\label{sec:intro}

The anomalous magnetic moment $g\!-\!2$ of the electron has played 
the central role in testing the validity of quantum electrodynamics (QED)
as well as the standard model.
The latest measurement of $a_e\equiv (g\!-\!2)/2$ by the Harvard group 
has reached the precision of $0.24\times 10^{-9}$ 
\cite{Hanneke:2008tm,Hanneke:2010au}:
\begin{eqnarray}
a_e(\text{HV08})= 1~159~652~180.73~ (0.28) \times 10^{-12} ~~~[0.24 \text{ppb}]
~.
\label{a_eHV08}
\end{eqnarray}
At present the best prediction of theory consists of 
QED corrections of up to the eighth order
\cite{Kinoshita:2005sm,Aoyama:2007dv,Aoyama:2007mn}, and
hadronic corrections \cite{Davier:2010nc,Teubner:2010ah,Krause:1996rf,
Melnikov:2003xd,Bijnens:2007pz,Prades:2009tw,Nyffeler:2009tw} 
and electro-weak corrections 
\cite{Czarnecki:1995sz,Knecht:2002hr,Czarnecki:2002nt} 
scaled down from their contributions to the muon $g\!-\!2$.
To compare the theoretical prediction with the experiment 
(\ref{a_eHV08}),
we also need  the value of the fine structure constant $\alpha$
determined by a method independent of $g\!-2\!$ .
The best value of such an $\alpha$ has been obtained recently
from the measurement of $h/m_{\text{Rb}}$, the ratio of the Planck constant
and the mass of Rb atom,  
combined with the very precisely known 
Rydberg constant and $m_\text{Rb}/m_e$\cite{Bouchendira:2010es}:
\begin{eqnarray}
\alpha^{-1} (\text{Rb10}) = 137.035~999~037~(91)~~~[0.66 \text{ppb}].
\label{alinvRb10}
\end{eqnarray}  
With this  $\alpha$   
the theoretical prediction of $a_e$ becomes 
\begin{eqnarray}
a_e(\text{theory}) = 1~159~652~181.13~(0.11)(0.37)(0.77) \times 10^{-12},
\label{a_etheory}
\end{eqnarray}
where the first, second, and third uncertainties come
from the calculated eighth-order QED term, the tenth-order estimate, and the
fine structure constant (\ref{alinvRb10}), respectively.
The theory (\ref{a_etheory})
is thus in good agreement with the
experiment (\ref{a_eHV08}):  
\begin{eqnarray}
a_e(\text{HV08}) - a_e(\text{theory}) = -0.40~ (0.88) \times 10^{-12},
\end{eqnarray}
proving that QED (standard model) is in good shape even at this very high
precision.

An alternative test of QED is to compare
the $\alpha$ of (\ref{alinvRb10}) with
the value of $\alpha$ determined from the
experiment and theory of $g\!-2\!$~:  
\begin{eqnarray}
\alpha^{-1}(a_e 08) = 137.035~999~085~(12)(37)(33)~~~[0.37 \text{ppb}],
\label{alinvae}
\end{eqnarray}
where the first, second, and third uncertainties come
from the eighth-order QED term, the tenth-order estimate, and the
measurement of $a_e(\text{HV}08)$, respectively.
Although the uncertainty of $\alpha^{-1}(a_e08)$ in (\ref{alinvae}) is a
factor 2 smaller than $\alpha^{-1}(\text{Rb}10)$, it is not a firm
factor since it depends on the estimate of the tenth-order term, which 
is only a crude guess \cite{Mohr:2008fa}.
For a more stringent test  of QED, it is obviously necessary to calculate
the actual value of the tenth-order term.
In anticipating of  this challenge we launched 
a systematic program 
several years ago
to evaluate the complete tenth-order term
\cite{Kinoshita:2004wi,Aoyama:2005kf,Aoyama:2007bs}.

The tenth-order QED contribution to the 
anomalous magnetic moment of the electron can be written as
\begin{eqnarray}
	a_e^{(10)} 
	= \left (\frac{\alpha}{\pi} \right )^5 
        \left [ A_1^{(10)}
	+ A_2^{(10)} (m_e/m_\mu) 
	+ A_2^{(10)} (m_e/m_\tau) 
	+ A_3^{(10)} (m_e/m_\mu, m_e/m_\tau) \right] ,
\label{eq:ae10th}
\end{eqnarray}
where the electron-muon mass ratio $m_e/m_\mu = 4.836~ 331~ 71~(12)
\times 10^{-3}$ 
and the electron-tau mass ratio $m_e/m_\tau = 2.875~ 64~ (47)\times 10^{-4}$
\cite{Mohr:2008fa}. The contribution to the mass-independent
The contribution to the mass-independent term $A_1^{(10)}$ may be
classified into six gauge-invariant sets, further divided into
32 gauge-invariant subsets depending on the nature of closed
lepton loop subdiagrams.
Thus far, 24 gauge-invariant subsets 
which consist of 2785 vertex diagrams, 
have been evaluated and published 
\cite{Kinoshita:2004wi,Aoyama:2008gy,Aoyama:2008hz,Aoyama:2010yt,Aoyama:2010pk}. 
Throughout the paper  the overall factor $(\alpha/\pi)^5$ is omitted for simplicity.

In this paper we report the value of $A_1^{(10)}$ contributed by 
a subset, called Set~I(i),
which consists of 105 Feynman diagrams obtained by insertion 
of proper eighth-order vacuum-polarization diagrams 
in the second-order anomalous magnetic moment $M_2$.
These diagrams can be represented by 39 independent integrals
taking account of various symmetry properties.
The evaluation of these integrals would be straightforward 
if the spectral function 
of the eighth-order vacuum-polarization were known.
Unfortunately, it is not available at present.
Thus we follow an alternative approach of 
expressing the eighth-order vacuum-polarization function $\Pi^{(8)} (q^2)$
as a set of Feynman-parametric integrals
and inserting them in the virtual photon line of 
the second-order anomalous magnetic moment $M_2$ 
\cite{Kinoshita:1990}.

Construction of the Feynman-parametric integral of the vacuum-polarization
function and removal of subdiagram
ultraviolet(UV) divergences by \textit{K}-operation \cite{Kinoshita:1990}
are described in Sec.~\ref{sec:vacpol}.
This scheme is implemented by an automated code generation
system {\sc gencodevp{\it N}} developed specifically for this purpose.
Incorporation of $\Pi^{(8)}$ in $M_2$ is carried out in Sec.~\ref{sec:m2p8}.
Since the \textit{K}-operation subtracts only the UV-divergent part of the renormalization
constant, additional removal of
UV-finite parts of renormalization constants must be carried out
to obtain the standard on-the-mass-shell renormalization.
This is shown explicitly in Sec.~\ref{sec:residual}.
Numerical evaluation of $M_{2,P_8}^{(ee)}$ is described in Sec.~\ref{sec:numerical}, 
where the first $e$ in the superscript $(ee)$ 
refers to the open electron line
and the second $e$ refers to the closed electron loop.
The contributions of the muon loop and tau-lepton loop to the electron $g\!-\!2$, 
namely $M_{2,P_8}^{(em)}$ and $M_{2,P_8}^{(et)}$, 
are described in Sec.~\ref{sec:numerical-2}.
The contribution of the Set~I(i) diagrams to the muon $g\!-\!2$
is described in Sec.~\ref{sec:muon}.
Section~\ref{sec:summary} is devoted to the summary and discussion of this work.
Especially, our result of the electron-loop contribution to the muon $g\!-\!2$
is compared to the prediction based on the  renormalization group  \cite{Kataev:1991cp} and to the result obtained by the analytic-asymtotic expansion
\cite{Baikov:2008si,Chetyrkin:2008}.

Appendix~\ref{sec:app:p4s} describes the construction of 
Feynman-parametric integrals for $M_{2,P_4^*}$.
Appendix~\ref{sec:app:onshell}
 describes the on-shell renormalization scheme for the vacuum-polarization
function.
Appendix~\ref{sec:app:kop}
gives intermediate renormalization of individual
diagrams by the \textit{K}-operation.
Appendix~\ref{sec:app:div.str.}
gives the divergence structure of quantities of sixth or lower orders. 

\begin{figure}
  \subfigure[Type f]{%
    \includegraphics[scale=.29]{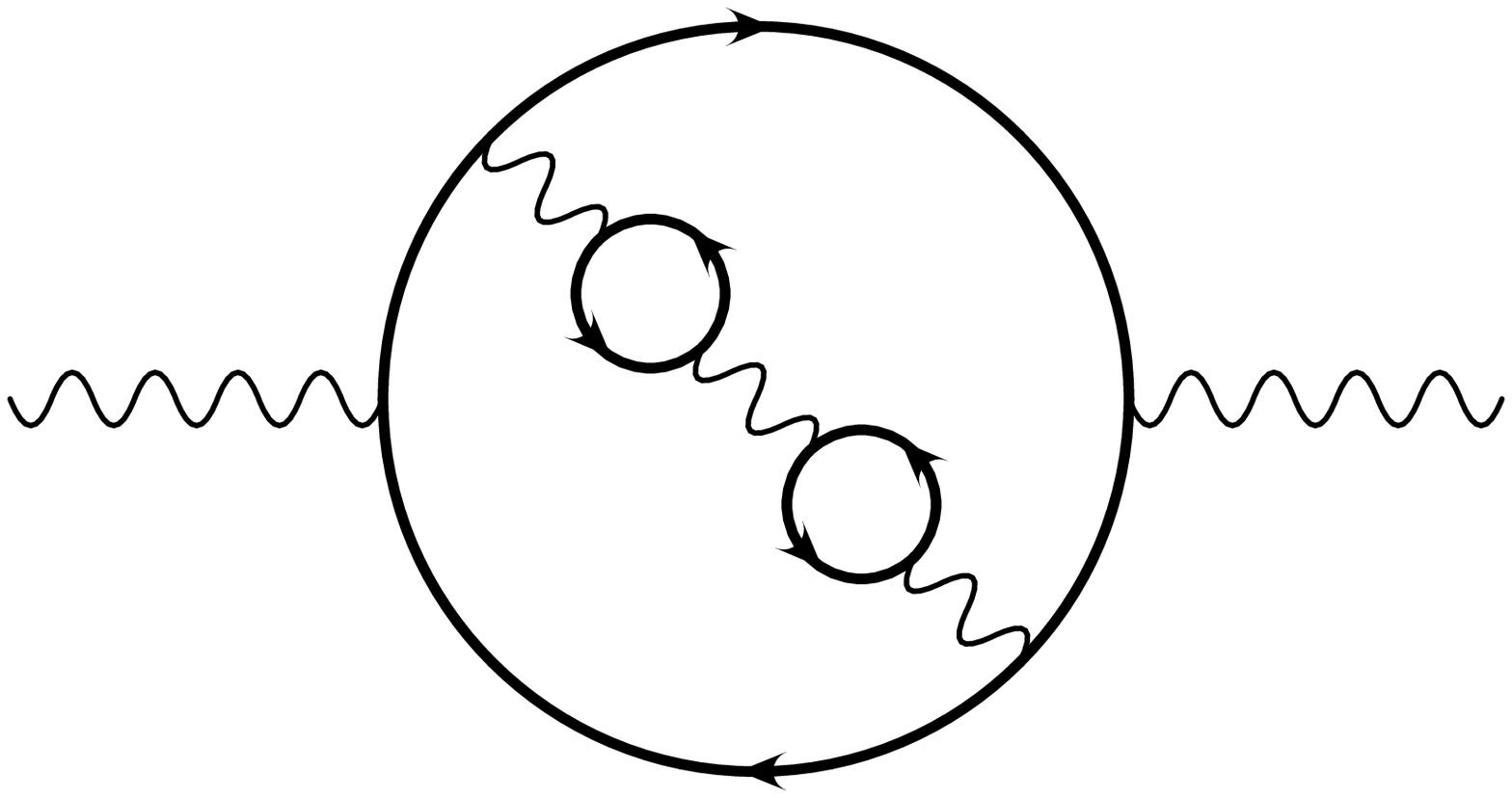}
    \label{fig:vp8:f}
  }
  \subfigure[Type g]{%
    \includegraphics[scale=.29]{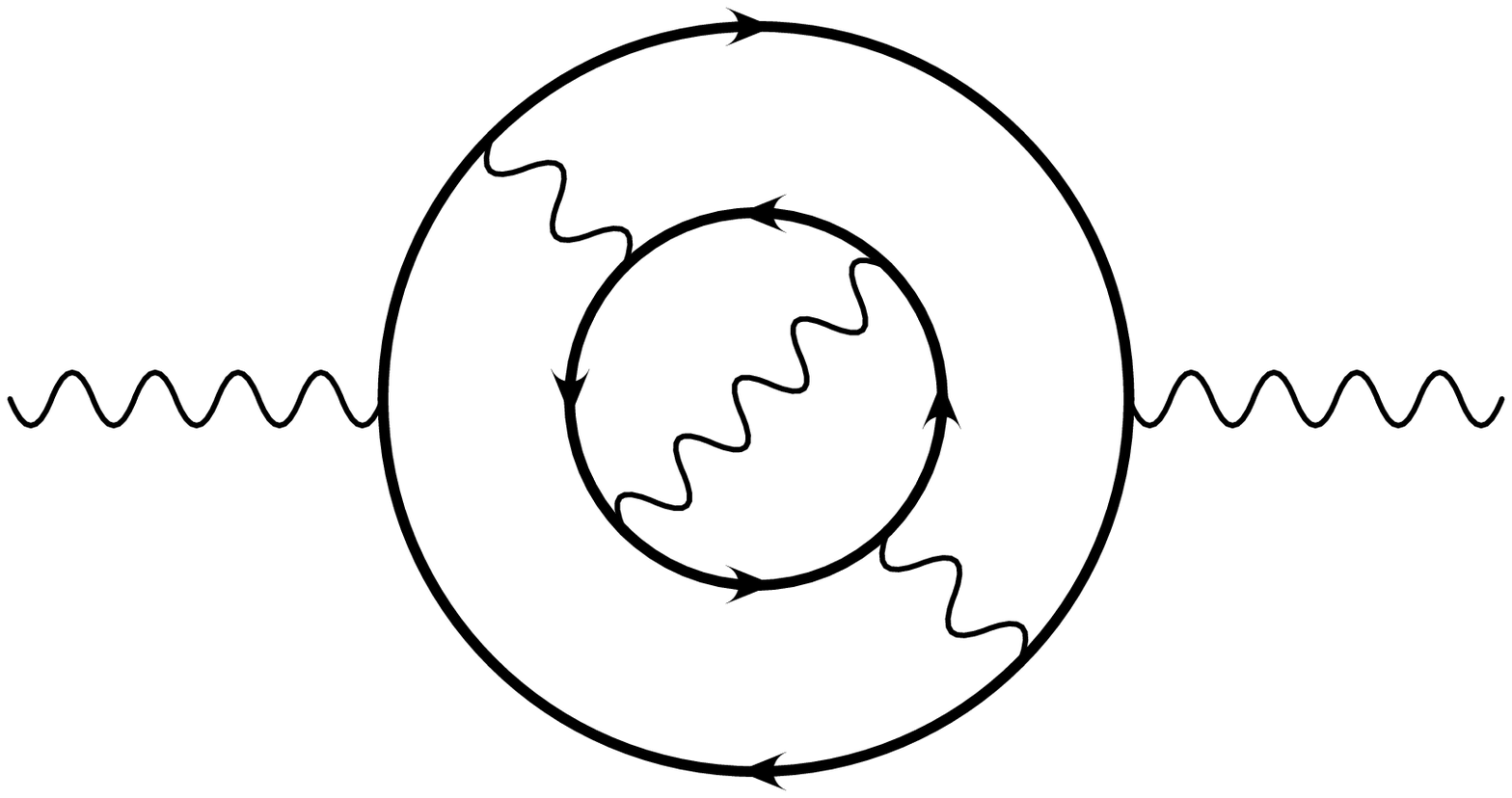}
    \label{fig:vp8:g}
  }
  \subfigure[Type h]{%
    \includegraphics[scale=.29]{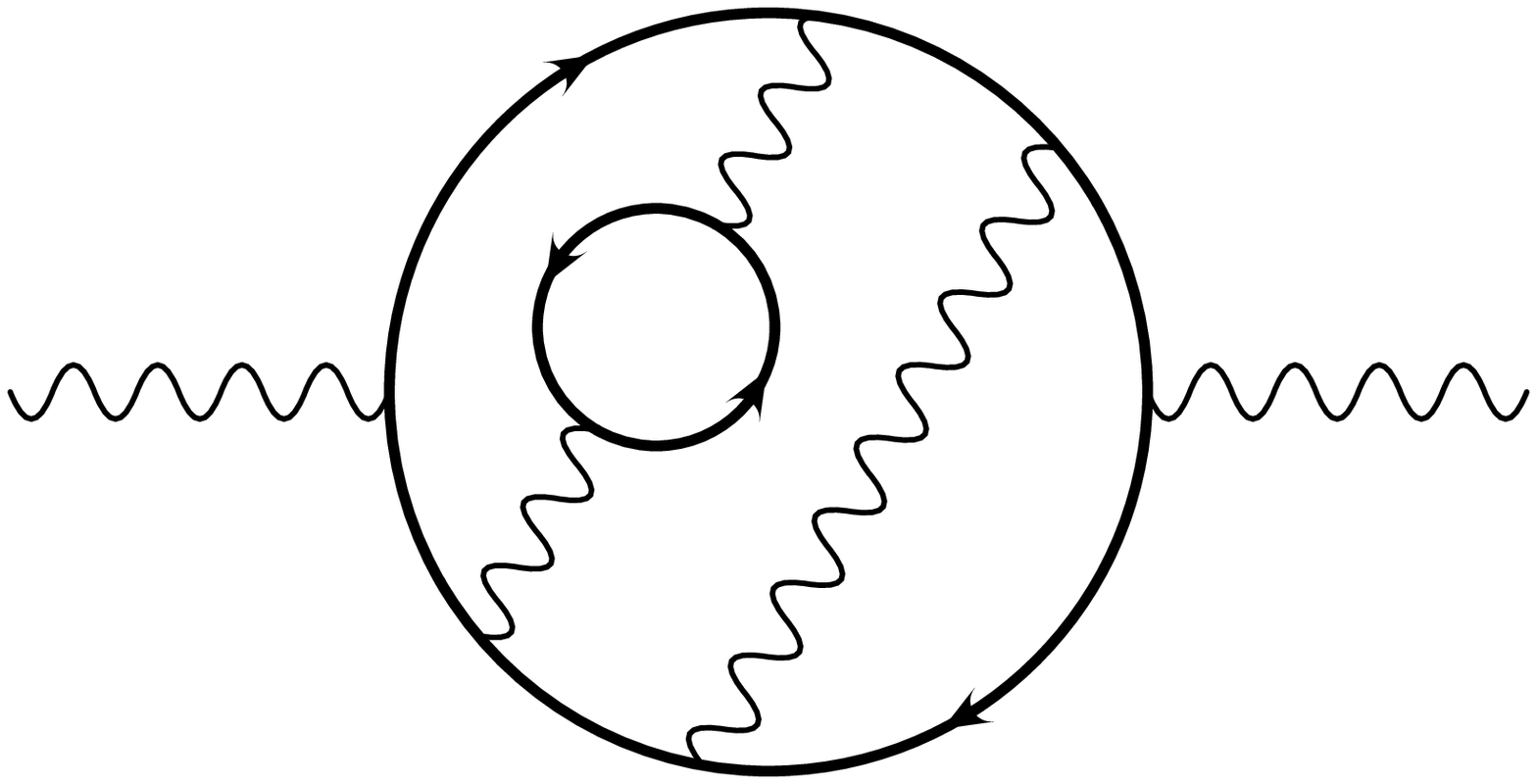}
    \label{fig:vp8:h}
  }
  \subfigure[Type i]{%
    \includegraphics[scale=.29]{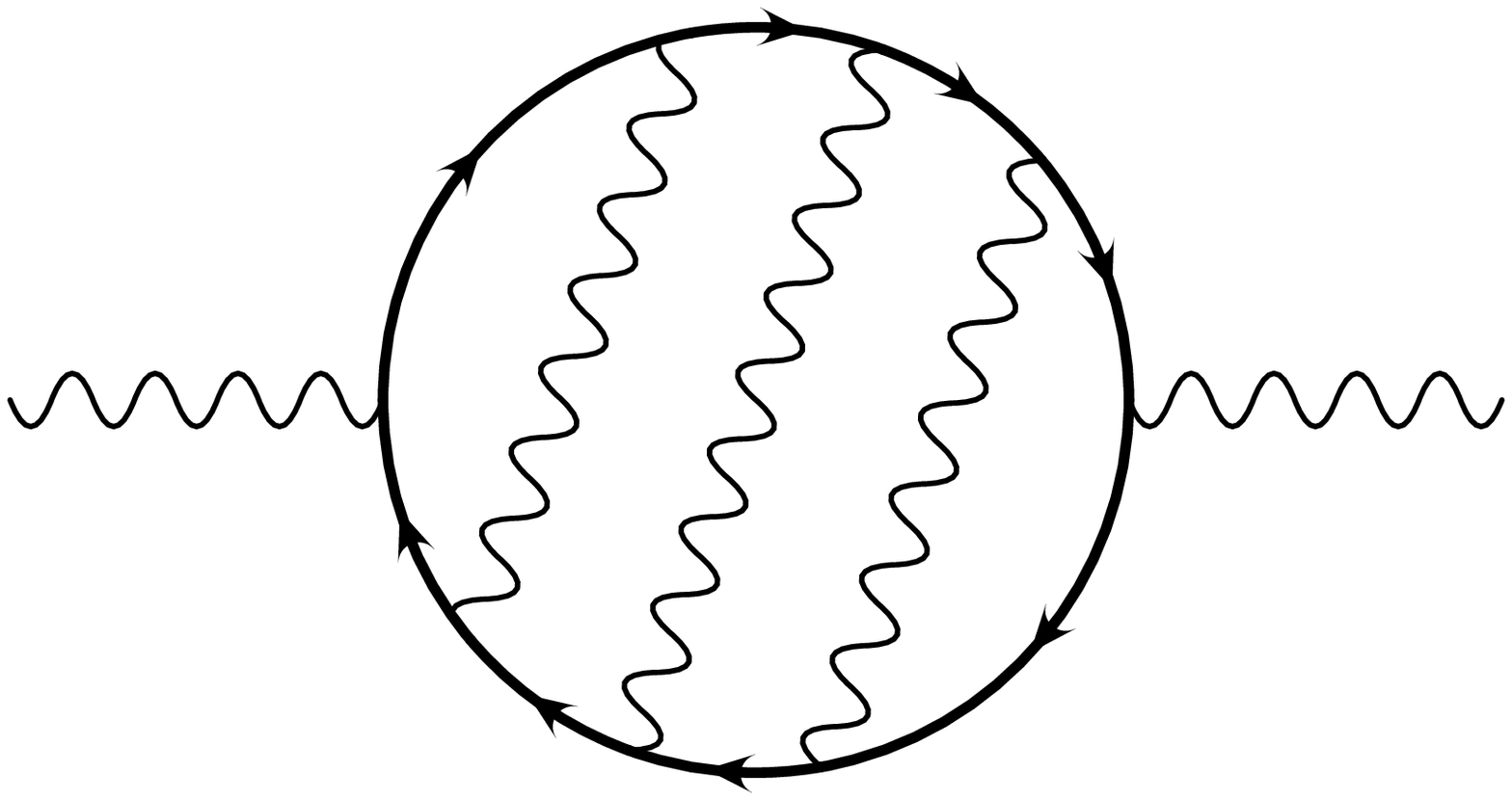}
    \label{fig:vp8:i}
  }
  \subfigure[Type j]{%
    \includegraphics[scale=.29]{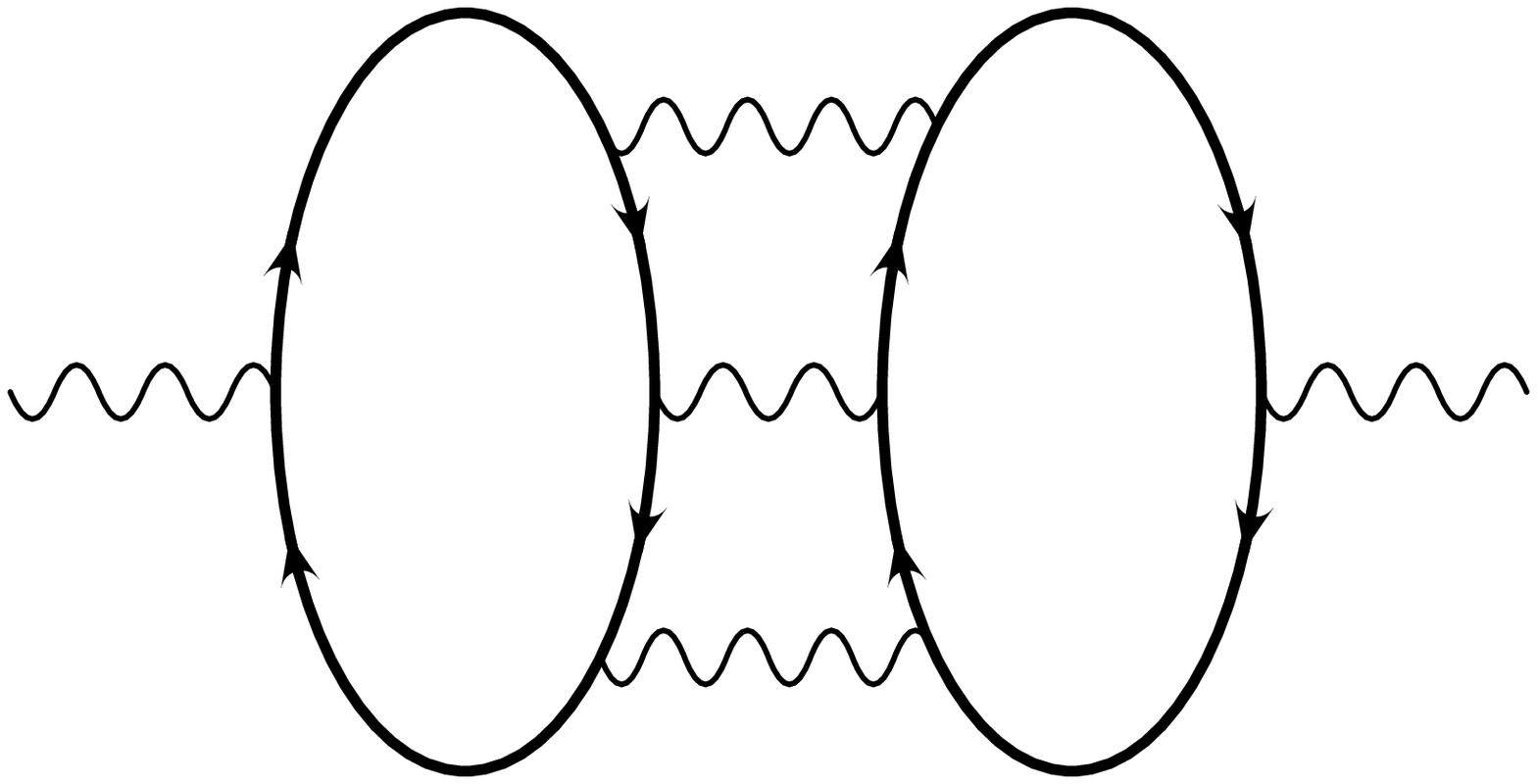}
    \label{fig:vp8:j}
  }
  \caption{%
    Five types of diagrams that contribute to the eighth-order 
    vacuum-polarization.
    \label{fig:vp8types}
  }
\end{figure}

\section{Parametric integral of vacuum-polarization function}
\label{sec:vacpol}

Diagrams that contribute to the eighth-order vacuum-polarization 
can be classified into five types according to their structures 
(See Fig.~\ref{fig:vp8types}).
Contributions from 
the diagrams of Types f, g and  h, and j to the tenth-order lepton $g\!-\!2$  
have been evaluated previously in Refs.~\cite{Kinoshita:2005sm,Aoyama:2008hz,Aoyama:2008gy}, respectively.
In this paper we focus our attention on 
the remaining Type i,  
a set of 105 proper eighth-order 
vacuum-polarization diagrams,  which 
is the most complicated one of diagrams shown in Fig.~\ref{fig:vp8types}
and evaluate its tenth-order contribution Set I(i) to $g\!-\!2$.

\subsection{Diagram representation}

In order to deal with diagrams which
contain closed lepton loops as well as open lepton paths, 
we have to generalize the rules
for the diagrams without closed lepton loop described in 
Ref.~\cite{Aoyama:2005kf}.

We begin by representing a diagram in terms of
a sequence of symbols that characterize 
the photon lines by the following rules: 
\begin{enumerate}[1)]
\item Assign indices to photon lines, e.g.  
by lower-case alphabets, `\texttt{a}', `\texttt{b}', \dots. 
\item
Identify a vertex by the index of photon line 
that is attached to the vertex. 
\item 
Read the indices of vertices 
along a lepton path (or loop) in a certain direction. 
(We adopt the reverse of the direction of the lepton propagator.)
\item 
Enclose the sequences of indices of closed lepton loops by parentheses
(but not indices of open lepton lines). 
\end{enumerate}

\begin{figure}
  \includegraphics[scale=1.0]{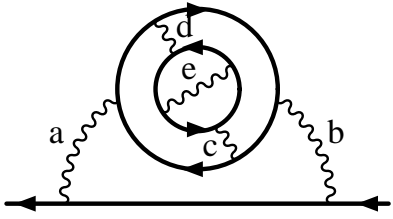}
  \caption{%
    \texttt{ab(acbd)(cede)}: 
    An example of sequential representation of a diagram of Set I(g).
    \label{fig:example}
  }
\end{figure}

For example, the tenth-order diagram with two lepton loops 
shown in Fig.~\ref{fig:example} (which belongs to Set~I(g))
may be represented by a sequence, ``\texttt{ab(acbd)(cede)}''. 

This representation is not unique because there are several possible choices
of assignment of photon line indices, 
cyclic permutations of vertices along lepton loops, 
and permutations of lepton loops and paths.
To reduce the ambiguity we adopt the convention: 
the sequence for the open lepton path comes first, 
followed by the lexicographical sequences of closed lepton loops.
The sequence within a loop is chosen also lexicographically,
e.g. the sequence \texttt{(dacb)} is rotated into \texttt{(acbd)}. 
The photon line indices are taken from `\texttt{a}' in order of appearance 
in the sequence.\footnote{ This convention may still have ambiguities.
However, it works for the diagrams with a single lepton loop 
that are discussed in the present article.}

For diagrams describing a vacuum-polarization loop, 
which is our main concern,
we adopt an additional rule
that the two photon lines external to the vacuum-polarization loop
are labeled by 
`\texttt{s}' and `\texttt{t}', 
whose Lorentz indices  are $\mu$ and $\nu$, respectively. 
We also assume that the external momentum $q$ flows 
in from the photon line `\texttt{t}' ($\nu$) and leaves 
from the photon line `\texttt{s}' ($\mu$). 
The sequence of lepton lines in the loop is chosen to 
start from the index `\texttt{s}'. 

\subsection{Algorithm to generate a proper lepton loop diagram}

We now present an algorithm for 
generating proper lepton loops of $2n$-th order. 
A diagram of this type has a single lepton loop 
that consists of $2n$ vertices, $2n$ lepton lines, 
two external photon lines, 
and $(n-1)$ internal photon lines attached to the lepton loop. 
All lepton lines are directed, and two external photon lines are 
distinguished. 

The algorithm is as follows: 
\begin{enumerate}[1)]
\item
A vertex to which an external photon line labeled by `\texttt{s}' is attached 
is chosen as the first element of the sequence.
Assign the index `0' to this vertex, and assign
numeric indices to other vertices sequentially along the loop 
in a certain direction. 
\item
Another vertex is chosen to which the other external photon line 
labeled by `\texttt{t}' is attached. 
There are $(2n-1)$ choices of vertices. 
\item
The remaining $(2n-2)$ vertices are made into $(n-1)$ pairs. 
Each pair corresponds to an internal photon line 
that connects the two vertices of that pair. 
There are $(2n-3)!!$ ways to construct $(n-1)$ pairs. 
\end{enumerate}
Therefore, the total number of diagrams is $(2n-1)\times(2n-3)!!$. 

\begin{figure}
  \includegraphics[scale=1.0]{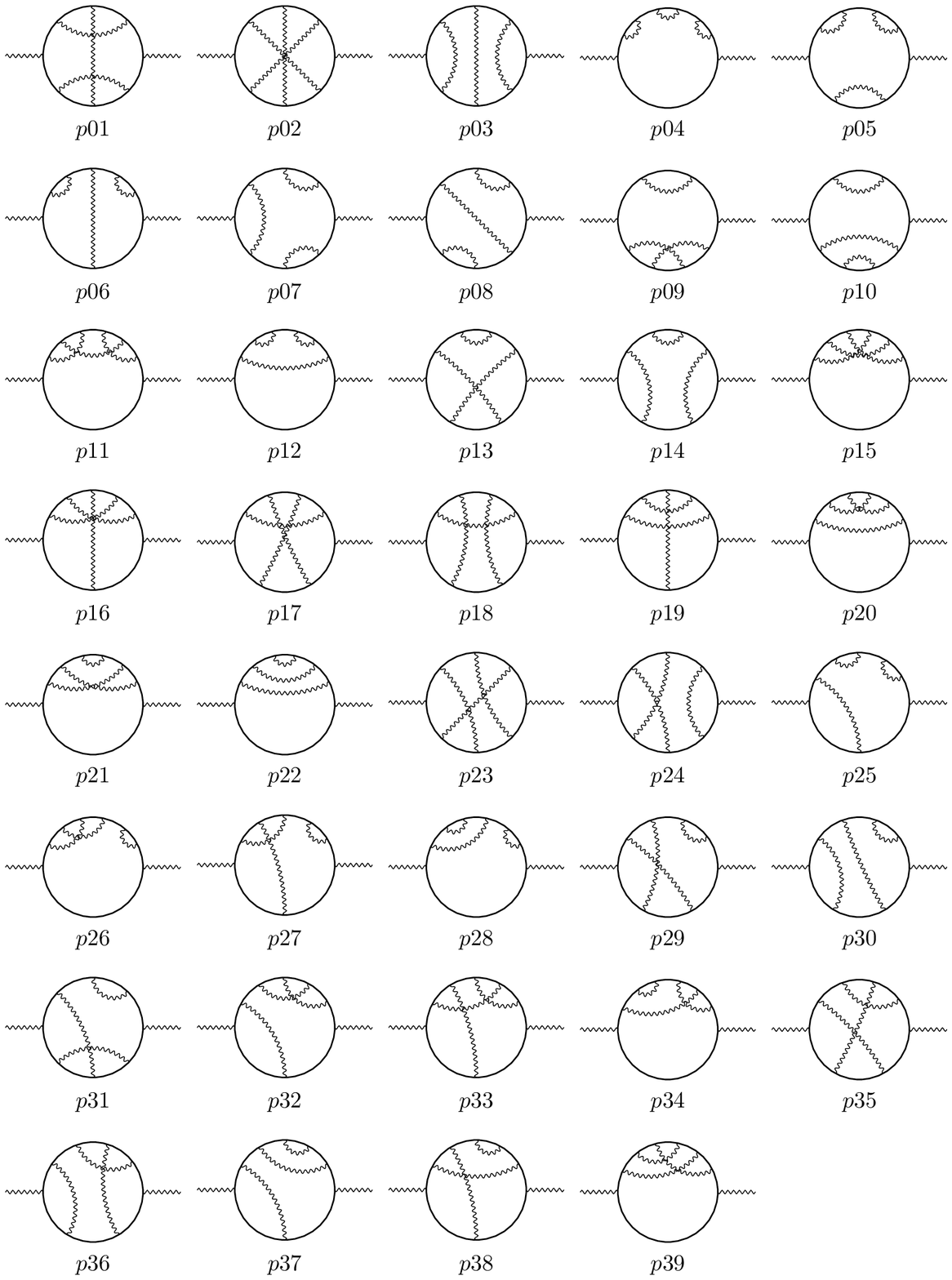}
  \caption{%
    Eighth-order vacuum-polarization diagrams containing 
    one closed lepton loop.
    \label{fig:vp8}
  }
\end{figure}

Taking into account the time-reversal symmetry of QED and the symmetry 
by the permutation of Lorentz indices of external photon lines, 
we identify the equivalent diagrams with respect to 
the reversal of sequences and exchange of symbols 
`\texttt{s}' and `\texttt{t}'. 
In this fashion we obtain a complete set of topologically distinct diagrams 
with an appropriate weight factor of the symmetry. 
In the case of the Set~I(i) we have 39 distinct diagrams,
which are shown in Fig.~\ref{fig:vp8}.

\subsection{Photon self-energy amplitude}
\label{sec:photonselfenergy}

The momentum representation of the $2n$th-order vacuum-polarization 
diagram \textit{G} has the form given by the Feynman-Dyson rule: 
\begin{multline}
  i \Pi_G^{\mu\nu}(q) 
  =
  (-1)(-ie)^{2n}
  \int \dfrac{d^4 l_1}{(2\pi)^4} \cdots \dfrac{d^4 l_n}{(2\pi)^4}\,
\\
  \Tr\left[
    \gamma^{\mu}
    \frac{i}{\pslash_1 - m}
    \gamma^{a}
    \cdots
    \gamma^\nu
    \frac{i}{\pslash_{i+1} - m}
    \gamma_{a}
    \cdots
  \right]
  \prod_{j} \left( \dfrac{-i}{k_j^2} \right)~,
  \label{eq:amplitude:pself}
\end{multline}
where $m$ is the rest mass of loop leptons,
$p_i$ is the momentum flowing on the lepton line $i$, and 
$k_j$ is the momentum flowing on the photon line $j$. 
These momenta are given as linear combinations of the loop momenta 
$l_1$, \dots, $l_n$, and the external momentum $q$. 

\begin{figure}
  \includegraphics[scale=0.4]{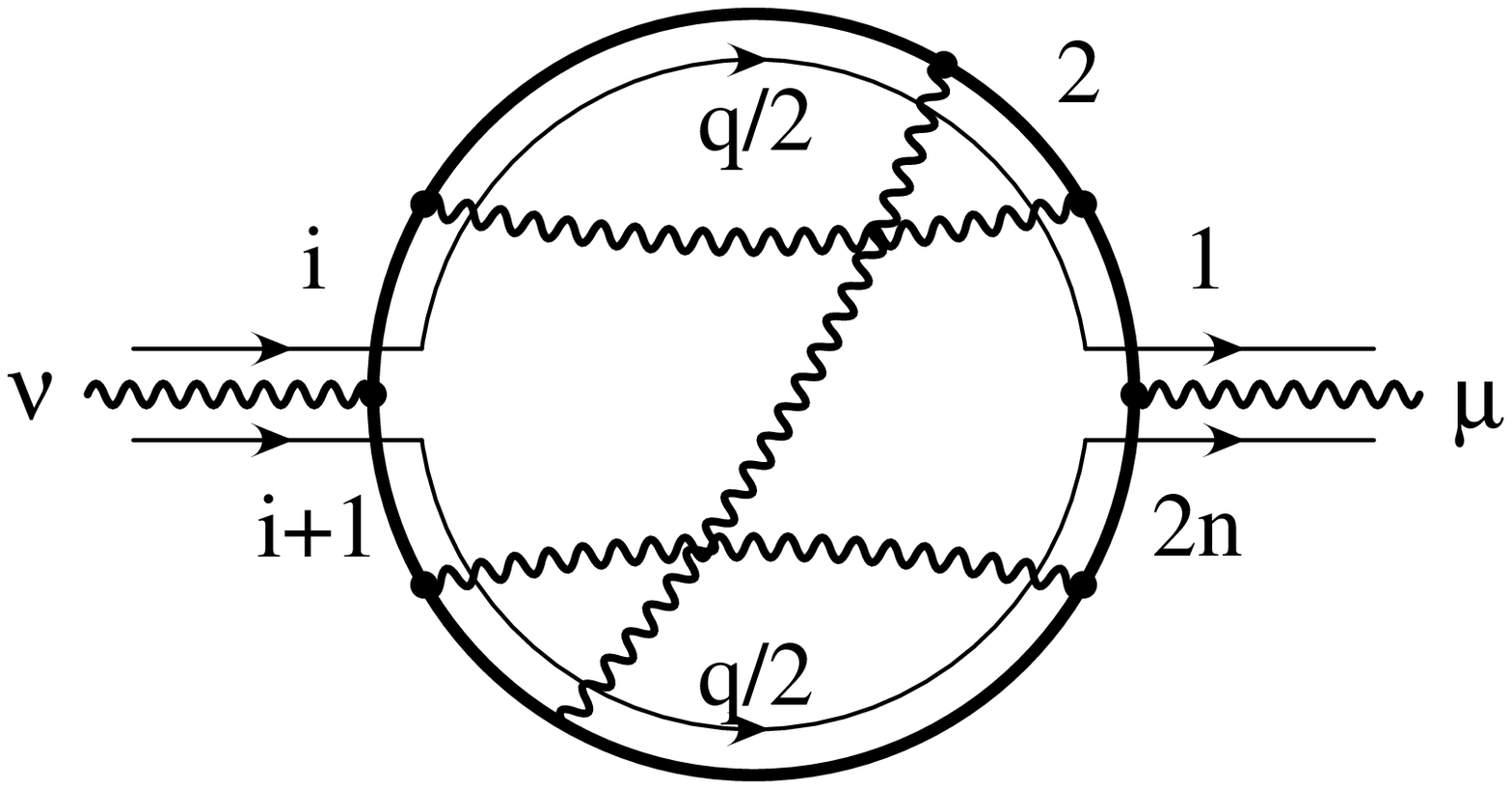}
  \caption{%
    Flow of extra momentum $q$. 
    \label{fig:qflow}
  }
\end{figure}

As a convention, the flow of the external momentum $q$ is chosen 
as shown in Fig.~\ref{fig:qflow}, where each fraction $q/2$ flows 
on the upper- (lower-) semicircle that consists of lepton lines $1,\dots,i$ 
($i+1,\dots,2n$), respectively. 
The function (\ref{eq:amplitude:pself}) is quadratically divergent, and 
we assume that the above expression is appropriately regularized 
by the Pauli-Villars regularization. 

We adopt here an approach that exploits gauge invariance  
of the sum $\Pi^{\mu\nu} = \sum_G \Pi_G^{\mu\nu}$
which allows us to ignore the gauge-dependent part of 
(\ref{eq:amplitude:pself}).
The gauge invariance ensures the identity 
\begin{eqnarray}
  q_\alpha \Pi^{\alpha\nu}(q) = 0,
  \label{eq:wt}
\end{eqnarray}
which holds for Pauli-Villars regularized function $\Pi^{\mu\nu}$. 
Differentiating it with respect to $q_\mu$, we obtain 
\begin{eqnarray}
  \Pi^{\mu\nu}(q) 
  = - q_\alpha \frac{\partial\Pi^{\alpha\nu}}{\partial q_\mu}.
  \label{eq:wt2}
\end{eqnarray}
Since $q_\mu$ is the external momentum, the order of $q_\mu$ and 
the integration over the loop momenta can be interchanged, 
as far as the integral is properly regularized. 
Thus we can write 
\begin{multline}
  i \Pi_G^{\mu\nu}(q) 
  = 
  - (-1)(-ie)^{2n} 
  q_\alpha 
  \int \dfrac{d^4 l_1}{(2\pi)^4} \cdots \dfrac{d^4 l_n}{(2\pi)^4}\ 
\\
  \frac{\partial}{\partial q_\mu}\!
  \left.
  \Tr\left[
    \gamma^{\alpha}
    \frac{i}{\pslash_1 - m}
    \cdots
    \gamma^\nu
    \frac{i}{\pslash_{i+1} - m}
    \cdots
  \right]
  \prod_{j} \left( \dfrac{-i}{k_j^2} \right)
  \right|_{\rm PV}. 
  \label{eq:amp1}
\end{multline}
Carrying out the differentiation with respect to $q_\mu$, we obtain 
\begin{multline}
  i \Pi_G^{\mu\nu}(q) 
  = 
  - (-1)(-ie)^{2n} i^{2n} 
  \int \dfrac{d^4 l_1}{(2\pi)^4} \cdots \dfrac{d^4 l_n}{(2\pi)^4}\ 
\\
  \sum_{j}
  \left.
  \Tr\left[
    \qslash
    \frac{1}{\pslash_1 - m}
    \cdots
    \left\{
      \frac{1}{\pslash_j - m}
      \left(
      \mp\frac{\gamma^\mu}{2}
      \right)
      \frac{1}{\pslash_j - m}
    \right\}
    \dots
    \gamma^\nu
    \frac{1}{\pslash_{i+1} - m}
    \cdots
  \right]
  \prod_{j} \left( \dfrac{-i}{k_j^2} \right)
  \right|_{\rm PV}, 
  \label{eq:amp2}
\end{multline}
where the minus (plus) sign in $(\mp\gamma^\mu/2)$ 
is taken when the line $j$ belongs to the upper (lower) semicircle 
of the diagram. 

The calculation can be simplified using the identity 
\cite{Kinoshita:1990} 
\begin{eqnarray}
  \frac{(\pslash_i + m_i) \gamma^\mu (\pslash_i + m_i)}{(p_i^2 - m_i^2)^2}
  =
  2 D_i^\mu (\Dslash_i + m_i) \frac{1}{(p_i^2 - m_i^2)^2}, 
  \label{eq:dop-deriv}
\end{eqnarray}
where $D_i^\mu$ is defined by 
\begin{eqnarray}
  D_i^\mu 
  = 
  \frac{1}{2}
  \int_{m_i^2}^{\infty}\! 
  dm_i^2\ \frac{\partial}{\partial q_{i\mu}}.
  \label{eq:dop}
\end{eqnarray}
We can now move the trace operation outside the $k$ integration. 
With the help of Feynman parameters $z_i$ associated with the line $i$, 
the denominators can be combined into one, 
and then the momentum integration can be carried out analytically. 
Now we bring back $D$-operators inside the $z$ integral and obtain 
\begin{eqnarray}
  \Pi^{\mu\nu}(q) = -\left(-\frac{\alpha}{4\pi}\right)^n
  \int
  \frac{(dz)}{U^2}\ 
  ( \sum_j \mp z_j D_j^\mu )
  \Tr\left[
    \qslash (\Dslash_i + m) \dots \gamma^\nu \dots
  \right]
  \frac{1}{V^n}, 
  \label{eq:amp3}
\end{eqnarray}
where 
\begin{eqnarray}
(dz) \equiv \delta (1 - \sum_i z_i ) \prod_i dz_i,~~~~i= 1, 2, \ldots, 3n-1.
\end{eqnarray}
$U$ is a homogeneous function of $z$'s determined 
from the topology of the diagram and $V$ is defined by 
\begin{eqnarray}
  V = V_0 - q^2 G, \qquad 
  V_0 = \sum_\text{all lepton lines} z_i m^2, \qquad
  G = \sum_{i\in\mathcal{P}(\mu,\nu)} z_i A_i,
\end{eqnarray}
where the path $\mathcal{P}(\mu,\nu)$ is arbitrarily taken
to run between two external photon lines.
In the present case, we choose the convention that
the path $\mathcal{P}$ runs on the upper half part of the loop
of diagrams shown in Fig.~\ref{fig:qflow}.

From Lorentz invariance and gauge invariance 
we have the general structure 
\begin{eqnarray}
  \Pi^{(2n)}_{G \mu\nu}(q)
  = 
  (q_\mu q_\nu - q^2 g_{\mu\nu})\,\widetilde{\Pi}_G^{(2n)}(q^2)\ + \ 
  \text{(gauge-dependent terms)}.
  \label{eq:amplitude:extract}
\end{eqnarray}
The Lorentz scalar $\widetilde{\Pi}^{(2n)}$ has only a logarithmic 
divergence so that the number of auxiliary masses in Pauli-Villars 
regularization can be reduced to one. 
The renormalization of subdiagram UV divergences are carried out 
independently of each other. 
Gauge-dependent terms cancel out when all diagrams of Set~I(i) 
are combined. 
The charge renormalization can be achieved by 
\begin{eqnarray}
  \Pi_G^{(2n)}(q^2) 
  = 
  \widetilde{\Pi}_G^{(2n)}(q^2) - \widetilde{\Pi}_G^{(2n)}(0).
  \label{eq:amplitude:renom}
\end{eqnarray}

For the case $n=4$, which is our concern, $\widetilde{\Pi}_G^{(2n)}(q^2)$
can be expressed in the form 
\begin{align}
  \widetilde{\Pi}_G^{(8)}(q^2)
  = 
  &
  \int (dz)\,
  \Biggl[
    \frac{D_0 + q^2 B_0 + q^4 C_0 + q^6 E_0}{U^2 V^3}
    +\frac{D_1 + q^2 B_1 + q^4 C_1}{U^3 V^2}
\nonumber \\ & \qquad\qquad
    +\frac{D_2 + q^2 B_2}{U^4 V} 
    +\frac{D_3}{U^5}\ln\left[\frac{V_0}{V}\right]
  \Biggr], 
  \label{eq:amplitude:decompose}
\end{align}
when the $D$-operation is carried out  
(omitting the factor $(\alpha/\pi)^4$ for simplicity).
The coefficients $D_l$, $B_l$, $C_l$, and $E_l$ are 
functions of building blocks, $B_{ij}$ and $A_j$ 
described in the following subsection. 
The suffix $l$ labels the number of contractions 
of $D$-operators. 


\subsection{Building blocks}
\label{sec:buildinglocks}

Building blocks, $U$ and $B_{ij}$, are homogeneous polynomials 
of degree $n$ and $n-1$ of the Feynman parameters $z_i$, 
respectively. 
They are determined by the topological structure of the diagram 
called the chain diagram 
that is derived by amputating all external lines and 
disregarding distinction of the types of lines \cite{Kinoshita:1990}. 

The fundamental set of circuits of the chain diagram 
that consists of $n$ independent self-nonintersecting closed loops 
are chosen in the following manner. 
One type of circuit is formed by an
internal photon line and consecutive lepton lines that connect 
the endpoints of the photon line. 
The direction of the circuit is chosen to be that of the lepton lines. 
We may assign the direction of the photon line accordingly. 
There are $n-1$ circuits of this sort. 
The $n$-th circuit is chosen to be the closed lepton loop itself. 

Then, for $i$ and $j$ that label the lines of chain diagram, 
$U$ and $B_{ij}$ are given by 
\begin{eqnarray}
  \label{eq:def-u}
  U_{st} &=& \sum_{k}\,z_k \xi_{k,s} \xi_{k,t}, 
  \quad
  U = \det_{st}\,U_{st}, \nonumber \\[1ex]
  \label{eq:def-bij}
  B_{ij} &=& U \sum_{s,t}\,\xi_{i,s} \xi_{j,t} (U^{-1})_{st}. 
\end{eqnarray}
where $s$ and $t$ refer to the circuits. 
The loop matrix $\xi_{k,c}$ takes $(1,-1,0)$ according to 
whether the line $i$ is (along, against, outside of) the circuit $c$. 

Once $U$ and $B_{ij}$ are obtained, another building block, the scalar current $A_j$, is given by 
\begin{eqnarray}
	A_j 
	= 
	\sum_{i\in\mathcal{P}(\mu,\nu)}\,
	\left ( \delta_{ij} - z_i B_{ij} / U \right ).
\label{eq:amplitude:aj}
\end{eqnarray}

\subsection{UV divergence}

The amplitude constructed thus far may have UV divergences 
when the sum of Feynman parameters 
of one or more internal loops tends to zero. 
We adopt subtractive renormalization here in a suitable way 
for numerical treatment. 
The subtraction terms are prepared as an integral over the same 
Feynman parameter space as the original unrenormalized amplitude 
so that the divergences of the amplitude are canceled pointwise. 
These subtraction terms are constructed by a simple algorithm 
called \textit{K}-operation \cite{Kinoshita:1990} for each 
occurrence of subdiagram UV divergences. 
The whole divergent structure of a diagram is recognized by 
Zimmermann's forest formula. 

The subdiagrams relevant for the UV divergence are of 
self-energy type or vertex type. 
For the proper lepton loops of the present concern, such a 
subdiagram is represented by an open segment of the lepton loop, 
which involves vertices and lepton lines in the segment, and 
photon lines whose endpoints are included in the segment. 
Therefore, we have to examine every segment of the lepton loop 
whenever it corresponds to a one-particle irreducible subdiagram 
of self-energy type or vertex type. 

The inclusion relation of subdiagrams are found by examining 
the inclusion relation of the segments: 
they are \textit{independent} or \textit{overlapping}, 
or one segment is completely \textit{included} in the other. 
Once the relation is known, the complete set of forests are 
constructed by finding all possible sets of subdiagrams 
whose elements are not overlapping with each other. 

Each forest corresponds to a particular emergence of UV divergence, 
and it is related to a subtraction term in the subtractive 
renormalization. 
For a vertex subdiagram $\mathcal{S}$ of a diagram $\mathcal{G}$, 
the subtraction term  defined by 
the \textit{K}-operation factorizes analytically by construction as 
\begin{eqnarray}
	L_\mathcal{S}^{\rm UV}  \Pi_{\mathcal{G}/\mathcal{S}},
\end{eqnarray}
where $L_\mathcal{S}^{\rm UV}$ is the UV-divergent part of vertex renormalization constant 
for the subdiagram $S$, and $\Pi_{\mathcal{G}/\mathcal{S}}$ is 
the amplitude of the reduced diagram $\mathcal{G}/\mathcal{S}$. 
When a subdiagram $\mathcal{S}$ is of self-energy type, 
the subtraction term factorizes analytically as
\begin{eqnarray}
	\delta m_\mathcal{S}^{\rm UV} \Pi_{\mathcal{G}/\mathcal{S}(i^*)}
	+ B_\mathcal{S}^{\rm UV}  \Pi_{\mathcal{G}/\mathcal{S},i^\prime},
\end{eqnarray}
where $\delta m_\mathcal{S}^{\rm UV}$ is the UV-divergent part of mass renormalization constant $\delta m_\mathcal{S}$, 
and $B_\mathcal{S}^{\rm UV}$ is the UV-divergent part of wave-function renormalization constant $B_\mathcal{S}$. 
When a forest consists of more than one subdiagram, 
the subtraction term becomes products of renormalization constants 
and reduced amplitudes. 

\section{Insertion of $\Pi^{(8)}$ in $M_2$}
\label{sec:m2p8}

The easiest way to insert the eighth-order vacuum-polarization loops in
$M_2$ is by the formula \cite{Lautrup:1968,Lautrup:1972}
\begin{eqnarray}
M_{2,P_8} = - \int_0^1 dy (1-y) \Pi^{(8)} \left ( \frac{-y^2}{1-y} \right ),
\end{eqnarray}
where $\Pi^{(8)}$ is given by 
Eqs.~(\ref{eq:amplitude:renom}) and~(\ref{eq:amplitude:decompose}).
It is straightforward to include this in the automated code generation system.

As a check of the integration codes, we have also derived the following formula
from Eq.~(\ref{eq:amplitude:decompose})
applying the method described in Ref.~\cite{Kinoshita:1990} which yields
\begin{align}
M_{2,P_8} &=  \int_0^1 dy (1-y) \int (dz)
\left [ \frac{D_0}{U^2 V_0^3 W^3} (W^3-(W-1)^3)
+ \frac{B_0}{U^2 G V_0^2 W^3} (W-1)^2 \right .
\nonumber \\
&- \left .  \frac{C_0}{U^2 G^2 V_0 W^3} (W-1) 
+\frac{E_0}{U^2 G^3 W^3}
+  \frac{D_1}{U^3 V_0^2 W^2}(W^2-(W-1)^2) 
\right .
\nonumber \\
&+  \left .  \frac{B_1}{U^3 G V_0 W^2}(W-1) 
-  \frac{C_1}{U^3 G^2 W^2} 
+  \frac{D_2}{U^4 V_0 W}
+  \frac{B_2}{U^4 G W}
+ \frac{D_3}{U^5} \ln \left ( \frac{W}{W-1} \right ) \right ], 
\end{align}
where 
\begin{eqnarray}
W=1+\frac{V_0}{G} \frac{1-y}{y^2} .
\end{eqnarray}
The $UV$ divergent terms of $M_{2, P_8}$
can be isolated by the \textit{K}-operation.
See Appendix~\ref{sec:app:kop} for details.

\section{Residual renormalization}
\label{sec:residual}

The standard on-the-mass-shell renormalization
of our vacuum-polarization function $\Pi^{(8)}$
is given explicitly in Appendix~\ref{sec:app:onshell}. 
Actually, it is not suitable for evaluation of these terms
on the computer, because
individual terms of these functions are UV-divergent.
Thus it is necessary to convert them into sums of UV-finite quantities.
This is achieved by an
intermediate renormalization procedure carried out by the \textit{K}-operation 
shown in Appendix~\ref{sec:app:kop}.
The \textit{K}-operation subtracts only the UV-divergent parts of
renormalization constants.
In order to obtain the standard on-the-mass-shell renormalization,
the remaining UV-finite terms must be
removed by a procedure called residual renormalization. 

Substituting expressions given in Appendix~\ref{sec:app:kop}
into corresponding expressions in Appendix~\ref{sec:app:onshell},
and making use of various subdiagram relations listed in  
Appendix~\ref{sec:app:div.str.},
we can convert the right-hand-side of equations of
Appendix~\ref{sec:app:onshell} into the sum of finite quantities.
Collecting all terms thus created we obtain
\begin{align}
  a_{l_1}^{(10)} [\text{I(i)}^{(l_1 l_2)}]
  & =   M_{2,\Delta \! P_8}^{(l_1 l_2)}
  \nonumber \\
  & - 6 \Delta \LB_2  M_{2,\Delta \! P_6}^{(l_1 l_2)}
  \nonumber \\
  & + \{ 14 (\Delta \LB_2)^2 - 4 \Delta \LB_4 \}  M_{2,\Delta \! P_4}^{(l_1 l_2)}
  \nonumber \\
  & + \{ -14 (\Delta \LB_2)^3 + 14 \Delta \LB_4 ~\Delta \LB_2 - 2 \Delta \LB_6 \} M_{2,P_2}^{(l_1 l_2)}
  \nonumber \\
  & - \Delta \delta m_4 M_{2,\Delta \! P_4^*}^{(l_1 l_2)}
  \nonumber \\
  & + ( 12 \Delta \LB_2 \Delta \delta m_4  + 2 \Delta \delta m_4 \Delta \delta m_2^* - 2 \Delta \delta m_6  ) 
   M_{2,P_{2^*}}^{(l_1 l_2)}.
\label{a10Ii}
\end{align}
Suppressing the superscript $(l_1l_2)$ for simplicity
the residual renormalizaton terms are defined as follows: 
\begin{eqnarray}
M_{2,\Delta \! P_8} &=& \sum_{i=p01}^{p39} n_{Fi} M_{2, i},
\nonumber \\
M_{2,\Delta \! P_6} &=& \sum_{\beta=A}^{H} \eta_\beta   M_{2,P_{6\beta}},
\nonumber \\
M_{2,\Delta \! P_4} &=&  M_{2,P_{4a}} + 2 M_{2,P_{4b}}, 
\nonumber \\
M_{2,\Delta \! P_{4^*}} &=& M_{2,P_{4a^*}} + 2 M_{2,P_{4b^*}}, 
\nonumber \\
\Delta \LB_{6} &=& \sum_{\beta=A}^{H} \lambda_\beta \Delta \LB_{6\beta},
\nonumber \\
\Delta \LB_{4} &=&  \sum_{i=1}^3 ( \Delta L_{4a,i} +\Delta L_{4b,i} ) 
                               +  \Delta B_{4a} +\Delta B_{4b},
\nonumber \\
\Delta \LB_{2} &=&  \Delta B_{2} 
\nonumber \\
\Delta \delta m_{6} &=& \sum_{\beta=A}^{H} \lambda_\beta \Delta \delta m_{6\beta},
\nonumber \\
\Delta \delta m_{4} &=& \Delta \delta m_{4a} + \Delta \delta m_{4b},
\end{eqnarray}
where $n_{Fi} =1$ for $i=p01, p02, p03$,
$n_{Fi} =2$ for $i=p04, ..., p22$,
and $n_{Fi} =4$ for $i=p23, ..., p39$,
$\eta_A=\eta_C=\eta_D=\eta_F=2$, $\eta_B=\eta_G=\eta_H=1$, $\eta_E=4$,
and $\lambda_A=\lambda_B= \lambda_C= \lambda_E= \lambda_F = \lambda_H =1$,
$\lambda_D =\lambda_G=2$. 
$\Delta \LB_{6\beta},~~\beta = A, \dots, H,$ are defined in 
Appendix~\ref{sec:app:div.str.}.
Numerical values of $\Delta \LB_2$, 
$\Delta \delta m_6$, 
$\Delta \delta m_4$, 
$M_{2,P_{2^*}}$, $\delta m_{2^*}$, etc., are listed in Table~\ref{table:renom}.

The coefficient $-6$ of $M_{2,\Delta \! P_6}$ in 
(\ref{a10Ii})
can be readily understood noting that the vacuum-polarization function
$\Pi_6$ has 6 fermion lines into which two-point vertex 
can be inserted.
This insertion is the source of 
the wave-function renormalization constant $B_2$ 
and the self-mass  $\delta m_2$ term. 
Since $\Delta \delta m_2 = 0$, however, only the $\Delta B_2$ survives
in (\ref{a10Ii}).
For convenience let us call this an insertion of $B_2$.

We find 14 different ways of insertion of two $B_2$'s in $\Pi_4$.
Ten comes from insertions of two disconnected second-order self-energy diagrams and four comes from insertions of the fourth-order self-energy diagram in which 
a second-order self-energy diagram is completely included in another second-order self-energy diagram. 
Three $B_2$'s can be inserted in $\Pi_2$ in 14 ways.
Insertion of one $B_4$ and one $B_2$ in $\Pi_2$ can also be made in 14
different ways.
All these three terms should be accompanied by terms proportional to
$\Delta \delta m_2$ which, however, vanish in our formulation based
on the \textit{K}-operation.

The factor $-4$ in $ - 4 \Delta \LB_4  M_{2,\Delta \! P_4}$ is due to the fact
that $\Pi_4$ has 4 fermion lines into which $B_4$ can be inserted.
The apparent absence of the coefficient $4$ 
in $ - \Delta \delta m_4 M_{2,\Delta \! P_{4^*}}$ can be accounted for
by the fact that $\Pi_4$ has four fermion lines into which
a two-point vertex can be inserted.  
Thus the coefficient 4 is absorbed in the definition of $\Pi_{4^*}$.
This term is present in (\ref{a10Ii}) since $\Delta \delta m_4$ is nonvanishing.

Finally one $B_6$ can be inserted in $\Pi_2$ in two ways.
The factor 2 in $ - 2\Delta \delta m_6 M_{2,P_{2^*}}$
is the same as that of $-2 \Delta B_6  M_{2,P_2}$.
The term $  2\Delta \delta m_4 \Delta \delta m_{2^*}  M_{2,P_{2^*}}$
is related to the subdiagram of $ - 2\Delta \delta m_6  M_{2,P_{2^*}}$
except for the factor $-1$.
Application of \textit{K}-operation on the second-order self-energy subdiagram of $\Pi_{4^*}$
yields $6 \Delta B_2 \Pi_{2^*}$.
Application of \textit{K}-operation on the second-order self-energy subdiagram
of $\delta m_6$ yields 6 $\Delta \delta m_4 \Delta B_2$.
Together they give $12 \Delta \LB_2 \Delta \delta m_4  M_{2,P_{2^*}}$.

A similar argument can be given starting from vertex renormalization 
subdiagrams although it does not give information on $\Delta \delta m$ term.
Consideration on vertex renormalization is not necessary, however, since 
$a_{l_1}^{(10)} [\text{I(i)}]$ is free from infrared(IR) divergence so 
that $L_n$
is always combined with $B_n$ to form an finite combination $\Delta \LB_n$.

The reason the coefficients of residual renormalization terms
can be determined by the argument described above
is that the UV-finite parts are not affected by \textit{K}-operation
which deals only with UV-divergent parts so that the coefficients
of residual renormalization terms inherit the structure
of the standard renormalization unaltered.

\begingroup
\renewcommand{\baselinestretch}{1.0}

\begin{table*}
  \caption{%
Contributions of diagrams of Set~I(\textit{i}) to $a_e$ 
for $(l_1l_2) = (ee)$. 
The superscript $(ee)$ is suppressed for simplicity. 
The multiplicity $n_F$ is the number of vertex diagrams 
represented by the integral and 
is incorporated in the numerical value. 
All integrals are evaluated initially with $10^8$ sampling points per iteration,iterated 150 times, followed by $10^9$ points, iterated 10 times. 
  \label{table:setI(i)_(ee)}
}

  \begin{ruledtabular}
    \begin{tabular}{lcdcc}
\multicolumn{1}{c}{Integral} & 
\multicolumn{1}{c}{$n_F$} &
\multicolumn{1}{c}{\hspace*{4em}\parbox[t]{10em}{Value (Error) \\[-1ex] including $n_F$}} &
\multicolumn{1}{c}{\parbox[t]{10em}{Sampling per \\[-1ex]  iteration}} &
\multicolumn{1}{c}{\parbox[t]{4em}{No. of \\[-1ex] iterations}} \\[3ex]
\hline
$M_{2, p01}$ & 1 &    0.035~760~8~( 60) & $ 1\times 10^8,~~1\times 10^9 $ & 150,~~10\\
$M_{2, p02}$ & 1 &    0.017~303~9~( 40) & $ 1\times 10^8,~~1\times 10^9 $ & 150,~~10\\
$M_{2, p03}$ & 1 &    0.039~757~3~( 73) & $ 1\times 10^8,~~1\times 10^9 $ & 150,~~10\\
$M_{2, p04}$ & 2 &    0.037~755~2~( 93) & $ 1\times 10^8,~~1\times 10^9 $ & 150,~~10\\
$M_{2, p05}$ & 2 &    0.062~796~0~(190) & $ 1\times 10^8,~~1\times 10^9 $ & 150,~~10\\
$M_{2, p06}$ & 2 &    0.129~748~0~(225) & $ 1\times 10^8,~~1\times 10^9 $ & 150,~~10\\
$M_{2, p07}$ & 2 &    0.128~655~4~(251) & $ 1\times 10^8,~~1\times 10^9 $ & 150,~~10\\
$M_{2, p08}$ & 2 &    0.103~304~6~(167) & $ 1\times 10^8,~~1\times 10^9 $ & 150,~~10\\
$M_{2, p09}$ & 2 &   -0.038~968~7~( 92) & $ 1\times 10^8,~~1\times 10^9 $ & 150,~~10\\
$M_{2, p10}$ & 2 &   -0.057~281~7~(105) & $ 1\times 10^8,~~1\times 10^9 $ & 150,~~10\\
$M_{2, p11}$ & 2 &    0.020~893~9~( 34) & $ 1\times 10^8,~~1\times 10^9 $ & 150,~~10\\
$M_{2, p12}$ & 2 &    0.038~800~8~( 54) & $ 1\times 10^8,~~1\times 10^9 $ & 150,~~10\\
$M_{2, p13}$ & 2 &    0.017~581~6~(166) & $ 1\times 10^8,~~1\times 10^9 $ & 150,~~10\\
$M_{2, p14}$ & 2 &    0.090~813~8~(165) & $ 1\times 10^8,~~1\times 10^9 $ & 150,~~10\\
$M_{2, p15}$ & 2 &    0.008~522~3~( 17) & $ 1\times 10^8,~~1\times 10^9 $ & 150,~~10\\
$M_{2, p16}$ & 2 &    0.023~211~8~( 28) & $ 1\times 10^8,~~1\times 10^9 $ & 150,~~10\\
$M_{2, p17}$ & 2 &   -0.009~190~1~( 34) & $ 1\times 10^8,~~1\times 10^9 $ & 150,~~10\\
$M_{2, p18}$ & 2 &    0.011~705~8~( 31) & $ 1\times 10^8,~~1\times 10^9 $ & 150,~~10\\
$M_{2, p19}$ & 2 &    0.024~548~0~( 48) & $ 1\times 10^8,~~1\times 10^9 $ & 150,~~10\\
$M_{2, p20}$ & 2 &    0.028~129~1~( 24) & $ 1\times 10^8,~~1\times 10^9 $ & 150,~~10\\
$M_{2, p21}$ & 2 &   -0.014~422~7~( 29) & $ 1\times 10^8,~~1\times 10^9 $ & 150,~~10\\
$M_{2, p22}$ & 2 &    0.012~984~7~( 39) & $ 1\times 10^8,~~1\times 10^9 $ & 150,~~10\\
$M_{2, p23}$ & 4 &    0.008~650~4~( 35) & $ 1\times 10^8,~~1\times 10^9 $ & 150,~~10\\
$M_{2, p24}$ & 4 &    0.037~873~8~( 95) & $ 1\times 10^8,~~1\times 10^9 $ & 150,~~10\\
$M_{2, p25}$ & 4 &    0.168~761~3~(302) & $ 1\times 10^8,~~1\times 10^9 $ & 150,~~10\\
$M_{2, p26}$ & 4 &   -0.061~567~5~(141) & $ 1\times 10^8,~~1\times 10^9 $ & 150,~~10\\
$M_{2, p27}$ & 4 &   -0.145~657~6~(325) & $ 1\times 10^8,~~1\times 10^9 $ & 150,~~10\\
$M_{2, p28}$ & 4 &   -0.078~819~8~(148) & $ 1\times 10^8,~~1\times 10^9 $ & 150,~~10\\
$M_{2, p29}$ & 4 &    0.110~765~8~(297) & $ 1\times 10^8,~~1\times 10^9 $ & 150,~~10\\
$M_{2, p30}$ & 4 &    0.217~591~7~(407) & $ 1\times 10^8,~~1\times 10^9 $ & 150,~~10\\
$M_{2, p31}$ & 4 &   -0.149~396~6~(349) & $ 1\times 10^8,~~1\times 10^9 $ & 150,~~10\\
$M_{2, p32}$ & 4 &   -0.122~439~0~(139) & $ 1\times 10^8,~~1\times 10^9 $ & 150,~~10\\
$M_{2, p33}$ & 4 &    0.043~600~9~(107) & $ 1\times 10^8,~~1\times 10^9 $ & 150,~~10\\
$M_{2, p34}$ & 4 &   -0.003~177~4~( 75) & $ 1\times 10^8,~~1\times 10^9 $ & 150,~~10\\
$M_{2, p35}$ & 4 &   -0.054~641~9~(129) & $ 1\times 10^8,~~1\times 10^9 $ & 150,~~10\\
$M_{2, p36}$ & 4 &   -0.138~680~8~(150) & $ 1\times 10^8,~~1\times 10^9 $ & 150,~~10\\
$M_{2, p37}$ & 4 &   -0.102~260~8~(164) & $ 1\times 10^8,~~1\times 10^9 $ & 150,~~10\\
$M_{2, p38}$ & 4 &   -0.060~892~7~(234) & $ 1\times 10^8,~~1\times 10^9 $ & 150,~~10\\
$M_{2, p39}$ & 4 &   -0.008~689~5~( 43) & $ 1\times 10^8,~~1\times 10^9 $ & 150,~~10\\
    \end{tabular}
  \end{ruledtabular}
\end{table*}

\endgroup

\begingroup
\renewcommand{\baselinestretch}{1.2}

\begin{table}
  \caption{%
Auxiliary integrals for  Set~I(\textit{i}). 
Some integrals are known exactly. 
Other integrals are obtained by the integration routine VEGAS.
The suserscript $(l_1~l_2)$ indicates that the open and closed
fermion lines consist of fermions $l_1$ and $l_2$, respectively.
The letters $e$, $m$, and $t$ stand for electron, muon, and tau-lepton,
respectively.
\label{table:renom}
}
\begin{ruledtabular}
\begin{tabular}{ldld}
\multicolumn{1}{c}{Integral} & 
\multicolumn{1}{c}{Value (error)} &
\multicolumn{1}{c}{Integral} & 
\multicolumn{1}{c}{Value (error)} \\[1ex]
\hline
$\Delta \delta m_2^*$  & -0.75           &
$\Delta \LB_2$         &  0.75           \\
$\Delta \delta m_4$    &  1.906~340~(21) &
$\Delta \LB_4$         &  0.027~930~(28) \\
$\Delta \delta m_6$    & -2.340~65~(48)  &
$\Delta \LB_6$         &  0.100~86~(77)  \\
$M_{2,P2}^{(ee)}$             &  0.015~687~421 \cdots  &
$M_{2,P2}^{(em)}$             &  0.519~762~(21) \times 10^{-6} \\
$M_{2,P2}^{(me)}$             &  1.094~258~28 \cdots  &
$M_{2,P2}^{(mt)}$             &  0.000~078~067~(4)  \\ 
$M_{2,P2^*}^{(ee)}$           & -0.012~702~383 \cdots  & 
$M_{2,P2^*}^{(em)}$           & -0.519~719~(17) \times 10^{-6}  \\
$M_{2,P2^*}^{(me)}$           & -0.161~084~05 \cdots  & 
$M_{2,P2^*}^{(mt)}$           & -0.000~077~655~(3)  \\
$M_{2,\Delta P4}^{(ee)}$      &  0.076~401~785 \cdots  &
$M_{2,\Delta P4}^{(em)}$      &  0.275~271~(1)\times 10^{-5}  \\ 
$M_{2,\Delta P4}^{(me)}$      &  3.135~059~01~(2)   &
$M_{2,\Delta P4}^{(mt)}$      &  0.000~412~61~(3)   \\ 
$M_{2,\Delta P4^*}^{(ee)}$    & -0.117~770~(12)  & 
$M_{2,\Delta P4^*}^{(em)}$    & -0.000~005~505~(1)  \\
$M_{2,\Delta P4^*}^{(me)}$    & -0.754~40~(13)  & 
$M_{2,\Delta P4^*}^{(mt)}$    & -0.000~819~49~(8)  \\
$M_{2,\Delta P6}^{(ee)}$      & 0.187~046~(7)  &
$M_{2,\Delta P6}^{(em)}$      & 0.731~632~(71) \times 10^{-5} \\
$M_{2,\Delta P6}^{(me)}$      & 5.543~94~(42)   &     
$M_{2,\Delta P6}^{(mt)}$      & 0.001~094~28~(7) 
    \end{tabular}
  \end{ruledtabular}
\end{table}

\endgroup

\begingroup
\renewcommand{\baselinestretch}{1.0}

\begin{table*}
  \caption{%
Contributions of diagrams of Set~I(i) to $a_e$ 
for $(l_1l_2) = (em)$. 
The multiplicity $n_F$ is the number of vertex diagrams 
represented by the integral and 
is incorporated in the numerical value. 
The superscript $(em)$ is omitted for simplicity.
All integrals are evaluated in double precision. 
  \label{table:setI(i)_(em)}
}

  \begin{ruledtabular}
    \begin{tabular}{lcdcc}
\multicolumn{1}{c}{Integral} & 
\multicolumn{1}{c}{$n_F$} &
\multicolumn{1}{c}{\hspace*{4em}\parbox[t]{10em}{Value (Error) \\[-1ex] including $n_F$}} &
\multicolumn{1}{c}{\parbox[t]{10em}{Sampling per \\[-1ex] No. of iteration}} &
\multicolumn{1}{c}{\parbox[t]{4em}{No. of \\[-1ex] iterations}} \\[3ex]
\hline
$M_{2, p01}$ & 1 &  0.115~63~(  3)\times 10^{-5}& $ 1\times 10^8 $ &  50\\
$M_{2, p02}$ & 1 &  0.077~01~(  3)\times 10^{-5}& $ 1\times 10^8 $ &  50\\
$M_{2, p03}$ & 1 &  0.201~24~(  5)\times 10^{-5}& $ 1\times 10^8 $ &  50\\
$M_{2, p04}$ & 2 &  0.147~75~(  6)\times 10^{-5}& $ 1\times 10^8 $ &  50\\
$M_{2, p05}$ & 2 &  0.236~81~( 14)\times 10^{-5}& $ 1\times 10^8 $ &  50\\
$M_{2, p06}$ & 2 &  0.484~29~( 16)\times 10^{-5}& $ 1\times 10^8 $ &  50\\
$M_{2, p07}$ & 2 &  0.488~27~( 19)\times 10^{-5}& $ 1\times 10^8 $ &  50\\
$M_{2, p08}$ & 2 &  0.400~43~( 12)\times 10^{-5}& $ 1\times 10^8 $ &  50\\
$M_{2, p09}$ & 2 & -0.148~04~(  6)\times 10^{-5}& $ 1\times 10^8 $ &  50\\
$M_{2, p10}$ & 2 & -0.248~34~(  8)\times 10^{-5} & $ 1\times 10^8 $ &  50\\
$M_{2, p11}$ & 2 &  0.071~31~(  2)\times 10^{-5} & $ 1\times 10^8 $ &  50\\
$M_{2, p12}$ & 2 &  0.168~94~(  4)\times 10^{-5} & $ 1\times 10^8 $ &  50\\
$M_{2, p13}$ & 2 &  0.052~85~( 12)\times 10^{-5} & $ 1\times 10^8 $ &  50\\
$M_{2, p14}$ & 2 &  0.409~41~( 12)\times 10^{-5} & $ 1\times 10^8 $ &  50\\
$M_{2, p15}$ & 2 &  0.043~04~(  1)\times 10^{-5} & $ 1\times 10^8 $ &  50\\
$M_{2, p16}$ & 2 &  0.064~53~(  1)\times 10^{-5} & $ 1\times 10^8 $ &  50\\
$M_{2, p17}$ & 2 & -0.044~76~(  2)\times 10^{-5} & $ 1\times 10^8 $ &  50\\
$M_{2, p18}$ & 2 &  0.037~80~(  2)\times 10^{-5} & $ 1\times 10^8 $ &  50\\
$M_{2, p19}$ & 2 &  0.088~50~(  3)\times 10^{-5} & $ 1\times 10^8 $ &  50\\
$M_{2, p20}$ & 2 &  0.114~22~(  1)\times 10^{-5} & $ 1\times 10^8 $ &  50\\
$M_{2, p21}$ & 2 & -0.062~59~(  2)\times 10^{-5} & $ 1\times 10^8 $ &  50\\
$M_{2, p22}$ & 2 &  0.047~81~(  2)\times 10^{-5} & $ 1\times 10^8 $ &  50\\
$M_{2, p23}$ & 4 &  0.034~72~(  2)\times 10^{-5} & $ 1\times 10^8 $ &  50\\
$M_{2, p24}$ & 4 &  0.148~88~(  7)\times 10^{-5} & $ 1\times 10^8 $ &  50\\
$M_{2, p25}$ & 4 &  0.652~31~( 23)\times 10^{-5} & $ 1\times 10^8 $ &  50\\
$M_{2, p26}$ & 4 & -0.235~83~( 10)\times 10^{-5} & $ 1\times 10^8 $ &  50\\
$M_{2, p27}$ & 4 & -0.491~34~( 21)\times 10^{-5} & $ 1\times 10^8 $ &  50\\
$M_{2, p28}$ & 4 & -0.347~15~( 11)\times 10^{-5} & $ 1\times 10^8 $ &  50\\
$M_{2, p29}$ & 4 &  0.389~88~( 21)\times 10^{-5} & $ 1\times 10^8 $ &  50\\
$M_{2, p30}$ & 4 &  0.944~16~( 33)\times 10^{-5} & $ 1\times 10^8 $ &  50\\
$M_{2, p31}$ & 4 & -0.512~63~( 25)\times 10^{-5} & $ 1\times 10^8 $ &  50\\
$M_{2, p32}$ & 4 & -0.477~71~(  9)\times 10^{-5} & $ 1\times 10^8 $ &  50\\
$M_{2, p33}$ & 4 &  0.146~82~(  6)\times 10^{-5} & $ 1\times 10^8 $ &  50\\
$M_{2, p34}$ & 4 & -0.000~10~(  5)\times 10^{-5} & $ 1\times 10^8 $ &  50\\
$M_{2, p35}$ & 4 & -0.171~64~(  9)\times 10^{-5} & $ 1\times 10^8 $ &  50\\
$M_{2, p36}$ & 4 & -0.548~36~( 11)\times 10^{-5} & $ 1\times 10^8 $ &  50\\
$M_{2, p37}$ & 4 & -0.498~39~( 13)\times 10^{-5} & $ 1\times 10^8 $ &  50\\
$M_{2, p38}$ & 4 & -0.179~07~( 15)\times 10^{-5} & $ 1\times 10^8 $ &  50\\
$M_{2, p39}$ & 4 & -0.044~15~(  3)\times 10^{-5} & $ 1\times 10^8 $ &  50\\
    \end{tabular}
  \end{ruledtabular}
\end{table*}

\endgroup

\begingroup
\renewcommand{\baselinestretch}{1.0}

\begin{table*}
  \caption{%
Contributions of diagrams of Set~I(\textit{i}) to $a_\mu$ 
for $(l_1l_2) = (me)$. 
The multiplicity $n_F$ is the number of vertex diagrams 
represented by the integral and 
is incorporated in the numerical value. 
The superscript $(me)$ is omitted for simplicity.
All integrals are evaluated initially with 
sampling points $10^8$ per iteration, iterated  50 times, followed by
$10^9$ points per iteration, iterated 200 times, 
and $10^{10}$ points, iterated 10 to 80 times.
  \label{table:setI(i)_(me)}
}

  \begin{ruledtabular}
    \begin{tabular}{lcdcc}
\multicolumn{1}{c}{Integral} & 
\multicolumn{1}{c}{$n_F$} &
\multicolumn{1}{c}{\hspace*{4em}\parbox[t]{10em}{Value (Error) \\[-1ex] including $n_F$}} &
\multicolumn{1}{c}{\parbox[t]{10em}{Sampling per \\[-1ex]  iteration}} &
\multicolumn{1}{c}{\parbox[t]{4em}{No. of \\[-1ex] iterations}} \\[3ex]
\hline
$M_{2, p01}$ & 1 &  5.475~765~( 50) & $ 1\times 10^8,~~1\times 10^9,~~1\times 10^{10}$ & 50,~~200,~~40\\
$M_{2, p02}$ & 1 & -3.639~035~( 13) & $ 1\times 10^8,~~1\times 10^9,~~1\times 10^{10}$ & 50,~~200,~~40\\
$M_{2, p03}$ & 1 &  1.014~582~( 56) & $ 1\times 10^8,~~1\times 10^9,~~1\times 10^{10}$ & 50,~~200,~~40\\
$M_{2, p04}$ & 2 &  9.957~281~( 77) & $ 1\times 10^8,~~1\times 10^9,~~1\times 10^{10}$ & 50,~~200,~~60\\
$M_{2, p05}$ & 2 & 11.130~636~(109) & $ 1\times 10^8,~~1\times 10^9,~~1\times 10^{10}$ & 50,~~200,~~80\\
$M_{2, p06}$ & 2 &  3.445~706~( 95) & $ 1\times 10^8,~~1\times 10^9,~~1\times 10^{10}$ & 50,~~200,~~60\\
$M_{2, p07}$ & 2 &  1.150~328~(136) & $ 1\times 10^8,~~1\times 10^9,~~1\times 10^{10}$ & 50,~~200,~~70\\
$M_{2, p08}$ & 2 &  2.431~621~( 98) & $ 1\times 10^8,~~1\times 10^9,~~1\times 10^{10}$ & 50,~~200,~~60\\
$M_{2, p09}$ & 2 & -6.305~904~( 87) & $ 1\times 10^8,~~1\times 10^9,~~1\times 10^{10}$ & 50,~~200,~~60\\
$M_{2, p10}$ & 2 &  3.576~267~( 69) & $ 1\times 10^8,~~1\times 10^9,~~1\times 10^{10}$ & 50,~~200,~~10\\
$M_{2, p11}$ & 2 &  3.087~991~( 41) & $ 1\times 10^8,~~1\times 10^9,~~1\times 10^{10}$ & 50,~~200,~~20\\
$M_{2, p12}$ & 2 &  2.681~026~( 32) & $ 1\times 10^8,~~1\times 10^9,~~1\times 10^{10}$ & 50,~~200,~~10\\
$M_{2, p13}$ & 2 &  8.166~698~( 70) & $ 1\times 10^8,~~1\times 10^9,~~1\times 10^{10}$ & 50,~~200,~~20\\
$M_{2, p14}$ & 2 & -2.042~862~( 97) & $ 1\times 10^8,~~1\times 10^9,~~1\times 10^{10}$ & 50,~~200,~~60\\
$M_{2, p15}$ & 2 & -0.014~213~(  9) & $ 1\times 10^8,~~1\times 10^9,~~1\times 10^{10}$ & 50,~~200,~~20\\
$M_{2, p16}$ & 2 &  3.556~341~( 11) & $ 1\times 10^8,~~1\times 10^9,~~1\times 10^{10}$ & 50,~~200,~~20\\
$M_{2, p17}$ & 2 &  3.279~641~(  8) & $ 1\times 10^8,~~1\times 10^9,~~1\times 10^{10}$ & 50,~~200,~~20\\
$M_{2, p18}$ & 2 & -1.044~365~(  5) & $ 1\times 10^8,~~1\times 10^9,~~1\times 10^{10}$ & 50,~~200,~~20\\
$M_{2, p19}$ & 2 &  3.238~053~( 31) & $ 1\times 10^8,~~1\times 10^9,~~1\times 10^{10}$ & 50,~~200,~~20\\
$M_{2, p20}$ & 2 & -0.804~464~( 15) & $ 1\times 10^8,~~1\times 10^9,~~1\times 10^{10}$ & 50,~~200,~~20\\
$M_{2, p21}$ & 2 & -2.002~941~( 28) & $ 1\times 10^8,~~1\times 10^9,~~1\times 10^{10}$ & 50,~~200,~~20\\
$M_{2, p22}$ & 2 &  1.439~287~( 14) & $ 1\times 10^8,~~1\times 10^9,~~1\times 10^{10}$ & 50,~~200,~~20\\
$M_{2, p23}$ & 4 &  2.246~387~(  9) & $ 1\times 10^8,~~1\times 10^9,~~1\times 10^{10}$ & 50,~~200,~~20\\
$M_{2, p24}$ & 4 & -1.943~020~( 59) & $ 1\times 10^8,~~1\times 10^9,~~1\times 10^{10}$ & 50,~~200,~~50\\
$M_{2, p25}$ & 4 &  0.717~756~(211) & $ 1\times 10^8,~~1\times 10^9,~~1\times 10^{10}$ & 50,~~200,~~80\\
$M_{2, p26}$ & 4 &  -11.597~221~(131) & $ 1\times 10^8,~~1\times 10^9,~~1\times 10^{10}$ & 50,~~200,~~80\\
$M_{2, p27}$ & 4 &  -16.497~990~(188) & $ 1\times 10^8,~~1\times 10^9,~~1\times 10^{10}$ & 50,~~200,~~80\\
$M_{2, p28}$ & 4 &  7.463~932~( 85) & $ 1\times 10^8,~~1\times 10^9,~~1\times 10^{10}$ & 50,~~200,~~60\\
$M_{2, p29}$ & 4 & 18.659~291~(133) & $ 1\times 10^8,~~1\times 10^9,~~1\times 10^{10}$ & 50,~~200,~~80\\
$M_{2, p30}$ & 4 & -3.240~940~(198) & $ 1\times 10^8,~~1\times 10^9,~~1\times 10^{10}$ & 50,~~200,~~60\\
$M_{2, p31}$ & 4 &  -16.369~751~(181) & $ 1\times 10^8,~~1\times 10^9,~~1\times 10^{10}$ & 50,~~200,~~80\\
$M_{2, p32}$ & 4 & -1.501~462~( 99) & $ 1\times 10^8,~~1\times 10^9,~~1\times 10^{10}$ & 50,~~200,~~70\\
$M_{2, p33}$ & 4 &  8.005~147~( 70) & $ 1\times 10^8,~~1\times 10^9,~~1\times 10^{10}$ & 50,~~200,~~20\\
$M_{2, p34}$ & 4 & -5.994~355~( 74) & $ 1\times 10^8,~~1\times 10^9,~~1\times 10^{10}$ & 50,~~200,~~60\\
$M_{2, p35}$ & 4 &  -13.319~800~( 61) & $ 1\times 10^8,~~1\times 10^9,~~1\times 10^{10}$ & 50,~~200,~~20\\
$M_{2, p36}$ & 4 &  6.372~372~( 64) & $ 1\times 10^8,~~1\times 10^9,~~1\times 10^{10}$ & 50,~~200,~~50\\
$M_{2, p37}$ & 4 & -0.889~232~( 94) & $ 1\times 10^8,~~1\times 10^9,~~1\times 10^{10}$ & 50,~~200,~~60\\
$M_{2, p38}$ & 4 &  -12.751~250~( 97) & $ 1\times 10^8,~~1\times 10^9,~~1\times 10^{10}$ & 50,~~200,~~60\\
$M_{2, p39}$ & 4 &  1.541~038~( 33) & $ 1\times 10^8,~~1\times 10^9,~~1\times 10^{10}$ & 50,~~200,~~20\\
  \end{tabular}
  \end{ruledtabular}
\end{table*}

\endgroup

\begingroup
\renewcommand{\baselinestretch}{1.0}

\begin{table*}
\caption{%
Contributions of diagrams of Set~I(\textit{i}) to $a_\mu$ 
for $(l_1l_2) = (mt)$. 
The multiplicity $n_F$ is the number of vertex diagrams 
represented by the integral and 
is incorporated in the numerical value. 
The superscript $(mt)$ is omitted for simplicity.
\label{table:setI(i)_(mt)}
}

\begin{ruledtabular}
\begin{tabular}{lcdcc}
\multicolumn{1}{c}{Integral} & 
\multicolumn{1}{c}{$n_F$} &
\multicolumn{1}{c}{\hspace*{4em}\parbox[t]{10em}{Value (Error) \\[-1ex] including $n_F$}} &
\multicolumn{1}{c}{\parbox[t]{10em}{Sampling per \\[-1ex] No. of iteration}} &
\multicolumn{1}{c}{\parbox[t]{4em}{No. of \\[-1ex] iterations}} \\[3ex]
\hline
$M_{2, p01}$ & 1 &  0.000~173~73~(  5) & $ 1\times 10^8 $ &  50\\
$M_{2, p02}$ & 1 &  0.000~115~02~(  4) & $ 1\times 10^8 $ &  50\\
$M_{2, p03}$ & 1 &  0.000~299~06~(  8) & $ 1\times 10^8 $ &  50\\
$M_{2, p04}$ & 2 &  0.000~220~89~(  9) & $ 1\times 10^8 $ &  50\\
$M_{2, p05}$ & 2 &  0.000~354~37~( 21) & $ 1\times 10^8 $ &  50\\
$M_{2, p06}$ & 2 &  0.000~725~06~( 24) & $ 1\times 10^8 $ &  50\\
$M_{2, p07}$ & 2 &  0.000~730~62~( 28) & $ 1\times 10^8 $ &  50\\
$M_{2, p08}$ & 2 &  0.000~599~02~( 19) & $ 1\times 10^8 $ &  50\\
$M_{2, p09}$ & 2 & -0.000~221~58~( 10) & $ 1\times 10^8 $ &  50\\
$M_{2, p10}$ & 2 & -0.000~370~54~( 12) & $ 1\times 10^8 $ &  50\\
$M_{2, p11}$ & 2 &  0.000~107~04~(  3) & $ 1\times 10^8 $ &  50\\
$M_{2, p12}$ & 2 &  0.000~252~11~(  6) & $ 1\times 10^8 $ &  50\\
$M_{2, p13}$ & 2 &  0.000~079~35~( 18) & $ 1\times 10^8 $ &  50\\
$M_{2, p14}$ & 2 &  0.000~610~34~( 18) & $ 1\times 10^8 $ &  50\\
$M_{2, p15}$ & 2 &  0.000~064~07~(  1) & $ 1\times 10^8 $ &  50\\
$M_{2, p16}$ & 2 &  0.000~097~35~(  2) & $ 1\times 10^8 $ &  50\\
$M_{2, p17}$ & 2 & -0.000~066~71~(  3) & $ 1\times 10^8 $ &  50\\
$M_{2, p18}$ & 2 &  0.000~056~95~(  3) & $ 1\times 10^8 $ &  50\\
$M_{2, p19}$ & 2 &  0.000~132~63~(  4) & $ 1\times 10^8 $ &  50\\
$M_{2, p20}$ & 2 &  0.000~170~72~(  2) & $ 1\times 10^8 $ &  50\\
$M_{2, p21}$ & 2 & -0.000~093~37~(  3) & $ 1\times 10^8 $ &  50\\
$M_{2, p22}$ & 2 &  0.000~071~59~(  4) & $ 1\times 10^8 $ &  50\\
$M_{2, p23}$ & 4 &  0.000~051~80~(  4) & $ 1\times 10^8 $ &  50\\
$M_{2, p24}$ & 4 &  0.000~222~52~( 11) & $ 1\times 10^8 $ &  50\\
$M_{2, p25}$ & 4 &  0.000~975~69~( 35) & $ 1\times 10^8 $ &  50\\
$M_{2, p26}$ & 4 & -0.000~352~91~( 15) & $ 1\times 10^8 $ &  50\\
$M_{2, p27}$ & 4 & -0.000~737~37~( 32) & $ 1\times 10^8 $ &  50\\
$M_{2, p28}$ & 4 & -0.000~517~81~( 17) & $ 1\times 10^8 $ &  50\\
$M_{2, p29}$ & 4 &  0.000~584~31~( 32) & $ 1\times 10^8 $ &  50\\
$M_{2, p30}$ & 4 &  0.001~408~74~( 49) & $ 1\times 10^8 $ &  50\\
$M_{2, p31}$ & 4 & -0.000~768~96~( 38) & $ 1\times 10^8 $ &  50\\
$M_{2, p32}$ & 4 & -0.000~714~49~( 14) & $ 1\times 10^8 $ &  50\\
$M_{2, p33}$ & 4 &  0.000~220~34~( 10) & $ 1\times 10^8 $ &  50\\
$M_{2, p34}$ & 4 & -0.000~000~59~(  8) & $ 1\times 10^8 $ &  50\\
$M_{2, p35}$ & 4 & -0.000~257~99~( 13) & $ 1\times 10^8 $ &  50\\
$M_{2, p36}$ & 4 & -0.000~820~21~( 17) & $ 1\times 10^8 $ &  50\\
$M_{2, p37}$ & 4 & -0.000~741~52~( 19) & $ 1\times 10^8 $ &  50\\
$M_{2, p38}$ & 4 & -0.000~269~58~( 22) & $ 1\times 10^8 $ &  50\\
$M_{2, p39}$ & 4 & -0.000~065~68~(  4) & $ 1\times 10^8 $ &  50\\

\end{tabular}
\end{ruledtabular}
\end{table*}

\endgroup


\section{Numerical evaluation of $M_{2,P_8}^{(ee)}$}
\label{sec:numerical}

FORTRAN codes of our diagrams are generated by 
{\sc gencodevp}{\it N} 
following the procedures described in Section \ref{sec:vacpol}.
The validity of 
{\sc gencodevp}{\it N} 
has been tested thoroughly for diagrams of Set II(d)
whose integrals are known by several other means \cite{Baikov:1995ui,Kinoshita:1998jf,Kinoshita:1998jg,Aoyama:2011rm}.
Numerical integration was carried out 
by an adaptive-iterative Monte-Carlo numerical integration routine 
VEGAS \cite{Lepage:1977sw} with $10^8$ sampling points per iteration
and 150 iterations followed by $10^9$ sampling points per iteration
and 10 iterations.
The results are summarized in 
Tables~\ref{table:setI(i)_(ee)} and~\ref{table:renom}.
 From these data we obtain
\begin{eqnarray}
  a_e^{(10)} [\text{I(i)}^{(ee)}] = 0.017~47~(11).
  \label{ae1i_ee}
\end{eqnarray}
%

\section{Numerical evaluation of $M_{2,P_8}^{(em)}$}
\label{sec:numerical-2}

Once FORTRAN programs for mass-independent diagrams are obtained,
it is straightforward to evaluate the contribution of mass-dependent
term $A_2^{(10)} (m_e/m_\mu)$.
We simply have to choose an appropriate loop fermion mass.
The results are summarized in Table~\ref{table:setI(i)_(em)}. From this table 
we obtain
\begin{eqnarray}
a_e^{(10)} [\text{I(i)}^{(em)}] = 0.000~001~666~(24).
\label{ae1i_em}
\end{eqnarray}
The numerical data used to obtain (\ref{ae1i_em}) are listed in
Tables~\ref{table:setI(i)_(em)} and \ref{table:renom}.
The contribution $A_2^{(10)} (m_e/m_\tau)$ is two orders of magnitude
smaller than (\ref{ae1i_em}) so that it is negligible at present.

\section{Contribution to the muon $g\!-\!2$}
\label{sec:muon}

The codes described above can also be applied to calculate
the contribution of the Set~I(i) to the muon $g\!-\!2$.
From Tables~\ref{table:setI(i)_(me)}  and \ref{table:renom}, 
we obtain
\begin{eqnarray}
a_\mu^{(10)} [\text{I(i)}^{(me)}] = 0.087~1~(59).
\label{amu1i_me}
\end{eqnarray}
We have also evaluated the tau-lepton contribution. The values listed in Tables~\ref{table:setI(i)_(mt)} and \ref{table:renom} lead  to
\begin{eqnarray}
a_\mu^{(10)} [\text{I(i)}^{(mt)}] = 0.000~237~(1).
\label{amu1i_mt}
\end{eqnarray}

The contribution of Set~I(i) diagrams to muon $g\!-\!2$ was 
first discussed in Eq.~(19) of Ref.~\cite{Kataev:1991cp} in which the
renormalization group was efficiently used  to pick up the leading logarithmic
contribution:
\begin{eqnarray}
a_\mu^{(10)}[\text{I(i)}^{(me)}] = - \frac{1}{2} a_4^{[1]} + \frac{115}{512}
-\frac{23}{128}\ln\left( \frac{m_\mu}{m_e} \right) + \mathrm{O}(m_e/m_\mu)
,
\label{asymptform}
\end{eqnarray}
where $a_4^{[1]}$ is the constant term of the asymptotic expansion of the
proper vacuum-polarization function $\Pi^{(8)}(q)$ of the eighth order, which 
was left undetermined.
This constant can be determined from
our numerical result  Eq.~(\ref{amu1i_me}):
\begin{eqnarray}
a_4^{[1]}[\text{numerical}] = - 1.641~0~(59)~.
\end{eqnarray}

Recently the asymptotic analytic form of $a_4^{[1]}$ was obtained directly 
together with other eighth-order vacuum-polarization 
diagrams \cite{Baikov:2008si}.
The explicit form of their $a_4^{[1]}$ was given only in 
the slide of the conference talk \cite{Chetyrkin:2008},  which reads
\begin{eqnarray}
a_4^{[1]}[\text{anal.-asympt.}] = -1.971~6~+\mathrm{O}(m_e/m_\mu).
\end{eqnarray}
Substituting this value to Eq.~(\ref{asymptform}), 
they obtained the asymptotic contribution of the Set~I(i) 
%
\begin{eqnarray}
a_\mu^{(10)} [\text{I(i)}:\text{anal.-asympt.}] = 0.252~37 
           + \mathrm{O}(m_e/m_\mu).
\label{amu1i_me_asym}
\end{eqnarray}
Whether this is in agreement with our result (\ref{amu1i_me})
or not is somewhat subtle and will be discussed in the next section.

\section{Summary}
\label{sec:summary}

In this paper we obtained the eighth-order vacuum-polarization function
$\Pi^{(8)} (q^2)$ as a sum of Feynman-parametric integrals.
It is then applied to the calculation of the tenth-order lepton $g\!-\!2$.
Collecting (\ref{ae1i_ee}) and (\ref{ae1i_em}) we obtain the contribution of
the gauge-invariant Set I(i) to the electron $g\!-\!2$
\begin{eqnarray}
a_e^{(10)} [\text{I(i)}^{(\text{all})}] = 0.017~47~(11).
\label{ae1i}
\end{eqnarray}
From (\ref{ae1i_ee}), (\ref{amu1i_me}) and (\ref{amu1i_mt}) we obtain
the contribution from Set~I(i) to the muon $g\!-\!2$
\begin{eqnarray}
a_\mu^{(10)} [\text{I(i)}^{(\text{all})}] = 0.104~8~(59).
\label{amu1i}
\end{eqnarray}

It is difficult to decide whether our result
(\ref{amu1i_me}) and asymptotic result
(\ref{amu1i_me_asym}) are in agreement or not.
In order to illuminate this problem it may be helpful to compare
$a_\mu$ of similar structure in lower orders.

For the 6th-order $a_\mu$ obtained by inserting a proper 4th-order 
vacuum-polarization $\Pi^{(4)}$ in the second order $M_2$ gives
\cite{Kinoshita:2004wi,Kallen:1955fb}
\begin{eqnarray}
a_\mu^{(6)}[\text{num.}] &=& 1.493~671~581~(8), \nonumber \\
a_\mu^{(6)}[\text{asym.}] &=& 1.517~3 + \mathrm{O}(m_e/m_\mu),
\label{amu6comp}
\end{eqnarray}
where the overall factor $(\alpha/\pi)^3$ is omitted for simplicity.

For the 8th-order $a_\mu$ obtained by inserting a proper 6th-order 
vacuum-polarization $\Pi^{(6)}$ in the second order $M_2$ gives
the coefficients of  $(\alpha/\pi)^4$: 
\begin{eqnarray}
a_\mu^{(8)}[\text{num.}] &=& -0.230~596~(416),   \nonumber \\
a_\mu^{(8)}[\text{Pade}] &=& -0.230~362~(5),     \nonumber \\
a_\mu^{(8)}[\text{asym.}] & = & -0.290~987 + \mathrm{O}(m_e/m_\mu),    \label{amu8comp}
\end{eqnarray}
where the numerical evaluation $a_\mu^{(8)}[\text{num.}]$ is from \cite{
Kinoshita:1998jg,Kinoshita:2005zr},
the Pad{\'e} approximation $a_\mu^{(8)}[\text{Pade}]$ is 
from \cite{Baikov:1995ui},
and the asymptotic result $a_\mu^{(8)}[\text{asym.}]$ 
is from \cite{Broadhurst:1992za}. Note that the asymptotic result of (\ref{amu8comp}) contains the leading logarithmic and next-to-leading constant terms.

The difference between the numerical and asymptotic results come from the
contribution of order $m_e/m_\mu$. From the sixth-order (\ref{amu6comp}), 
eighth-order (\ref{amu8comp}), and tenth-order cases  we find 
\begin{eqnarray}
&& a_\mu^{(6)}[\text{num.}] - a_\mu^{(6)}[\text{asym.}] =-0.024,
\nonumber
\\
&& a_\mu^{(8)}[\text{num.}] - a_\mu^{(8)}[\text{asym.}] = 0.061,
\nonumber
\\
&& a_\mu^{(10)}[\text{num.}] - a_\mu^{(10)}[\text{asym.}] = 0.17~.
\end{eqnarray}

Note that the difference increases as the order of perturbation increases.
Nevertheless, we cannot exclude the possibility that 
the difference between (\ref{amu1i_me}) and (\ref{amu1i_me_asym}) 
is caused by some error.
One possible cause is that the uncertainty of (\ref{amu1i_me})
is gross underestimate because of insufficient data sampling.
The situation might be similar to the case of
$a_\mu^{(8)}[\text{num.}]$ in early calculations
\cite{Kinoshita:1990wp,Kinoshita:1993pq,Baikov:1995ui}
where poor sampling of integrands 
 was found to be the cause
of large discrepancy with the Pad{\'e} result. 
This problem was finally settled by going to a much larger sampling, 
which led to (\ref{amu8comp}).

In order to examine the possibility of gross underestimate of errors 
in (\ref{amu1i_me})
we evaluated the integrals with the sampling points per iteration ${\cal N}$
of $10^8$, $10^9$, and even with $10^{10}$.
The results show no sign of the central values drifting beyond
the error bars estimated by VEGAS as ${\cal N}$ increases.
We are therefore confident that our result (\ref{amu1i_me}) 
is free from the problem caused by insufficient samplings.


\begin{acknowledgments}

This work is supported in part by the JSPS Grant-in-Aid for Scientific Research
(C)19540322 and (C)20540261.
T. K.'s work is supported in part
by the U. S. National Science Foundation under Grant NSF-PHY-0757868,
and the International Exchange Support Grants (FY2010) of RIKEN.
T. K. thanks RIKEN for the hospitality extended to him
while a part of this work was carried out.
Numerical calculations  are conducted in part 
on the RIKEN Super Combined Cluster System (RSCC)
and the RIKEN Integrated Cluster of Clusters (RICC)
supercomputing systems.
\end{acknowledgments}


\appendix





\vspace*{6mm}


%
\begin{figure}
\subfigure[$\Pi_{4a}$]{%
  \includegraphics[scale=0.35]{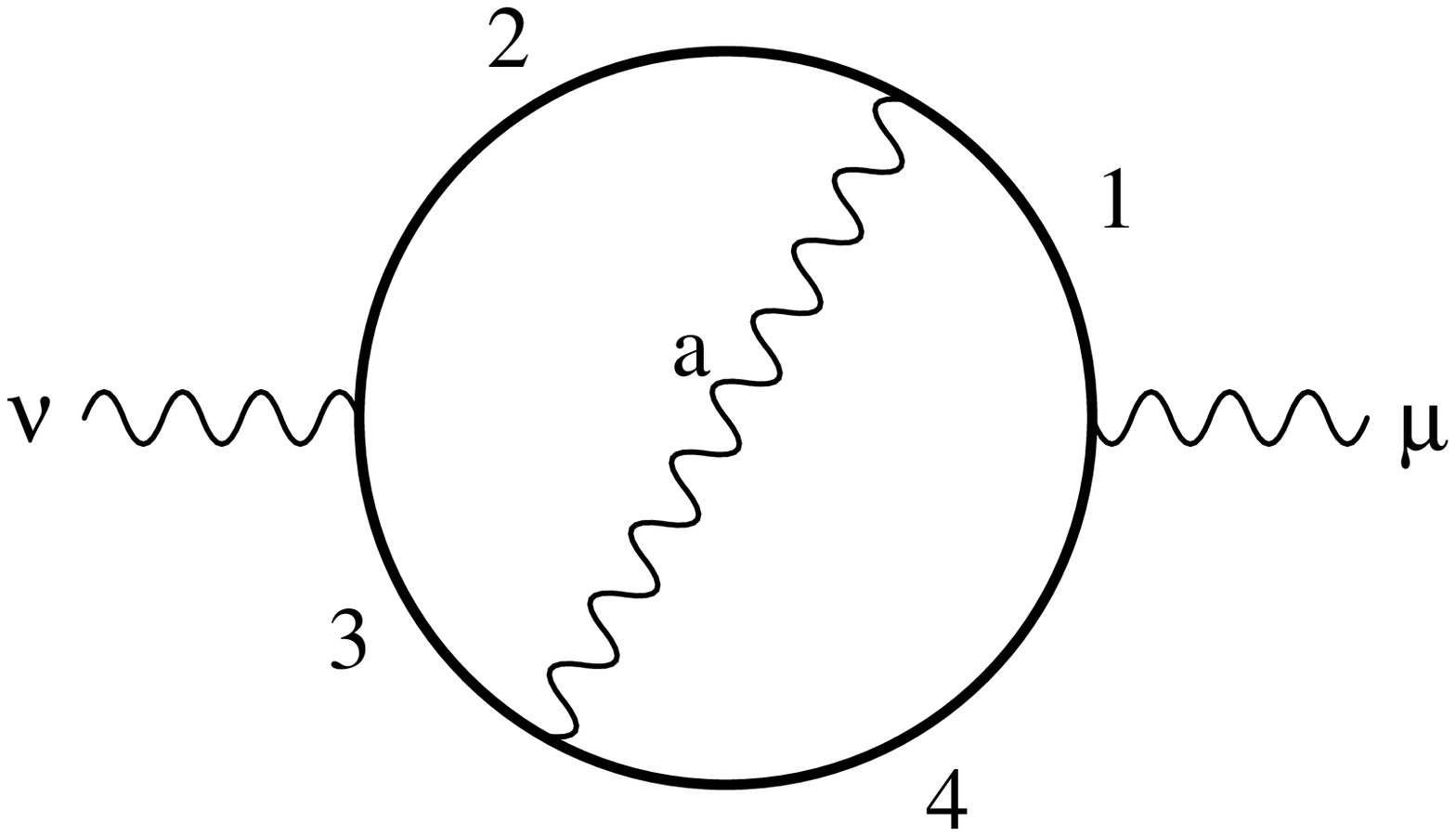}
\label{fig:p4:p4a}
}
\hspace{3em}
\subfigure[$\Pi_{4b}$]{%
  \includegraphics[scale=0.35]{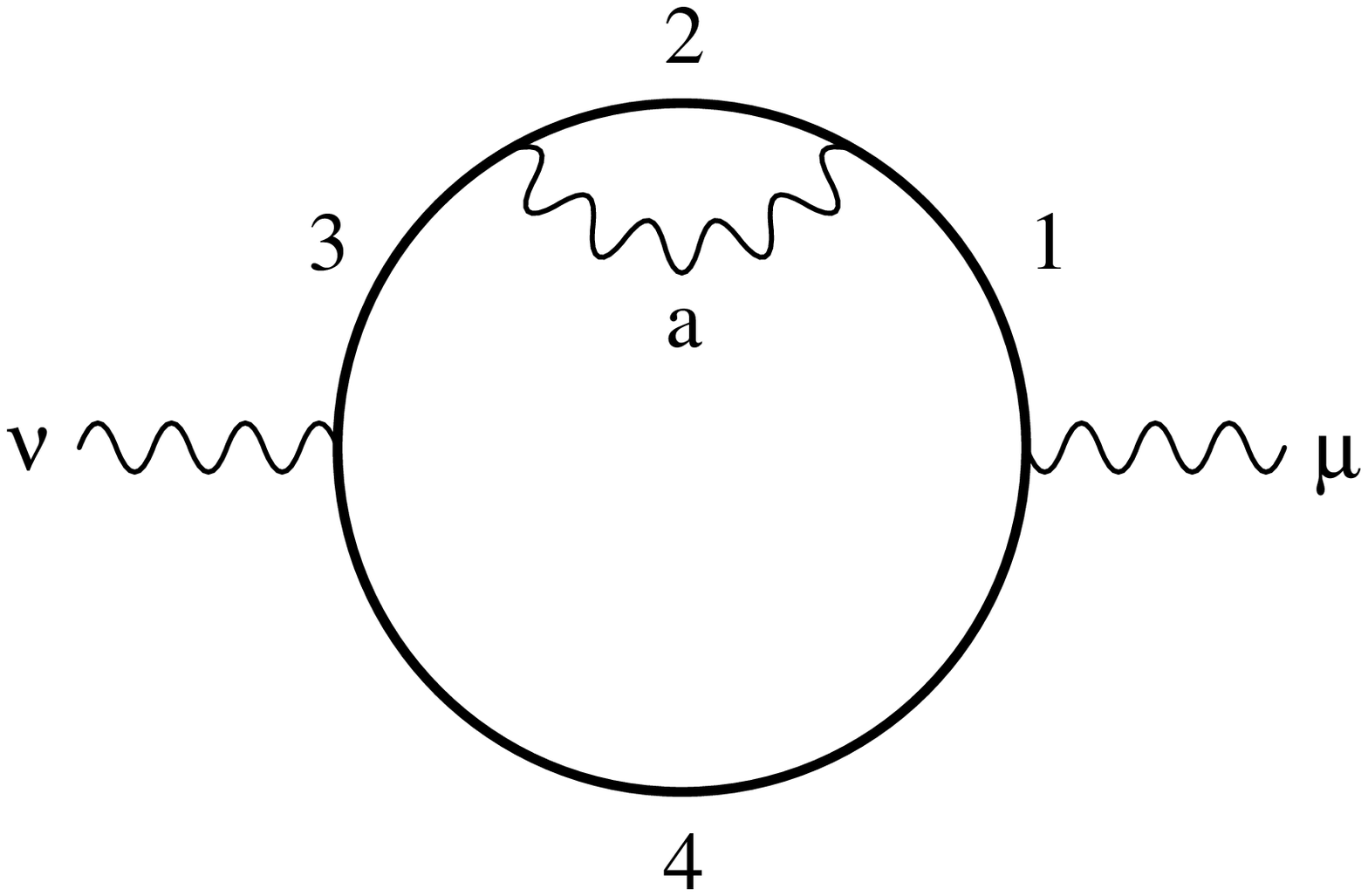}
\label{fig:p4:p4b}
}
\caption{%
Fourth-order vacuum polarization diagrams. 
}
\label{fig:p4}
\end{figure}
\begin{figure}
\subfigure[$\Pi_{4a,1^*}$]{%
  \includegraphics[scale=0.21]{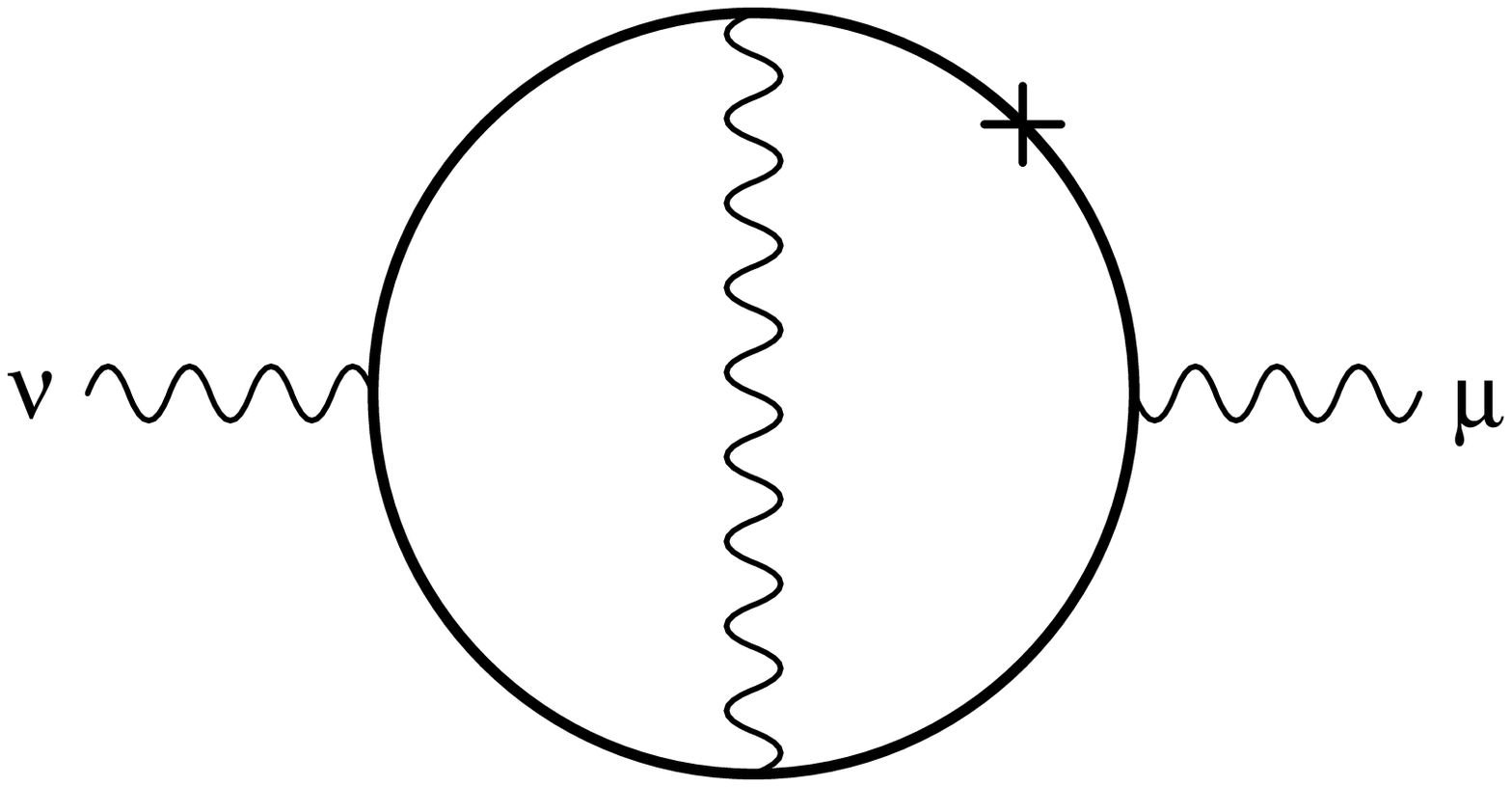}
\label{fig:p4s:p4a1s}
}
\hspace{0.5em}
\subfigure[$\Pi_{4b,1^*}$]{%
  \includegraphics[scale=0.21]{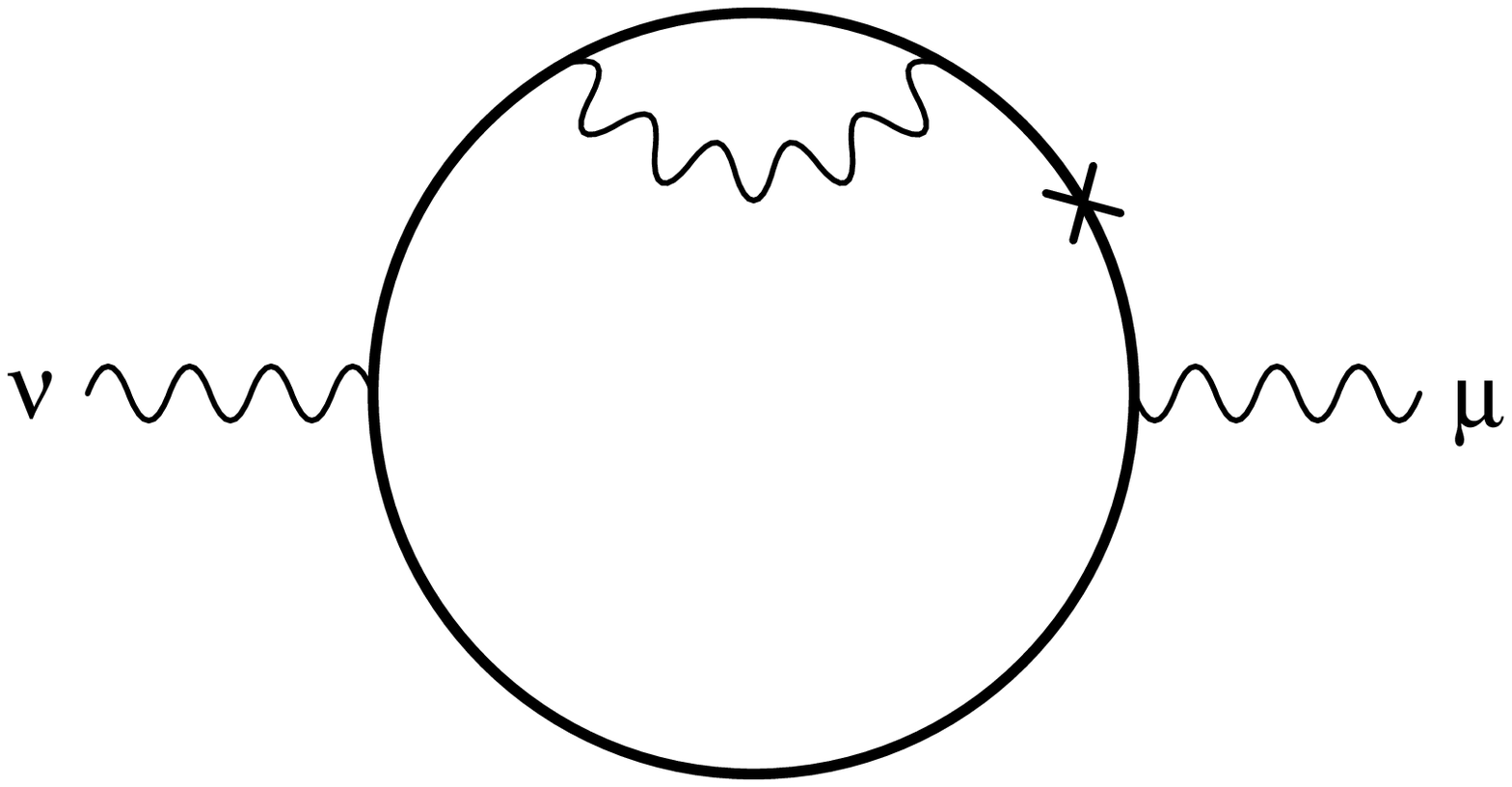}
\label{fig:p4s:p4b1s}
}
\hspace{0.5em}
\subfigure[$\Pi_{4b,2^*}$]{%
  \includegraphics[scale=0.21]{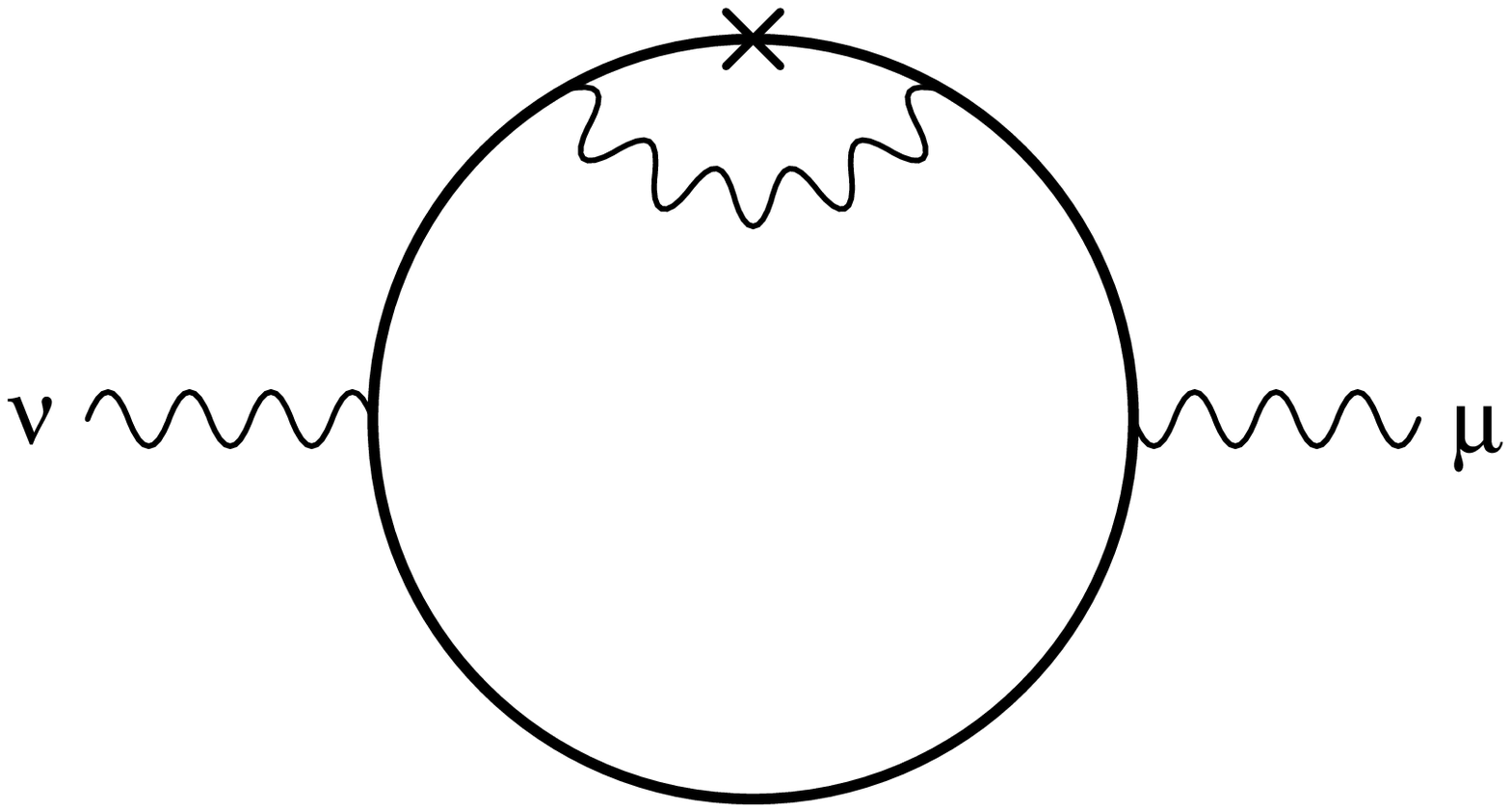}
\label{fig:p4s:p4b2s}
}
\hspace{0.5em}
\subfigure[$\Pi_{4b,4^*}$]{%
  \includegraphics[scale=0.21]{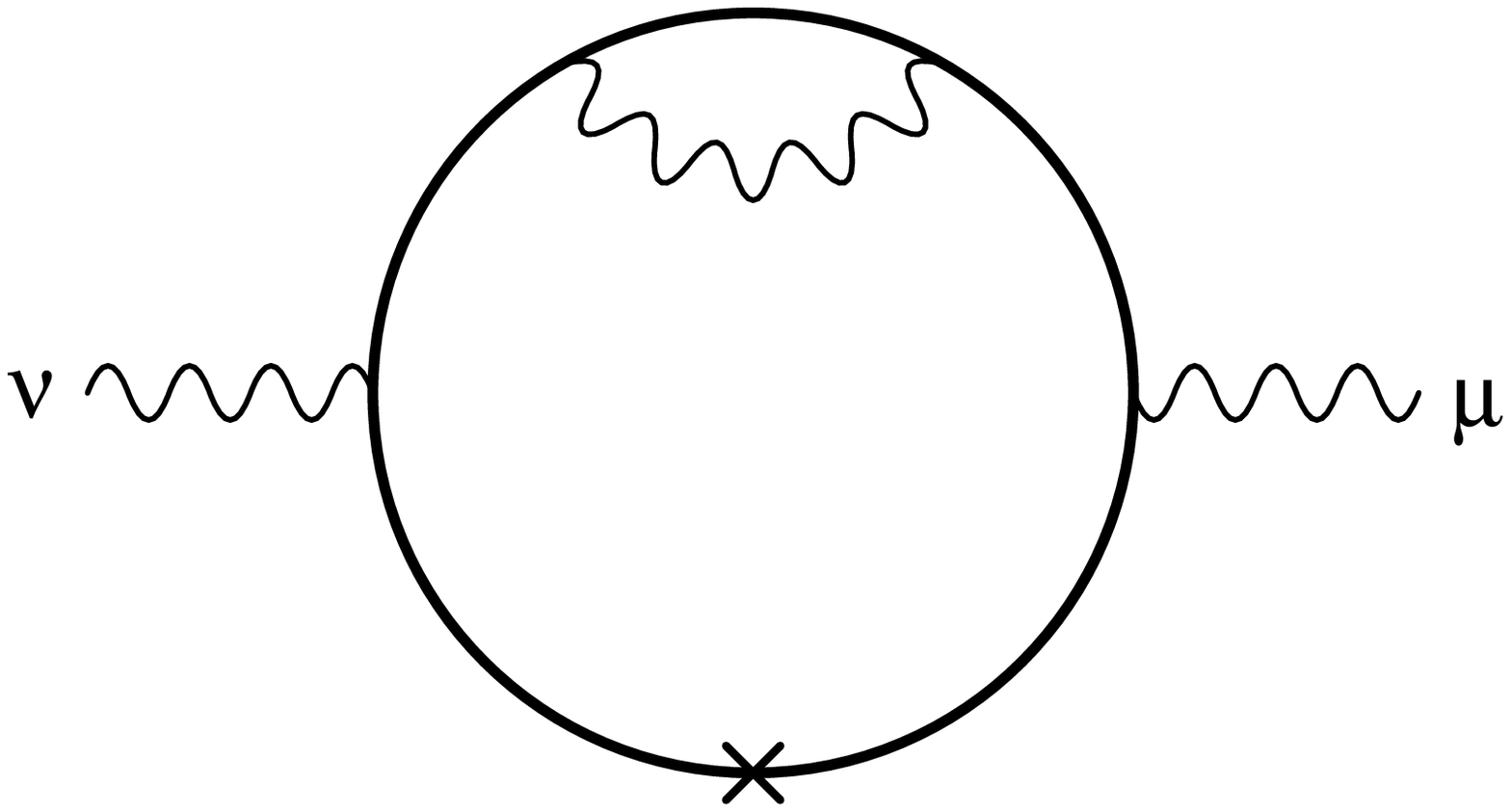}
\label{fig:p4s:p4b4s}
}
\caption{%
Fourth-order vacuum polarization diagrams with mass insertions. 
}
\label{fig:p4s}
\end{figure}
%

\section{Fourth-order vacuum-polarization functions with mass insertion
and their contribution to $M_{2,P_{4^*}}$ }
\label{sec:app:p4s}

Since $M_{2,\Delta P_{4^*}}$ does not appear except in Set~I(i) in our study of
the tenth-order $g\!-\!2$, we shall derive parametric formulas for the diagrams
$\Pi_{4a^*}$, $\Pi_{4b^*}$, and $M_{2,P_{4^*}}$ in this Appendix.

Our derivation follows closely the treatment of
diagrams $\Pi_{4a}^{\mu \nu}$ and $\Pi_{4b}^{\mu \nu}$ 
(without mass insertion)
which consist of four lepton lines forming a closed loop
and an internal photon line $a$ as shown in Fig.~\ref{fig:p4}.
Following the steps leading to Eq.~(\ref{eq:amp3}) of Sec.~\ref{sec:photonselfenergy} we obtain
\begin{eqnarray}
  \Pi_{4^*}^{\mu \nu} &=& 
  + 2 \left( \frac{-1}{4} \right )^2
  \int \frac{dz)}{U^2} \left( \sum_j \pm z_j D_j^\mu \right)
  \Tr\left[
    \qslash (\Dslash_i +m) \ldots \gamma^\nu \ldots \right] \frac{1}{V^3},
\end{eqnarray}
From Lorentz invariance and gauge invariance, we have the general structure
\begin{eqnarray}
\Pi_{4^*}^{\mu \nu} &=& (q_\mu q_\nu - q^2 g_{\mu \nu}) {\widetilde \Pi}_{4^*}
+ (\text{gauge-dependent terms}).
\end{eqnarray}
Charge renormalization is achieved by
\begin{eqnarray}
\Pi_{4^*} (q^2) &=& {\widetilde \Pi}_{4^*}(q^2) - {\widetilde \Pi}_{4^*}(0).
\end{eqnarray}

When the \textit{D}-operation is carried out in ${\widetilde \Pi}_{4^*}$,
the result can be expressed in the form
\begin{eqnarray}
  \Pi_{4^*}(q^2) &=& \int (dz) \left[
    \frac{D_0}{U^2} \left ( \frac{1}{V^2}- \frac{1}{V_0^2} \right)
    + \frac{q^2 B_0}{V^2} + \frac{D_1}{U^3} \left( \frac{1}{V} -\frac{1}{V_0}\right)
    \right]
\end{eqnarray}
where $V_0 = z_{1234} m^2$ and $D_0, B_0, D_1, U, V$ and $(dz)$ 
 are diagram-specific.

\subsection{Diagram $M_{2,P_{4a^*}}$ }

This diagram has 
four fermion lines into which mass vertex can be inserted.
They all give the same contribution to $M_{2,P_{4a^*}}$ so that we have
to evaluate only one of them such as $\Pi_{4a,1^*}$
of Fig.~\ref{fig:p4s:p4a1s}.
For this diagram we find
\begin{eqnarray}
D_0 &=& r^4 (-2 A_3 A_4 -2 A_2 A_4 + A_2 A_3 +2 A_1 A_4 - A_1 A_3 -A_1 A_2),          \nonumber \\
B_0 &=& r^2 A_4^2 (A_2 A_3 -A_1 A_3 -A_1 A_2),          \nonumber \\
D_1 &=& r^2 B_{11} (A_3 A_4 + A_2 A_4 +2 A_1 A_3 +2 A_1 A_2) \nonumber \\       
    &+& r^2 B_{12} ( -3 A_3 A_4 -A_2 A_4 +2 A_1 A_4 -2 A_1 A_2) \nonumber\\
    &+& r^2 B_{22} ( 2 A_1 A_4 ) 
\end{eqnarray}
where $r$ is the ratio of the mass $m$ of the loop lepton and the mass of 
the lepton of $M_2$ and
\begin{eqnarray}
A_1 &=& 1-z_1 B_{11}-z_2 B_{12},~~~ A_2 = 1-z_1 B_{12}-z_2 B_{22}, \nonumber \\ 
A_3 &=& -z_1 B_{13}-z_2 B_{23},~~~ A_4 = -z_1 B_{14}-z_2 B_{24}, \nonumber \\ 
B_{11} &=&  z_{23a},~~~B_{12} = z_{a},~~~B_{22} = z_{14a}, \nonumber \\
U &=& z_{14a} z_{23}+ z_{14} z_a,~~~G = z_1 A_1 +z_2 A_2,~~~ V= V_0 -q^2 G , \nonumber \\
(dz) &=& z_1 dz_1 dz_2 dz_3 dz_4 dz_a \delta ( 1 - z_{1234a}),~~~ z_i \geq 0.
\end{eqnarray}

This diagram has a UV divergence from the subvertex \{2,3,a,\},
which can be isolated by the $K_{23}$ operation.
Subtraction of this term yields a UV-finite value 
\begin{eqnarray}
\Delta \Pi_{4a^*} &=& 4 (1-K_{23}) \Pi_{4a,1^*},
\end{eqnarray}
and a finite contribution $ M_{2,P_{4a^*}}$ to $g\!-\!2$.
By numerical integration we obtain the value $-0.066~907~(7)$
for the $(ee)$ case, and $-0.443~935~(69)$ for the $(me)$ case.

\subsection{Diagram $M_{2,P_{4b^*}}$ }

The diagrams $\Pi_{4b,1^*}$ and $\Pi_{4b,3^*}$ 
give identical contribution to $M_{2,P_{4b^*}}$, 
whereas the diagrams $\Pi_{4b,2^*}$ and $\Pi_{4b,4^*}$ 
must be treated separately.

For the diagrams $\Pi_{4b,1^*}$ 
(see Fig.~\ref{fig:p4s:p4b1s}) we find
\begin{eqnarray}
D_0 &=& r^4 (2 A_2 A_4 -6 A_1 A_4 ),          \nonumber \\
B_0 &=& r^2 (3 A_1^2 A_2 A_4 -2A_1^3 A_4),          \nonumber \\
D_1 &=& r^2 B_{11} (6 A_1 A_4 - A_1 A_2 +2 A_1^2) \nonumber \\       
    &+& r^2 B_{12} ( -9 A_1 A_4  -2 A_1^2),
\end{eqnarray}
where
\begin{eqnarray}
A_1 &=& 1-z_1 B_{11}-z_2 B_{12}-z_3 B_{13}, \nonumber \\
A_2 &=& 1-z_1 B_{12}-z_2 B_{22}-z_3 B_{23}, \nonumber \\ 
A_3 &=& A_1, ~~~ A_4 = A_1 -1, \nonumber \\ 
B_{11} &=& B_{13}=B_{14}=B_{33}=B_{34}=B_{44}=z_{2a}, \nonumber \\
~~~B_{12} &=& B_{23}=B_{24}= z_{a}, ~~~B_{22} = z_{134a}, \nonumber \\
U &=& z_{134} z_{2a}+ z_{2} z_a,~~~G = z_1 A_1 +z_2 A_2+z_3 A_3,~~~ V= V_0 -q^2 G , \nonumber \\
(dz) &=& z_1 dz_1 dz_2 dz_3 dz_4 dz_a \delta ( 1 - z_{1234a}),~~~ z_i \geq 0.
\end{eqnarray}

This diagram has a UV-divergence from the self-energy 
subdiagram \{2,a\} which can be subtracted
by the $K_2$-operation
\begin{eqnarray}
\Delta \Pi_{4b,1^*} &=&  (1-K_{2}) \Pi_{4b,1^*}.
\end{eqnarray}

For the diagrams $\Pi_{4b,2^*}$ 
(see Fig.~\ref{fig:p4s:p4b2s}) we find
\begin{eqnarray}
D_0 &=& r^4 (2 A_2 A_4 -4 A_1 A_4 ),          \nonumber \\
B_0 &=& r^2 (-4 A_1 A_2^2 A_4 +2A_1^2 A_2 A_4),          \nonumber \\
D_1 &=& r^2 B_{22} (8 A_1 A_4 ) \nonumber \\       
    &+& r^2 B_{11} (2 A_2 A_4 ) \nonumber \\       
    &+& r^2 B_{12} (4 A_2 A_4  -8 A_1 A_4 +4 A_1 A_2 - 2 A_1^2),
\end{eqnarray}
where
\begin{eqnarray}
A_1 &=& 1-z_1 B_{11}-z_2 B_{12}-z_3 B_{13},
~~~ A_2 = 1-z_1 B_{12}-z_2 B_{22}-z_3 B_{23}, \nonumber \\ 
~~~ A_3 &=& 1-z_1 B_{13}-z_2 B_{23}-z_3 B_{33}, \nonumber \\ 
~~~ A_4 &=&  -z_1 B_{14}-z_2 B_{24}-z_3 B_{34}, \nonumber \\ 
B_{11} &=& B_{13}=B_{14}=B_{33}=B_{34}=B_{44}=z_{2a}, \nonumber \\
~~~B_{12} &=& B_{23}=B_{24}= z_{a}, ~~~B_{22} = z_{134a}, \nonumber \\
U &=& z_{134} z_{2a}+ z_{2} z_a,~~~G = z_1 A_1 +z_2 A_2+z_3 A_3,~~~V= V_0 -q^2 G , \nonumber \\
(dz) &=& z_2 dz_1 dz_2 dz_3 dz_4 dz_a \delta ( 1 - z_{1234a}),~~~ z_i \geq 0.
\end{eqnarray}

This diagram has a UV-divergence from the 
subdiagram \{$2^*$,a\} which can be subtracted
by the $K_{2^*}$-operation
\begin{eqnarray}
\Delta \Pi_{4b,2^*} &=&  (1-K_{2^*}) \Pi_{4b,2^*}.
\end{eqnarray}

For the diagrams $\Pi_{4b,4^*}$ 
(see Fig.~\ref{fig:p4s:p4b4s}) we find
\begin{eqnarray}
D_0 &=& r^4 (2 A_2 A_4 -8 A_1 A_4 ),          \nonumber \\
B_0 &=& r^2 (2 A_1^2 A_2 A_4 ),          \nonumber \\
D_1 &=& r^2 B_{11} (2 A_2 A_4 ) \nonumber \\       
    &+& r^2 B_{12} ( -8 A_1 A_4  -2 A_1^2),
\end{eqnarray}
where
\begin{eqnarray}
A_1 &=& 1-z_1 B_{11}-z_2 B_{12}-z_3 B_{13},
~~~ A_2 = 1-z_1 B_{12}-z_2 B_{22}-z_3 B_{23}, \nonumber \\ 
~~~ A_3 &=& 1-z_1 B_{13}-z_2 B_{23}-z_3 B_{33}, \nonumber \\ 
~~~ A_4 &=&  -z_1 B_{14}-z_2 B_{24}-z_3 B_{34}, \nonumber \\ 
B_{11} &=& B_{13}=B_{14}=B_{33}=B_{34}=B_{44}=z_{2a},  \nonumber \\
~~~B_{12} &=&B_{23}=B_{24}= z_{a}, ~~~B_{22} = z_{134a}, \nonumber \\
U &=& z_{134} z_{2a}+ z_{2} z_a,~~~G = z_1 A_1 +z_2 A_2+z_3 A_3,~~~ V= V_0 -q^2 G , \nonumber \\
(dz) &=& z_4 dz_1 dz_2 dz_3 dz_4 dz_a \delta ( 1 - z_{1234a}),~~~ z_i \geq 0.
\end{eqnarray}

As is for the diagram $\Pi_{4b,1^*}$ this diagram has a UV-divergence 
from the self-energy subdiagram \{2,a\} which can be subtracted
by the $K_2$-operation.

By numerical integration we obtain for the $(ee)$ case
\begin{eqnarray}
2  M_{2,\Delta \! P_{4b^*}}^{(ee)} &=& -0.046~309~(7) + 0.033~507~(4) -0.038~061~(6) 
 \nonumber \\
                   &=& -0.050~863~(9) ,
\end{eqnarray}
where the right-hand-side of the first line is listed  in order 
of $P_{4b,1^*}$, $P_{4b,2^*}$, and $P_{4b,4^*}$.


The result for the $(me)$ case is
\begin{eqnarray}
 2 M_{2, \Delta \! P_{4b^*}}^{(me)}& =& -0.409~550~(61) + 0.528~759~(51) -0.429~683~(62) \nonumber \\
                           & =&  -0.310~464~(101).
\end{eqnarray}

The sums $ M_{2,\Delta \!  P_{4^*}} 
       \equiv  M_{2, \Delta \!P_{4a^*}}+  2 M_{2, \Delta \! P_{4b^*}}$
for the $(ee)$, $(em)$, $(me)$ and $(mt)$ cases  are listed in Table \ref{table:renom}.





\section{Standard on-the-mass-shell renormalization}
\label{sec:app:onshell}

This Appendix describes the standard on-the-mass-shell renormalization
of vacuum-polarization function $\Pi^{(n)}$, where $n = 2, 4, 6, 8$.
$\Pi^{(2)}$ consists of only one diagram, but higher order functions
consist of several diagrams, which must be distinguished by an additional
symbol.  For instance $\Pi^{(4i)}$ with $i=a, b$,
$\Pi^{(6j)}$ with $j=A, B, ..., H$.
However, the eighth-order functions are denoted as $\Pi^{(k)}, 
k=p01, p02,...,p39$ to avoid overcrowding.

Renormalization terms include functions such as $\Pi^{(2^*)}$
which means insertion of a two-point vertex (such as a mass vertex)
in the fermion line.

Quantities $L_n, B_n, \delta m_n$ denote vertex renormalization constant,
wave function renormalization constant, mass renormalization constant
of $n$-th order of the standard on-the-mass-shell renormalization, respectively.
We must also deal with renormalization constants with mass insertion.
For instance $L_2$, which contains two electron lines, 
it is necessary to distinguish the lines into which two-point vertex
insertion is made.  Suppose we name them line~1 and line~2.
Then $L_{2(1^* 1^*)}$ implies that two two-point vertices are inserted
in the fermion line~1 of $L_2$, while $L_{2(1^* 2^*)}$ means that
one two-point vertex is inserted in line~1 while another is inserted in line~2.
(Previously
 \cite{Kinoshita:1990}
 we used the notations $L_{2^{**\dagger}}$ and $L_{2^{*\dagger *}}$
for these quantities.)

\subsection{Standard renormalization of fourth-order vacuum-polarization}
%
%
\begin{align*}
\Pi_{\rm ren}^{(4a)}&=
\Pi^{(4a)}-2 L_{2} \Pi^{(2)}
\end{align*}
%
%
\begin{align*}
\Pi_{\rm ren}^{(4b)}&=
\Pi^{(4b)}-{\delta m}_{2} \Pi^{(2\ast)}-B_{2} \Pi^{(2)}
\end{align*}

\subsection{Standard renormalization of sixth-order vacuum-polarization}
%
%
\begin{align*}
\Pi_{\rm ren}^{(6A)}&=
\Pi^{(6A)}+2 {\delta m}_{2} B_{2} \Pi^{(2\ast)}-2 {\delta m}_{2} \Pi^{(4b,1\ast)}+({\delta m}_{2})^{2} \Pi^{(2\ast\ast)}-2 B_{2} \Pi^{(4b)} +(B_{2})^{2} \Pi^{(2)}
\end{align*}
%
%
\begin{align*}
\Pi_{\rm ren}^{(6B)}&=
\Pi^{(6B)}+2 {\delta m}_{2} B_{2} \Pi^{(2\ast)}-2 {\delta m}_{2} \Pi^{(4b,4\ast)}+({\delta m}_{2})^{2} \Pi^{(2\ast\ast)}-2 B_{2} \Pi^{(4b)} +(B_{2})^{2} \Pi^{(2)}
\end{align*}
%
%
\begin{align*}
\Pi_{\rm ren}^{(6C)}&=
\Pi^{(6C)}-{\delta m}_{4b} \Pi^{(2\ast)}+{\delta m}_{2\ast} {\delta m}_{2} \Pi^{(2\ast)}+{\delta m}_{2} B_{2\ast} \Pi^{(2)}+{\delta m}_{2} B_{2} \Pi^{(2\ast)} \\
& -{\delta m}_{2} \Pi^{(4b,2\ast)}-B_{4b} \Pi^{(2)}-B_{2} \Pi^{(4b)}+(B_{2})^{2} \Pi^{(2)}
\end{align*}
%
%
\begin{align*}
\Pi_{\rm ren}^{(6D)}&=
\Pi^{(6D)}-{\delta m}_{4a} \Pi^{(2\ast)}+2 {\delta m}_{2} L_{2} \Pi^{(2\ast)}-B_{4a} \Pi^{(2)}+2 B_{2} L_{2} \Pi^{(2)} -2 L_{2} \Pi^{(4b)}
\end{align*}
%
%
\begin{align*}
\Pi_{\rm ren}^{(6E)}&=
\Pi^{(6E)}+{\delta m}_{2} L_{2\ast} \Pi^{(2)}+{\delta m}_{2} L_{2} \Pi^{(2\ast)}-{\delta m}_{2} \Pi^{(4a,1\ast)}+2 B_{2} L_{2} \Pi^{(2)} \\
& -B_{2} \Pi^{(4a)}-L_{4b,1} \Pi^{(2)}-L_{2} \Pi^{(4b)}
\end{align*}
%
%
\begin{align*}
\Pi_{\rm ren}^{(6F)}&=
\Pi^{(6F)}-2 L_{4a,1} \Pi^{(2)}-L_{2} \Pi^{(4a)}+2 (L_{2})^{2} \Pi^{(2)}
\end{align*}
%
%
\begin{align*}
\Pi_{\rm ren}^{(6G)}&=
\Pi^{(6G)}-2 L_{4b,2} \Pi^{(2)}-2 L_{2} \Pi^{(4a)}+3 (L_{2})^{2} \Pi^{(2)}
\end{align*}
%
%
\begin{align*}
\Pi_{\rm ren}^{(6H)}&=
\Pi^{(6H)}-2 L_{4a,2} \Pi^{(2)}
\end{align*}

\subsection{Standard renormalization of eighth-order vacuum-polarization}
%
%
\begin{align*}
\Pi_{\rm ren}^{(p01)}&=
\Pi^{(p01)}-2 L_{6F,3} \Pi^{(2)}+4 L_{4a,1} L_{2} \Pi^{(2)}-2 L_{2} \Pi^{(6F)}+(L_{2})^{2} \Pi^{(4a)} -2 (L_{2})^{3} \Pi^{(2)}
\end{align*}
%
%
\begin{align*}
\Pi_{\rm ren}^{(p02)}&=
\Pi^{(p02)}-2 L_{6H,3} \Pi^{(2)}
\end{align*}
%
%
\begin{align*}
\Pi_{\rm ren}^{(p03)}&=
\Pi^{(p03)}-2 L_{6B,3} \Pi^{(2)}+6 L_{4b,2} L_{2} \Pi^{(2)}-2 L_{4b,2} \Pi^{(4a)}-2 L_{2} \Pi^{(6G)} \\
& +3 (L_{2})^{2} \Pi^{(4a)}-4 (L_{2})^{3} \Pi^{(2)}
\end{align*}
%
%
\begin{align*}
\Pi_{\rm ren}^{(p04)}&=
\Pi^{(p04)}+6 {\delta m}_{2} B_{2} \Pi^{(4b,1\ast)}-3 {\delta m}_{2} (B_{2})^{2} \Pi^{(2\ast)}-{\delta m}_{2} \Pi^{(6A,3\ast)}-2 {\delta m}_{2} \Pi^{(6A,1\ast)} \\
& -3 ({\delta m}_{2})^{2} B_{2} \Pi^{(2\ast\ast)}+({\delta m}_{2})^{2} \Pi^{(4b,1\ast3\ast)}+2 ({\delta m}_{2})^{2} \Pi^{(4b,1\ast1\ast)}-({\delta m}_{2})^{3} \Pi^{(2\ast\ast\ast)}-3 B_{2} \Pi^{(6A)} \\
& +3 (B_{2})^{2} \Pi^{(4b)}-(B_{2})^{3} \Pi^{(2)}
\end{align*}
%
%
\begin{align*}
\Pi_{\rm ren}^{(p05)}&=
\Pi^{(p05)}+4 {\delta m}_{2} B_{2} \Pi^{(4b,4\ast)}+2 {\delta m}_{2} B_{2} \Pi^{(4b,1\ast)}-3 {\delta m}_{2} (B_{2})^{2} \Pi^{(2\ast)}-2 {\delta m}_{2} \Pi^{(6B,1\ast)} \\
& -{\delta m}_{2} \Pi^{(6A,6\ast)}-2 ({\delta m}_{2})^{2} B_{2} \Pi^{(2\ast\ast)}-({\delta m}_{2})^{2} B_{2} \Pi^{(2\ast\ast)}+({\delta m}_{2})^{2} \Pi^{(4b,4\ast4\ast)}+2 ({\delta m}_{2})^{2} \Pi^{(4b,1\ast4\ast)} \\
& -({\delta m}_{2})^{3} \Pi^{(2\ast\ast\ast)}-2 B_{2} \Pi^{(6B)}-B_{2} \Pi^{(6A)}+3 (B_{2})^{2} \Pi^{(4b)}-(B_{2})^{3} \Pi^{(2)}
\end{align*}
%
%
\begin{align*}
\Pi_{\rm ren}^{(p06)}&=
\Pi^{(p06)}-2 {\delta m}_{2} B_{2} L_{2\ast} \Pi^{(2)}-2 {\delta m}_{2} B_{2} L_{2} \Pi^{(2\ast)}+2 {\delta m}_{2} B_{2} \Pi^{(4a,1\ast)}+2 {\delta m}_{2} L_{4b,1} \Pi^{(2\ast)} \\
& +2 {\delta m}_{2} L_{2\ast} \Pi^{(4b)}-2 {\delta m}_{2} \Pi^{(6E,1\ast)}-2 ({\delta m}_{2})^{2} L_{2\ast} \Pi^{(2\ast)}+({\delta m}_{2})^{2} \Pi^{(4a,1\ast2\ast)}+2 B_{2} L_{4b,1} \Pi^{(2)} \\
& +2 B_{2} L_{2} \Pi^{(4b)}-2 B_{2} \Pi^{(6E)}-2 (B_{2})^{2} L_{2} \Pi^{(2)}+(B_{2})^{2} \Pi^{(4a)}-2 L_{4b,1} \Pi^{(4b)}
\end{align*}
%
%
\begin{align*}
\Pi_{\rm ren}^{(p07)}&=
\Pi^{(p07)}-2 {\delta m}_{2} B_{2} L_{2\ast} \Pi^{(2)}-2 {\delta m}_{2} B_{2} L_{2} \Pi^{(2\ast)}+2 {\delta m}_{2} B_{2} \Pi^{(4a,1\ast)}+2 {\delta m}_{2} L_{4b,1(1\ast)} \Pi^{(2)} \\
& +2 {\delta m}_{2} L_{2} \Pi^{(4b,4\ast)}-2 {\delta m}_{2} \Pi^{(6E,5\ast)}-({\delta m}_{2})^{2} L_{2(1\ast2\ast)} \Pi^{(2)}-({\delta m}_{2})^{2} L_{2} \Pi^{(2\ast\ast)}+({\delta m}_{2})^{2} \Pi^{(4a,1\ast4\ast)} \\
& +2 B_{2} L_{4b,1} \Pi^{(2)}+2 B_{2} L_{2} \Pi^{(4b)}-2 B_{2} \Pi^{(6E)}-2 (B_{2})^{2} L_{2} \Pi^{(2)}+(B_{2})^{2} \Pi^{(4a)} \\
& -L_{6A,3} \Pi^{(2)}-L_{2} \Pi^{(6B)}
\end{align*}
%
%
\begin{align*}
\Pi_{\rm ren}^{(p08)}&=
\Pi^{(p08)}-2 {\delta m}_{2} B_{2} L_{2\ast} \Pi^{(2)}-2 {\delta m}_{2} B_{2} L_{2} \Pi^{(2\ast)}+2 {\delta m}_{2} B_{2} \Pi^{(4a,1\ast)}+2 {\delta m}_{2} L_{4b,1} \Pi^{(2\ast)} \\
& +2 {\delta m}_{2} L_{2\ast} \Pi^{(4b)}-2 {\delta m}_{2} \Pi^{(6E,6\ast)}-2 ({\delta m}_{2})^{2} L_{2\ast} \Pi^{(2\ast)}+({\delta m}_{2})^{2} \Pi^{(4a,1\ast3\ast)}+2 B_{2} L_{4b,1} \Pi^{(2)} \\
& +2 B_{2} L_{2} \Pi^{(4b)}-2 B_{2} \Pi^{(6E)}-2 (B_{2})^{2} L_{2} \Pi^{(2)}+(B_{2})^{2} \Pi^{(4a)}-2 L_{4b,1} \Pi^{(4b)}
\end{align*}
%
%
\begin{align*}
\Pi_{\rm ren}^{(p09)}&=
\Pi^{(p09)}+{\delta m}_{4a} {\delta m}_{2} \Pi^{(2\ast\ast)}+{\delta m}_{4a} B_{2} \Pi^{(2\ast)}-{\delta m}_{4a} \Pi^{(4b,4\ast)}+{\delta m}_{2} B_{4a} \Pi^{(2\ast)} \\
& -4 {\delta m}_{2} B_{2} L_{2} \Pi^{(2\ast)}+4 {\delta m}_{2} L_{2} \Pi^{(4b,4\ast)}-{\delta m}_{2} \Pi^{(6D,6\ast)}-2 ({\delta m}_{2})^{2} L_{2} \Pi^{(2\ast\ast)}+B_{4a} B_{2} \Pi^{(2)} \\
& -B_{4a} \Pi^{(4b)}+4 B_{2} L_{2} \Pi^{(4b)}-B_{2} \Pi^{(6D)}-2 (B_{2})^{2} L_{2} \Pi^{(2)}-2 L_{2} \Pi^{(6B)}
\end{align*}
%
%
\begin{align*}
\Pi_{\rm ren}^{(p10)}&=
\Pi^{(p10)}+{\delta m}_{4b} {\delta m}_{2} \Pi^{(2\ast\ast)}+{\delta m}_{4b} B_{2} \Pi^{(2\ast)}-{\delta m}_{4b} \Pi^{(4b,4\ast)}-{\delta m}_{2\ast} {\delta m}_{2} B_{2} \Pi^{(2\ast)} \\
& +{\delta m}_{2\ast} {\delta m}_{2} \Pi^{(4b,4\ast)}-{\delta m}_{2\ast} ({\delta m}_{2})^{2} \Pi^{(2\ast\ast)}+{\delta m}_{2} B_{4b} \Pi^{(2\ast)}-{\delta m}_{2} B_{2\ast} B_{2} \Pi^{(2)}+{\delta m}_{2} B_{2\ast} \Pi^{(4b)} \\
& +2 {\delta m}_{2} B_{2} \Pi^{(4b,4\ast)}+{\delta m}_{2} B_{2} \Pi^{(4b,2\ast)}-2 {\delta m}_{2} (B_{2})^{2} \Pi^{(2\ast)}-{\delta m}_{2} \Pi^{(6C,6\ast)}-{\delta m}_{2} \Pi^{(6B,2\ast)} \\
& -({\delta m}_{2})^{2} B_{2\ast} \Pi^{(2\ast)}-({\delta m}_{2})^{2} B_{2} \Pi^{(2\ast\ast)}+({\delta m}_{2})^{2} \Pi^{(4b,2\ast4\ast)}+B_{4b} B_{2} \Pi^{(2)}-B_{4b} \Pi^{(4b)} \\
& -B_{2} \Pi^{(6C)}-B_{2} \Pi^{(6B)}+2 (B_{2})^{2} \Pi^{(4b)}-(B_{2})^{3} \Pi^{(2)}
\end{align*}
%
%
\begin{align*}
\Pi_{\rm ren}^{(p11)}&=
\Pi^{(p11)}-{\delta m}_{6F} \Pi^{(2\ast)}+2 {\delta m}_{4a} L_{2} \Pi^{(2\ast)}+2 {\delta m}_{2} L_{4a,1} \Pi^{(2\ast)}-3 {\delta m}_{2} (L_{2})^{2} \Pi^{(2\ast)} \\
& -B_{6F} \Pi^{(2)}+2 B_{4a} L_{2} \Pi^{(2)}+2 B_{2} L_{4a,1} \Pi^{(2)}-3 B_{2} (L_{2})^{2} \Pi^{(2)}-2 L_{4a,1} \Pi^{(4b)} \\
& -2 L_{2} \Pi^{(6D)}+3 (L_{2})^{2} \Pi^{(4b)}
\end{align*}
%
%
\begin{align*}
\Pi_{\rm ren}^{(p12)}&=
\Pi^{(p12)}-{\delta m}_{6A} \Pi^{(2\ast)}+2 {\delta m}_{4b(1\ast)} {\delta m}_{2} \Pi^{(2\ast)}+2 {\delta m}_{4b} B_{2} \Pi^{(2\ast)}-{\delta m}_{2\ast\ast} ({\delta m}_{2})^{2} \Pi^{(2\ast)} \\
& -2 {\delta m}_{2\ast} {\delta m}_{2} B_{2} \Pi^{(2\ast)}+2 {\delta m}_{2} B_{4b(1\ast)} \Pi^{(2)}-2 {\delta m}_{2} B_{2\ast} B_{2} \Pi^{(2)}+2 {\delta m}_{2} B_{2} \Pi^{(4b,2\ast)}-{\delta m}_{2} (B_{2})^{2} \Pi^{(2\ast)} \\
& -2 {\delta m}_{2} \Pi^{(6C,2\ast)}-({\delta m}_{2})^{2} B_{2\ast\ast} \Pi^{(2)}+({\delta m}_{2})^{2} \Pi^{(4b,2\ast2\ast)}-B_{6A} \Pi^{(2)}+2 B_{4b} B_{2} \Pi^{(2)} \\
& -2 B_{2} \Pi^{(6C)}+(B_{2})^{2} \Pi^{(4b)}-(B_{2})^{3} \Pi^{(2)}
\end{align*}
%
%
\begin{align*}
\Pi_{\rm ren}^{(p13)}&=
\Pi^{(p13)}+2 {\delta m}_{2} L_{4a,2(1\ast)} \Pi^{(2)}-{\delta m}_{2} \Pi^{(6H,2\ast)}+2 B_{2} L_{4a,2} \Pi^{(2)}-B_{2} \Pi^{(6H)} -2 L_{6D,4} \Pi^{(2)}
\end{align*}
%
%
\begin{align*}
\Pi_{\rm ren}^{(p14)}&=
\Pi^{(p14)}+2 {\delta m}_{2} L_{4b,2(1\ast)} \Pi^{(2)}-2 {\delta m}_{2} L_{2\ast} L_{2} \Pi^{(2)}+2 {\delta m}_{2} L_{2} \Pi^{(4a,1\ast)}-{\delta m}_{2} (L_{2})^{2} \Pi^{(2\ast)} \\
& -{\delta m}_{2} \Pi^{(6G,2\ast)}+2 B_{2} L_{4b,2} \Pi^{(2)}+2 B_{2} L_{2} \Pi^{(4a)}-3 B_{2} (L_{2})^{2} \Pi^{(2)}-B_{2} \Pi^{(6G)} \\
& -2 L_{6A,2} \Pi^{(2)}+2 L_{4b,1} L_{2} \Pi^{(2)}-2 L_{2} \Pi^{(6E)}+(L_{2})^{2} \Pi^{(4b)}
\end{align*}
%
%
\begin{align*}
\Pi_{\rm ren}^{(p15)}&=
\Pi^{(p15)}-{\delta m}_{6H} \Pi^{(2\ast)}+2 {\delta m}_{2} L_{4a,2} \Pi^{(2\ast)}-B_{6H} \Pi^{(2)}+2 B_{2} L_{4a,2} \Pi^{(2)} -2 L_{4a,2} \Pi^{(4b)}
\end{align*}
%
%
\begin{align*}
\Pi_{\rm ren}^{(p16)}&=
\Pi^{(p16)}-2 L_{6H,1} \Pi^{(2)}+2 L_{4a,2} L_{2} \Pi^{(2)}-L_{4a,2} \Pi^{(4a)}
\end{align*}
%
%
\begin{align*}
\Pi_{\rm ren}^{(p17)}&=
\Pi^{(p17)}-2 L_{6H,2} \Pi^{(2)}
\end{align*}
%
%
\begin{align*}
\Pi_{\rm ren}^{(p18)}&=
\Pi^{(p18)}-2 L_{6G,2} \Pi^{(2)}
\end{align*}
%
%
\begin{align*}
\Pi_{\rm ren}^{(p19)}&=
\Pi^{(p19)}-2 L_{6G,5} \Pi^{(2)}+2 L_{4b,2} L_{2} \Pi^{(2)}-L_{4b,2} \Pi^{(4a)}+2 L_{4a,1} L_{2} \Pi^{(2)} \\
& -L_{2} \Pi^{(6F)}+(L_{2})^{2} \Pi^{(4a)}-2 (L_{2})^{3} \Pi^{(2)}
\end{align*}
%
%
\begin{align*}
\Pi_{\rm ren}^{(p20)}&=
\Pi^{(p20)}-{\delta m}_{6C} \Pi^{(2\ast)}+2 {\delta m}_{4b} L_{2} \Pi^{(2\ast)}+{\delta m}_{4a} {\delta m}_{2\ast} \Pi^{(2\ast)}+{\delta m}_{4a} B_{2\ast} \Pi^{(2)} \\
& -{\delta m}_{4a} \Pi^{(4b,2\ast)}-2 {\delta m}_{2\ast} {\delta m}_{2} L_{2} \Pi^{(2\ast)}+{\delta m}_{2} B_{4a} \Pi^{(2\ast)}-2 {\delta m}_{2} B_{2\ast} L_{2} \Pi^{(2)}-2 {\delta m}_{2} B_{2} L_{2} \Pi^{(2\ast)} \\
& +2 {\delta m}_{2} L_{2} \Pi^{(4b,2\ast)}-B_{6C} \Pi^{(2)}+2 B_{4b} L_{2} \Pi^{(2)}+B_{4a} B_{2} \Pi^{(2)}-B_{4a} \Pi^{(4b)} \\
& +2 B_{2} L_{2} \Pi^{(4b)}-2 (B_{2})^{2} L_{2} \Pi^{(2)}-2 L_{2} \Pi^{(6C)}
\end{align*}
%
%
\begin{align*}
\Pi_{\rm ren}^{(p21)}&=
\Pi^{(p21)}-{\delta m}_{6E} \Pi^{(2\ast)}+{\delta m}_{4a(2\ast)} {\delta m}_{2} \Pi^{(2\ast)}+{\delta m}_{4a} B_{2} \Pi^{(2\ast)}+{\delta m}_{2} B_{4a(2\ast)} \Pi^{(2)} \\
& -2 {\delta m}_{2} B_{2} L_{2\ast} \Pi^{(2)}-2 {\delta m}_{2} B_{2} L_{2} \Pi^{(2\ast)}+2 {\delta m}_{2} L_{4b,1} \Pi^{(2\ast)}+2 {\delta m}_{2} L_{2\ast} \Pi^{(4b)}-{\delta m}_{2} \Pi^{(6D,3\ast)} \\
& -2 ({\delta m}_{2})^{2} L_{2\ast} \Pi^{(2\ast)}-B_{6E} \Pi^{(2)}+B_{4a} B_{2} \Pi^{(2)}+2 B_{2} L_{4b,1} \Pi^{(2)}+2 B_{2} L_{2} \Pi^{(4b)} \\
& -B_{2} \Pi^{(6D)}-2 (B_{2})^{2} L_{2} \Pi^{(2)}-2 L_{4b,1} \Pi^{(4b)}
\end{align*}
%
%
\begin{align*}
\Pi_{\rm ren}^{(p22)}&=
\Pi^{(p22)}-{\delta m}_{6B} \Pi^{(2\ast)}+{\delta m}_{4b(2\ast)} {\delta m}_{2} \Pi^{(2\ast)}+{\delta m}_{4b} {\delta m}_{2\ast} \Pi^{(2\ast)}+{\delta m}_{4b} B_{2\ast} \Pi^{(2)} \\
& +{\delta m}_{4b} B_{2} \Pi^{(2\ast)}-{\delta m}_{4b} \Pi^{(4b,2\ast)}-{\delta m}_{2\ast} {\delta m}_{2} B_{2\ast} \Pi^{(2)}-{\delta m}_{2\ast} {\delta m}_{2} B_{2} \Pi^{(2\ast)}+{\delta m}_{2\ast} {\delta m}_{2} \Pi^{(4b,2\ast)} \\
& -({\delta m}_{2\ast})^{2} {\delta m}_{2} \Pi^{(2\ast)}+{\delta m}_{2} B_{4b(2\ast)} \Pi^{(2)}+{\delta m}_{2} B_{4b} \Pi^{(2\ast)}-2 {\delta m}_{2} B_{2\ast} B_{2} \Pi^{(2)}+{\delta m}_{2} B_{2\ast} \Pi^{(4b)} \\
& +{\delta m}_{2} B_{2} \Pi^{(4b,2\ast)}-{\delta m}_{2} (B_{2})^{2} \Pi^{(2\ast)}-{\delta m}_{2} \Pi^{(6C,3\ast)}-({\delta m}_{2})^{2} B_{2\ast} \Pi^{(2\ast)}-B_{6B} \Pi^{(2)} \\
& +2 B_{4b} B_{2} \Pi^{(2)}-B_{4b} \Pi^{(4b)}-B_{2} \Pi^{(6C)}+(B_{2})^{2} \Pi^{(4b)}-(B_{2})^{3} \Pi^{(2)}
\end{align*}
%
%
\begin{align*}
\Pi_{\rm ren}^{(p23)}&=
\Pi^{(p23)}-2 L_{6G,3} \Pi^{(2)}
\end{align*}
%
%
\begin{align*}
\Pi_{\rm ren}^{(p24)}&=
\Pi^{(p24)}-L_{6E,3} \Pi^{(2)}-L_{6C,3} \Pi^{(2)}+3 L_{4a,2} L_{2} \Pi^{(2)}-L_{4a,2} \Pi^{(4a)} -L_{2} \Pi^{(6H)}
\end{align*}
%
%
\begin{align*}
\Pi_{\rm ren}^{(p25)}&=
\Pi^{(p25)}-2 {\delta m}_{2} B_{2} L_{2\ast} \Pi^{(2)}-2 {\delta m}_{2} B_{2} L_{2} \Pi^{(2\ast)}+2 {\delta m}_{2} B_{2} \Pi^{(4a,1\ast)}+{\delta m}_{2} L_{4b,1(4\ast)} \Pi^{(2)} \\
& +{\delta m}_{2} L_{4b,1(2\ast)} \Pi^{(2)}+2 {\delta m}_{2} L_{2} \Pi^{(4b,1\ast)}-{\delta m}_{2} \Pi^{(6E,4\ast)}-{\delta m}_{2} \Pi^{(6E,2\ast)}-({\delta m}_{2})^{2} L_{2(1\ast1\ast)} \Pi^{(2)} \\
& -({\delta m}_{2})^{2} L_{2} \Pi^{(2\ast\ast)}+({\delta m}_{2})^{2} \Pi^{(4a,1\ast1\ast)}+2 B_{2} L_{4b,1} \Pi^{(2)}+2 B_{2} L_{2} \Pi^{(4b)}-2 B_{2} \Pi^{(6E)} \\
& -2 (B_{2})^{2} L_{2} \Pi^{(2)}+(B_{2})^{2} \Pi^{(4a)}-L_{6A,1} \Pi^{(2)}-L_{2} \Pi^{(6A)}
\end{align*}
%
%
\begin{align*}
\Pi_{\rm ren}^{(p26)}&=
\Pi^{(p26)}+{\delta m}_{4a} {\delta m}_{2} \Pi^{(2\ast\ast)}+{\delta m}_{4a} B_{2} \Pi^{(2\ast)}-{\delta m}_{4a} \Pi^{(4b,1\ast)}+{\delta m}_{2} B_{4a} \Pi^{(2\ast)} \\
& -4 {\delta m}_{2} B_{2} L_{2} \Pi^{(2\ast)}+4 {\delta m}_{2} L_{2} \Pi^{(4b,1\ast)}-{\delta m}_{2} \Pi^{(6D,1\ast)}-2 ({\delta m}_{2})^{2} L_{2} \Pi^{(2\ast\ast)}+B_{4a} B_{2} \Pi^{(2)} \\
& -B_{4a} \Pi^{(4b)}+4 B_{2} L_{2} \Pi^{(4b)}-B_{2} \Pi^{(6D)}-2 (B_{2})^{2} L_{2} \Pi^{(2)}-2 L_{2} \Pi^{(6A)}
\end{align*}
%
%
\begin{align*}
\Pi_{\rm ren}^{(p27)}&=
\Pi^{(p27)}+{\delta m}_{2} L_{4a,1(2\ast)} \Pi^{(2)}+{\delta m}_{2} L_{4a,1} \Pi^{(2\ast)}-{\delta m}_{2} L_{2\ast} L_{2} \Pi^{(2)}+{\delta m}_{2} L_{2} \Pi^{(4a,1\ast)} \\
& -{\delta m}_{2} (L_{2})^{2} \Pi^{(2\ast)}-{\delta m}_{2} \Pi^{(6F,1\ast)}+2 B_{2} L_{4a,1} \Pi^{(2)}+B_{2} L_{2} \Pi^{(4a)}-2 B_{2} (L_{2})^{2} \Pi^{(2)} \\
& -B_{2} \Pi^{(6F)}-L_{6D,1} \Pi^{(2)}+L_{4b,1} L_{2} \Pi^{(2)}-L_{4a,1} \Pi^{(4b)}-L_{2} \Pi^{(6E)} \\
& +(L_{2})^{2} \Pi^{(4b)}
\end{align*}
%
%
\begin{align*}
\Pi_{\rm ren}^{(p28)}&=
\Pi^{(p28)}+{\delta m}_{4b} {\delta m}_{2} \Pi^{(2\ast\ast)}+{\delta m}_{4b} B_{2} \Pi^{(2\ast)}-{\delta m}_{4b} \Pi^{(4b,1\ast)}-{\delta m}_{2\ast} {\delta m}_{2} B_{2} \Pi^{(2\ast)} \\
& +{\delta m}_{2\ast} {\delta m}_{2} \Pi^{(4b,1\ast)}-{\delta m}_{2\ast} ({\delta m}_{2})^{2} \Pi^{(2\ast\ast)}+{\delta m}_{2} B_{4b} \Pi^{(2\ast)}-{\delta m}_{2} B_{2\ast} B_{2} \Pi^{(2)}+{\delta m}_{2} B_{2\ast} \Pi^{(4b)} \\
& +{\delta m}_{2} B_{2} \Pi^{(4b,2\ast)}+2 {\delta m}_{2} B_{2} \Pi^{(4b,1\ast)}-2 {\delta m}_{2} (B_{2})^{2} \Pi^{(2\ast)}-{\delta m}_{2} \Pi^{(6C,1\ast)}-{\delta m}_{2} \Pi^{(6A,2\ast)} \\
& -({\delta m}_{2})^{2} B_{2\ast} \Pi^{(2\ast)}-({\delta m}_{2})^{2} B_{2} \Pi^{(2\ast\ast)}+({\delta m}_{2})^{2} \Pi^{(4b,1\ast2\ast)}+B_{4b} B_{2} \Pi^{(2)}-B_{4b} \Pi^{(4b)} \\
& -B_{2} \Pi^{(6C)}-B_{2} \Pi^{(6A)}+2 (B_{2})^{2} \Pi^{(4b)}-(B_{2})^{3} \Pi^{(2)}
\end{align*}
%
%
\begin{align*}
\Pi_{\rm ren}^{(p29)}&=
\Pi^{(p29)}+{\delta m}_{2} L_{4a,2(2\ast)} \Pi^{(2)}+{\delta m}_{2} L_{4a,2} \Pi^{(2\ast)}-{\delta m}_{2} \Pi^{(6H,1\ast)}+2 B_{2} L_{4a,2} \Pi^{(2)} \\
& -B_{2} \Pi^{(6H)}-L_{6E,2} \Pi^{(2)}-L_{4a,2} \Pi^{(4b)}
\end{align*}
%
%
\begin{align*}
\Pi_{\rm ren}^{(p30)}&=
\Pi^{(p30)}+{\delta m}_{2} L_{4b,2(2\ast)} \Pi^{(2)}+{\delta m}_{2} L_{4b,2} \Pi^{(2\ast)}-2 {\delta m}_{2} L_{2\ast} L_{2} \Pi^{(2)}+{\delta m}_{2} L_{2\ast} \Pi^{(4a)} \\
& +{\delta m}_{2} L_{2} \Pi^{(4a,1\ast)}-{\delta m}_{2} (L_{2})^{2} \Pi^{(2\ast)}-{\delta m}_{2} \Pi^{(6G,1\ast)}+2 B_{2} L_{4b,2} \Pi^{(2)}+2 B_{2} L_{2} \Pi^{(4a)} \\
& -3 B_{2} (L_{2})^{2} \Pi^{(2)}-B_{2} \Pi^{(6G)}-L_{6B,2} \Pi^{(2)}-L_{4b,2} \Pi^{(4b)}+2 L_{4b,1} L_{2} \Pi^{(2)} \\
& -L_{4b,1} \Pi^{(4a)}-L_{2} \Pi^{(6E)}+(L_{2})^{2} \Pi^{(4b)}
\end{align*}
%
%
\begin{align*}
\Pi_{\rm ren}^{(p31)}&=
\Pi^{(p31)}+{\delta m}_{2} L_{4a,1(1\ast)} \Pi^{(2)}+{\delta m}_{2} L_{4a,1} \Pi^{(2\ast)}-{\delta m}_{2} L_{2\ast} L_{2} \Pi^{(2)}+{\delta m}_{2} L_{2} \Pi^{(4a,1\ast)} \\
& -{\delta m}_{2} (L_{2})^{2} \Pi^{(2\ast)}-{\delta m}_{2} \Pi^{(6F,5\ast)}+2 B_{2} L_{4a,1} \Pi^{(2)}+B_{2} L_{2} \Pi^{(4a)}-2 B_{2} (L_{2})^{2} \Pi^{(2)} \\
& -B_{2} \Pi^{(6F)}-L_{6D,3} \Pi^{(2)}+L_{4b,1} L_{2} \Pi^{(2)}-L_{4a,1} \Pi^{(4b)}-L_{2} \Pi^{(6E)} \\
& +(L_{2})^{2} \Pi^{(4b)}
\end{align*}
%
%
\begin{align*}
\Pi_{\rm ren}^{(p32)}&=
\Pi^{(p32)}+{\delta m}_{4a} L_{2\ast} \Pi^{(2)}+{\delta m}_{4a} L_{2} \Pi^{(2\ast)}-{\delta m}_{4a} \Pi^{(4a,1\ast)}-2 {\delta m}_{2} L_{2\ast} L_{2} \Pi^{(2)} \\
& +2 {\delta m}_{2} L_{2} \Pi^{(4a,1\ast)}-2 {\delta m}_{2} (L_{2})^{2} \Pi^{(2\ast)}+2 B_{4a} L_{2} \Pi^{(2)}-B_{4a} \Pi^{(4a)}+2 B_{2} L_{2} \Pi^{(4a)} \\
& -4 B_{2} (L_{2})^{2} \Pi^{(2)}-L_{6C,1} \Pi^{(2)}+2 L_{4b,1} L_{2} \Pi^{(2)}-2 L_{2} \Pi^{(6E)}-L_{2} \Pi^{(6D)} \\
& +2 (L_{2})^{2} \Pi^{(4b)}
\end{align*}
%
%
\begin{align*}
\Pi_{\rm ren}^{(p33)}&=
\Pi^{(p33)}-L_{6G,1} \Pi^{(2)}-L_{6F,1} \Pi^{(2)}+4 L_{4a,1} L_{2} \Pi^{(2)}-L_{4a,1} \Pi^{(4a)} \\
& -L_{2} \Pi^{(6F)}+(L_{2})^{2} \Pi^{(4a)}-2 (L_{2})^{3} \Pi^{(2)}
\end{align*}
%
%
\begin{align*}
\Pi_{\rm ren}^{(p34)}&=
\Pi^{(p34)}-{\delta m}_{6D} \Pi^{(2\ast)}+{\delta m}_{4b} L_{2} \Pi^{(2\ast)}+{\delta m}_{4a(1\ast)} {\delta m}_{2} \Pi^{(2\ast)}+{\delta m}_{4a} B_{2} \Pi^{(2\ast)} \\
& -{\delta m}_{2\ast} {\delta m}_{2} L_{2} \Pi^{(2\ast)}+{\delta m}_{2} B_{4a(1\ast)} \Pi^{(2)}-{\delta m}_{2} B_{2\ast} L_{2} \Pi^{(2)}-{\delta m}_{2} B_{2} L_{2\ast} \Pi^{(2)}-2 {\delta m}_{2} B_{2} L_{2} \Pi^{(2\ast)} \\
& +{\delta m}_{2} L_{4b,1} \Pi^{(2\ast)}+{\delta m}_{2} L_{2\ast} \Pi^{(4b)}+{\delta m}_{2} L_{2} \Pi^{(4b,2\ast)}-{\delta m}_{2} \Pi^{(6D,2\ast)}-({\delta m}_{2})^{2} L_{2\ast} \Pi^{(2\ast)} \\
& -B_{6D} \Pi^{(2)}+B_{4b} L_{2} \Pi^{(2)}+B_{4a} B_{2} \Pi^{(2)}+B_{2} L_{4b,1} \Pi^{(2)}+2 B_{2} L_{2} \Pi^{(4b)} \\
& -B_{2} \Pi^{(6D)}-2 (B_{2})^{2} L_{2} \Pi^{(2)}-L_{4b,1} \Pi^{(4b)}-L_{2} \Pi^{(6C)}
\end{align*}
%
%
\begin{align*}
\Pi_{\rm ren}^{(p35)}&=
\Pi^{(p35)}-L_{6G,4} \Pi^{(2)}-L_{6F,2} \Pi^{(2)}+2 L_{4a,2} L_{2} \Pi^{(2)}-L_{2} \Pi^{(6H)}
\end{align*}
%
%
\begin{align*}
\Pi_{\rm ren}^{(p36)}&=
\Pi^{(p36)}-L_{6D,2} \Pi^{(2)}-L_{6C,2} \Pi^{(2)}+2 L_{4b,2} L_{2} \Pi^{(2)}+3 L_{4a,1} L_{2} \Pi^{(2)} \\
& -L_{4a,1} \Pi^{(4a)}-L_{2} \Pi^{(6G)}-L_{2} \Pi^{(6F)}+2 (L_{2})^{2} \Pi^{(4a)}-3 (L_{2})^{3} \Pi^{(2)}
\end{align*}
%
%
\begin{align*}
\Pi_{\rm ren}^{(p37)}&=
\Pi^{(p37)}+{\delta m}_{4b} L_{2\ast} \Pi^{(2)}+{\delta m}_{4b} L_{2} \Pi^{(2\ast)}-{\delta m}_{4b} \Pi^{(4a,1\ast)}-{\delta m}_{2\ast} {\delta m}_{2} L_{2\ast} \Pi^{(2)} \\
& -{\delta m}_{2\ast} {\delta m}_{2} L_{2} \Pi^{(2\ast)}+{\delta m}_{2\ast} {\delta m}_{2} \Pi^{(4a,1\ast)}-2 {\delta m}_{2} B_{2\ast} L_{2} \Pi^{(2)}+{\delta m}_{2} B_{2\ast} \Pi^{(4a)}-{\delta m}_{2} B_{2} L_{2\ast} \Pi^{(2)} \\
& -{\delta m}_{2} B_{2} L_{2} \Pi^{(2\ast)}+{\delta m}_{2} B_{2} \Pi^{(4a,1\ast)}+{\delta m}_{2} L_{4b,1(3\ast)} \Pi^{(2)}+{\delta m}_{2} L_{2} \Pi^{(4b,2\ast)}-{\delta m}_{2} \Pi^{(6E,3\ast)} \\
& +2 B_{4b} L_{2} \Pi^{(2)}-B_{4b} \Pi^{(4a)}+B_{2} L_{4b,1} \Pi^{(2)}+B_{2} L_{2} \Pi^{(4b)}-B_{2} \Pi^{(6E)} \\
& -2 (B_{2})^{2} L_{2} \Pi^{(2)}+(B_{2})^{2} \Pi^{(4a)}-L_{6B,1} \Pi^{(2)}-L_{2} \Pi^{(6C)}
\end{align*}
%
%
\begin{align*}
\Pi_{\rm ren}^{(p38)}&=
\Pi^{(p38)}+{\delta m}_{2} L_{4a,1(4\ast)} \Pi^{(2)}+{\delta m}_{2} L_{4a,1(3\ast)} \Pi^{(2)}-2 {\delta m}_{2} L_{2\ast} L_{2} \Pi^{(2)}+{\delta m}_{2} L_{2\ast} \Pi^{(4a)} \\
& -{\delta m}_{2} \Pi^{(6F,2\ast)}+2 B_{2} L_{4a,1} \Pi^{(2)}+B_{2} L_{2} \Pi^{(4a)}-2 B_{2} (L_{2})^{2} \Pi^{(2)}-B_{2} \Pi^{(6F)} \\
& -L_{6E,1} \Pi^{(2)}-L_{6D,5} \Pi^{(2)}+2 L_{4b,1} L_{2} \Pi^{(2)}-L_{4b,1} \Pi^{(4a)}
\end{align*}
%
%
\begin{align*}
\Pi_{\rm ren}^{(p39)}&=
\Pi^{(p39)}-{\delta m}_{6G} \Pi^{(2\ast)}+{\delta m}_{4a} L_{2} \Pi^{(2\ast)}+{\delta m}_{2} L_{4b,2} \Pi^{(2\ast)}+{\delta m}_{2} L_{4a,1} \Pi^{(2\ast)} \\
& -2 {\delta m}_{2} (L_{2})^{2} \Pi^{(2\ast)}-B_{6G} \Pi^{(2)}+B_{4a} L_{2} \Pi^{(2)}+B_{2} L_{4b,2} \Pi^{(2)}+B_{2} L_{4a,1} \Pi^{(2)} \\
& -2 B_{2} (L_{2})^{2} \Pi^{(2)}-L_{4b,2} \Pi^{(4b)}-L_{4a,1} \Pi^{(4b)}-L_{2} \Pi^{(6D)}+2 (L_{2})^{2} \Pi^{(4b)}
\end{align*}


\section{Intermediate renormalization by \textit{K}-operation}
\label{sec:app:kop}

This Appendix describes how UV-divergent subdiagrams obtained by applying
the \textit{K}-operation on the original unrenormalized 
vacuum-polarization functions are separated out.

$L_2^{\rm UV}$, $B_2^{\rm UV}$, etc. denote UV-divergent parts
of renormalization constants $L_2$, $B_2$, etc.~defined 
by the \textit{K}-operation.
Quantities with $\Delta$ attached in front are \textit{finite parts}
of the quantities.
Note that $\Delta \delta m_2 = 0$ so that $\delta {m}_2^{\rm UV}$
can be replaced by $\delta m_2$.
Derivative amplitudes \cite{Kinoshita:1990}
are denoted as $L_2^{'}, B_2^{'}$, etc.

\subsection{Fourth-order vacuum-polarization}
%
%
\begin{align*}
\Pi^{(4a)} &=
\Delta \Pi^{(4a)}+2\,L_{2}^{\,{\rm UV}}\,\Pi^{(2)}
\end{align*}
%
%
\begin{align*}
\Pi^{(4b)} &=
\Delta \Pi^{(4b)}+B_{2}^{\,{\rm UV}}\,\Pi^{(2)}+{\delta m}_{2}^{}\,\Pi^{(2\ast)}
\end{align*}

\subsection{Sixth-order vacuum-polarization}
%
%
\begin{align*}
\Pi^{(6A)} &=
\Delta \Pi^{(6A)}+2\,B_{2}^{\,{\rm UV}}\,\Pi^{(4b)}-(B_{2}^{\,{\rm UV}})^{2}\,\Pi^{(2)}+2\,{\delta m}_{2}^{}\,\Pi^{(4b,1\ast)}-2\,{\delta m}_{2}^{}\,B_{2}^{\,{\rm UV}}\,\Pi^{(2\ast)} \\
& -({\delta m}_{2}^{})^{2}\,\Pi^{(2\ast\ast)}
\end{align*}
%
%
\begin{align*}
\Pi^{(6B)} &=
\Delta \Pi^{(6B)}+2\,B_{2}^{\,{\rm UV}}\,\Pi^{(4b)}-(B_{2}^{\,{\rm UV}})^{2}\,\Pi^{(2)}+2\,{\delta m}_{2}^{}\,\Pi^{(4b,4\ast)}-2\,{\delta m}_{2}^{}\,B_{2}^{\,{\rm UV}}\,\Pi^{(2\ast)} \\
& -({\delta m}_{2}^{})^{2}\,\Pi^{(2\ast\ast)}
\end{align*}
%
%
\begin{align*}
\Pi^{(6C)} &=
\Delta \Pi^{(6C)}+B_{2}^{\,{\rm UV}}\,\Pi^{(4b)}-B_{2}^{\,{\rm UV}}\,B_{2}^{\prime\,{\rm UV}}\,\Pi^{(2)}+B_{4b}^{\rm UV}\,\Pi^{(2)}-{\delta m}_{2}^{\prime\,{\rm UV}}\,B_{2}^{\,{\rm UV}}\,\Pi^{(2\ast)} \\
& +{\delta m}_{2}^{}\,\Pi^{(4b,2\ast)}-{\delta m}_{2}^{}\,{\delta m}_{2\ast}^{\,{\rm UV}}\,\Pi^{(2\ast)}+{\delta m}_{4b}^{\rm UV}\,\Pi^{(2\ast)}
\end{align*}
%
%
\begin{align*}
\Pi^{(6D)} &=
\Delta \Pi^{(6D)}+B_{4a}^{\rm UV}\,\Pi^{(2)}+2\,L_{2}^{\,{\rm UV}}\,\Pi^{(4b)}-2\,L_{2}^{\,{\rm UV}}\,B_{2}^{\,{\rm UV}}\,\Pi^{(2)}-2\,{\delta m}_{2}^{}\,L_{2}^{\,{\rm UV}}\,\Pi^{(2\ast)} \\
& +{\delta m}_{4a}^{\rm UV}\,\Pi^{(2\ast)}
\end{align*}
%
%
\begin{align*}
\Pi^{(6E)} &=
\Delta \Pi^{(6E)}+B_{2}^{\,{\rm UV}}\,\Pi^{(4a)}-L_{2}^{\prime\,{\rm UV}}\,B_{2}^{\,{\rm UV}}\,\Pi^{(2)}+L_{2}^{\,{\rm UV}}\,\Pi^{(4b)}-L_{2}^{\,{\rm UV}}\,B_{2}^{\,{\rm UV}}\,\Pi^{(2)} \\
& +L_{4b,1}^{\rm UV}\,\Pi^{(2)}+{\delta m}_{2}^{}\,\Pi^{(4a,1\ast)}-{\delta m}_{2}^{}\,L_{2}^{\,{\rm UV}}\,\Pi^{(2\ast)}
\end{align*}
%
%
\begin{align*}
\Pi^{(6F)} &=
\Delta \Pi^{(6F)}+L_{2}^{\,{\rm UV}}\,\Pi^{(4a)}-2\,(L_{2}^{\,{\rm UV}})^{2}\,\Pi^{(2)}+2\,L_{4a,1}^{\rm UV}\,\Pi^{(2)}
\end{align*}
%
%
\begin{align*}
\Pi^{(6G)} &=
\Delta \Pi^{(6G)}+2\,L_{2}^{\,{\rm UV}}\,\Pi^{(4a)}-3\,(L_{2}^{\,{\rm UV}})^{2}\,\Pi^{(2)}+2\,L_{4b,2}^{\rm UV}\,\Pi^{(2)}
\end{align*}
%
%
\begin{align*}
\Pi^{(6H)} &=
\Delta \Pi^{(6H)}+2\,L_{4a,2}^{\rm UV}\,\Pi^{(2)}
\end{align*}

\subsection{Eighth-order vacuum-polarization}
%
%
\begin{align*}
\Pi^{(p01)} &=
\Delta \Pi^{(p01)}+2\,L_{2}^{\,{\rm UV}}\,\Pi^{(6F)}-(L_{2}^{\,{\rm UV}})^{2}\,\Pi^{(4a)}+2\,(L_{2}^{\,{\rm UV}})^{3}\,\Pi^{(2)}-4\,L_{4a,1}^{\rm UV}\,L_{2}^{\,{\rm UV}}\,\Pi^{(2)} \\
& +2\,L_{6F,3}^{\rm UV}\,\Pi^{(2)}
\end{align*}
%
%
\begin{align*}
\Pi^{(p02)} &=
\Delta \Pi^{(p02)}+2\,L_{6H,3}^{\rm UV}\,\Pi^{(2)}
\end{align*}
%
%
\begin{align*}
\Pi^{(p03)} &=
\Delta \Pi^{(p03)}+2\,L_{2}^{\,{\rm UV}}\,\Pi^{(6G)}-3\,(L_{2}^{\,{\rm UV}})^{2}\,\Pi^{(4a)}+4\,(L_{2}^{\,{\rm UV}})^{3}\,\Pi^{(2)}+2\,L_{4b,2}^{\rm UV}\,\Pi^{(4a)} \\
& -6\,L_{4b,2}^{\rm UV}\,L_{2}^{\,{\rm UV}}\,\Pi^{(2)}+2\,L_{6B,3}^{\rm UV}\,\Pi^{(2)}
\end{align*}
%
%
\begin{align*}
\Pi^{(p04)} &=
\Delta \Pi^{(p04)}+3\,B_{2}^{\,{\rm UV}}\,\Pi^{(6A)}-3\,(B_{2}^{\,{\rm UV}})^{2}\,\Pi^{(4b)}+(B_{2}^{\,{\rm UV}})^{3}\,\Pi^{(2)}+{\delta m}_{2}^{}\,\Pi^{(6A,3\ast)} \\
& +2\,{\delta m}_{2}^{}\,\Pi^{(6A,1\ast)}-6\,{\delta m}_{2}^{}\,B_{2}^{\,{\rm UV}}\,\Pi^{(4b,1\ast)}+3\,{\delta m}_{2}^{}\,(B_{2}^{\,{\rm UV}})^{2}\,\Pi^{(2\ast)}-({\delta m}_{2}^{})^{2}\,\Pi^{(4b,1\ast3\ast)} \\
& -2\,({\delta m}_{2}^{})^{2}\,\Pi^{(4b,1\ast1\ast)}+3\,({\delta m}_{2}^{})^{2}\,B_{2}^{\,{\rm UV}}\,\Pi^{(2\ast\ast)}+({\delta m}_{2}^{})^{3}\,\Pi^{(2\ast\ast\ast)}
\end{align*}
%
%
\begin{align*}
\Pi^{(p05)} &=
\Delta \Pi^{(p05)}+2\,B_{2}^{\,{\rm UV}}\,\Pi^{(6B)}+B_{2}^{\,{\rm UV}}\,\Pi^{(6A)}-3\,(B_{2}^{\,{\rm UV}})^{2}\,\Pi^{(4b)}+(B_{2}^{\,{\rm UV}})^{3}\,\Pi^{(2)} \\
& +2\,{\delta m}_{2}^{}\,\Pi^{(6B,1\ast)}+{\delta m}_{2}^{}\,\Pi^{(6A,6\ast)}-4\,{\delta m}_{2}^{}\,B_{2}^{\,{\rm UV}}\,\Pi^{(4b,4\ast)}-2\,{\delta m}_{2}^{}\,B_{2}^{\,{\rm UV}}\,\Pi^{(4b,1\ast)} \\
& +3\,{\delta m}_{2}^{}\,(B_{2}^{\,{\rm UV}})^{2}\,\Pi^{(2\ast)}-({\delta m}_{2}^{})^{2}\,\Pi^{(4b,4\ast4\ast)}-2\,({\delta m}_{2}^{})^{2}\,\Pi^{(4b,1\ast4\ast)}+2\,({\delta m}_{2}^{})^{2}\,B_{2}^{\,{\rm UV}}\,\Pi^{(2\ast\ast)} \\
& +({\delta m}_{2}^{})^{2}\,B_{2}^{\,{\rm UV}}\,\Pi^{(2\ast\ast)}+({\delta m}_{2}^{})^{3}\,\Pi^{(2\ast\ast\ast)}
\end{align*}
%
%
\begin{align*}
\Pi^{(p06)} &=
\Delta \Pi^{(p06)}+2\,B_{2}^{\,{\rm UV}}\,\Pi^{(6E)}-(B_{2}^{\,{\rm UV}})^{2}\,\Pi^{(4a)}-L_{2}^{\prime\,{\rm UV}}\,B_{2}^{\,{\rm UV}}\,\Pi^{(4b)}+L_{2}^{\prime\,{\rm UV}}\,(B_{2}^{\,{\rm UV}})^{2}\,\Pi^{(2)} \\
& -L_{2}^{\prime\,{\rm UV}}\,B_{2}^{\,{\rm UV}}\,\Pi^{(4b)}+L_{2}^{\prime\,{\rm UV}}\,(B_{2}^{\,{\rm UV}})^{2}\,\Pi^{(2)}+2\,L_{4b,1}^{\rm UV}\,\Pi^{(4b)}-2\,L_{4b,1}^{\rm UV}\,B_{2}^{\,{\rm UV}}\,\Pi^{(2)} \\
& +2\,{\delta m}_{2}^{}\,\Pi^{(6E,1\ast)}-2\,{\delta m}_{2}^{}\,B_{2}^{\,{\rm UV}}\,\Pi^{(4a,1\ast)}+{\delta m}_{2}^{}\,L_{2}^{\prime\,{\rm UV}}\,B_{2}^{\,{\rm UV}}\,\Pi^{(2\ast)} \\
& +{\delta m}_{2}^{}\,L_{2}^{\prime\,{\rm UV}}\,B_{2}^{\,{\rm UV}}\,\Pi^{(2\ast)}-2\,{\delta m}_{2}^{}\,L_{4b,1}^{\rm UV}\,\Pi^{(2\ast)}-({\delta m}_{2}^{})^{2}\,\Pi^{(4a,1\ast2\ast)}
\end{align*}
%
%
\begin{align*}
\Pi^{(p07)} &=
\Delta \Pi^{(p07)}+2\,B_{2}^{\,{\rm UV}}\,\Pi^{(6E)}-(B_{2}^{\,{\rm UV}})^{2}\,\Pi^{(4a)}+L_{2}^{\prime\prime\,{\rm UV}}\,(B_{2}^{\,{\rm UV}})^{2}\,\Pi^{(2)}+L_{2}^{\,{\rm UV}}\,\Pi^{(6B)} \\
& -2\,L_{2}^{\,{\rm UV}}\,B_{2}^{\,{\rm UV}}\,\Pi^{(4b)}+L_{2}^{\,{\rm UV}}\,(B_{2}^{\,{\rm UV}})^{2}\,\Pi^{(2)}-2\,L_{4b,1(1^{\prime})}^{\rm UV}\,B_{2}^{\,{\rm UV}}\,\Pi^{(2)} \\
& +L_{6A,3}^{\rm UV}\,\Pi^{(2)}+2\,{\delta m}_{2}^{}\,\Pi^{(6E,5\ast)}-2\,{\delta m}_{2}^{}\,B_{2}^{\,{\rm UV}}\,\Pi^{(4a,1\ast)}-2\,{\delta m}_{2}^{}\,L_{2}^{\,{\rm UV}}\,\Pi^{(4b,4\ast)} \\
& +2\,{\delta m}_{2}^{}\,L_{2}^{\,{\rm UV}}\,B_{2}^{\,{\rm UV}}\,\Pi^{(2\ast)}-({\delta m}_{2}^{})^{2}\,\Pi^{(4a,1\ast4\ast)}+({\delta m}_{2}^{})^{2}\,L_{2}^{\,{\rm UV}}\,\Pi^{(2\ast\ast)}
\end{align*}
%
%
\begin{align*}
\Pi^{(p08)} &=
\Delta \Pi^{(p08)}+2\,B_{2}^{\,{\rm UV}}\,\Pi^{(6E)}-(B_{2}^{\,{\rm UV}})^{2}\,\Pi^{(4a)}-2\,L_{2}^{\prime\,{\rm UV}}\,B_{2}^{\,{\rm UV}}\,\Pi^{(4b)}+2\,L_{2}^{\prime\,{\rm UV}}\,(B_{2}^{\,{\rm UV}})^{2}\,\Pi^{(2)} \\
& +2\,L_{4b,1}^{\rm UV}\,\Pi^{(4b)}-2\,L_{4b,1}^{\rm UV}\,B_{2}^{\,{\rm UV}}\,\Pi^{(2)}+2\,{\delta m}_{2}^{}\,\Pi^{(6E,6\ast)}-2\,{\delta m}_{2}^{}\,B_{2}^{\,{\rm UV}}\,\Pi^{(4a,1\ast)} \\
& +2\,{\delta m}_{2}^{}\,L_{2}^{\prime\,{\rm UV}}\,B_{2}^{\,{\rm UV}}\,\Pi^{(2\ast)}-2\,{\delta m}_{2}^{}\,L_{4b,1}^{\rm UV}\,\Pi^{(2\ast)}-({\delta m}_{2}^{})^{2}\,\Pi^{(4a,1\ast3\ast)}
\end{align*}
%
%
\begin{align*}
\Pi^{(p09)} &=
\Delta \Pi^{(p09)}+B_{2}^{\,{\rm UV}}\,\Pi^{(6D)}+B_{4a}^{\rm UV}\,\Pi^{(4b)}-B_{4a}^{\rm UV}\,B_{2}^{\,{\rm UV}}\,\Pi^{(2)}+2\,L_{2}^{\,{\rm UV}}\,\Pi^{(6B)} \\
& -4\,L_{2}^{\,{\rm UV}}\,B_{2}^{\,{\rm UV}}\,\Pi^{(4b)}+2\,L_{2}^{\,{\rm UV}}\,(B_{2}^{\,{\rm UV}})^{2}\,\Pi^{(2)}+{\delta m}_{2}^{}\,\Pi^{(6D,6\ast)}-{\delta m}_{2}^{}\,B_{4a}^{\rm UV}\,\Pi^{(2\ast)} \\
& -4\,{\delta m}_{2}^{}\,L_{2}^{\,{\rm UV}}\,\Pi^{(4b,4\ast)}+4\,{\delta m}_{2}^{}\,L_{2}^{\,{\rm UV}}\,B_{2}^{\,{\rm UV}}\,\Pi^{(2\ast)}+2\,({\delta m}_{2}^{})^{2}\,L_{2}^{\,{\rm UV}}\,\Pi^{(2\ast\ast)} \\
& +{\delta m}_{4a}^{\rm UV}\,\Pi^{(4b,4\ast)}-{\delta m}_{4a}^{\rm UV}\,B_{2}^{\,{\rm UV}}\,\Pi^{(2\ast)}-{\delta m}_{4a}^{\rm UV}\,{\delta m}_{2}^{}\,\Pi^{(2\ast\ast)}
\end{align*}
%
%
\begin{align*}
\Pi^{(p10)} &=
\Delta \Pi^{(p10)}+B_{2}^{\,{\rm UV}}\,\Pi^{(6C)}+B_{2}^{\,{\rm UV}}\,\Pi^{(6B)}-B_{2}^{\,{\rm UV}}\,B_{2}^{\prime\,{\rm UV}}\,\Pi^{(4b)}-(B_{2}^{\,{\rm UV}})^{2}\,\Pi^{(4b)} \\
& +(B_{2}^{\,{\rm UV}})^{2}\,B_{2}^{\prime\,{\rm UV}}\,\Pi^{(2)}+B_{4b}^{\rm UV}\,\Pi^{(4b)}-B_{4b}^{\rm UV}\,B_{2}^{\,{\rm UV}}\,\Pi^{(2)}-{\delta m}_{2}^{\prime\,{\rm UV}}\,B_{2}^{\,{\rm UV}}\,\Pi^{(4b,4\ast)} \\
& +{\delta m}_{2}^{\prime\,{\rm UV}}\,(B_{2}^{\,{\rm UV}})^{2}\,\Pi^{(2\ast)}+{\delta m}_{2}^{}\,\Pi^{(6C,6\ast)}+{\delta m}_{2}^{}\,\Pi^{(6B,2\ast)}-{\delta m}_{2}^{}\,B_{2}^{\,{\rm UV}}\,\Pi^{(4b,4\ast)} \\
& -{\delta m}_{2}^{}\,B_{2}^{\,{\rm UV}}\,\Pi^{(4b,2\ast)}+{\delta m}_{2}^{}\,B_{2}^{\,{\rm UV}}\,B_{2}^{\prime\,{\rm UV}}\,\Pi^{(2\ast)}-{\delta m}_{2}^{}\,B_{4b}^{\rm UV}\,\Pi^{(2\ast)} \\
& -{\delta m}_{2}^{}\,{\delta m}_{2\ast}^{\,{\rm UV}}\,\Pi^{(4b,4\ast)}+{\delta m}_{2}^{}\,{\delta m}_{2\ast}^{\,{\rm UV}}\,B_{2}^{\,{\rm UV}}\,\Pi^{(2\ast)}+{\delta m}_{2}^{}\,{\delta m}_{2}^{\prime\,{\rm UV}}\,B_{2}^{\,{\rm UV}}\,\Pi^{(2\ast\ast)} \\
& -({\delta m}_{2}^{})^{2}\,\Pi^{(4b,2\ast4\ast)}+({\delta m}_{2}^{})^{2}\,{\delta m}_{2\ast}^{\,{\rm UV}}\,\Pi^{(2\ast\ast)}+{\delta m}_{4b}^{\rm UV}\,\Pi^{(4b,4\ast)}-{\delta m}_{4b}^{\rm UV}\,B_{2}^{\,{\rm UV}}\,\Pi^{(2\ast)} \\
& -{\delta m}_{4b}^{\rm UV}\,{\delta m}_{2}^{}\,\Pi^{(2\ast\ast)}
\end{align*}
%
%
\begin{align*}
\Pi^{(p11)} &=
\Delta \Pi^{(p11)}+B_{6F}^{\rm UV}\,\Pi^{(2)}+2\,L_{2}^{\,{\rm UV}}\,\Pi^{(6D)}-2\,L_{2}^{\,{\rm UV}}\,B_{4a}^{\rm UV}\,\Pi^{(2)}-3\,(L_{2}^{\,{\rm UV}})^{2}\,\Pi^{(4b)} \\
& +3\,(L_{2}^{\,{\rm UV}})^{2}\,B_{2}^{\,{\rm UV}}\,\Pi^{(2)}+2\,L_{4a,1}^{\rm UV}\,\Pi^{(4b)}-2\,L_{4a,1}^{\rm UV}\,B_{2}^{\,{\rm UV}}\,\Pi^{(2)}+3\,{\delta m}_{2}^{}\,(L_{2}^{\,{\rm UV}})^{2}\,\Pi^{(2\ast)} \\
& -2\,{\delta m}_{2}^{}\,L_{4a,1}^{\rm UV}\,\Pi^{(2\ast)}-2\,{\delta m}_{4a}^{\rm UV}\,L_{2}^{\,{\rm UV}}\,\Pi^{(2\ast)}+{\delta m}_{6F}^{\rm UV}\,\Pi^{(2\ast)}
\end{align*}
%
%
\begin{align*}
\Pi^{(p12)} &=
\Delta \Pi^{(p12)}+2\,B_{2}^{\,{\rm UV}}\,\Pi^{(6C)}-(B_{2}^{\,{\rm UV}})^{2}\,\Pi^{(4b)}+(B_{2}^{\,{\rm UV}})^{2}\,B_{2}^{\prime\prime\,{\rm UV}}\,\Pi^{(2)}-B_{4b(3^{\prime})}^{\rm UV}\,B_{2}^{\,{\rm UV}}\,\Pi^{(2)} \\
& -B_{4b(1^{\prime})}^{\rm UV}\,B_{2}^{\,{\rm UV}}\,\Pi^{(2)}+B_{6A}^{\rm UV}\,\Pi^{(2)}+{\delta m}_{2}^{\prime\prime\,{\rm UV}}\,(B_{2}^{\,{\rm UV}})^{2}\,\Pi^{(2\ast)}+2\,{\delta m}_{2}^{}\,\Pi^{(6C,2\ast)} \\
& -2\,{\delta m}_{2}^{}\,B_{2}^{\,{\rm UV}}\,\Pi^{(4b,2\ast)}+{\delta m}_{2}^{}\,{\delta m}_{2\ast}^{\prime\,{\rm UV}}\,B_{2}^{\,{\rm UV}}\,\Pi^{(2\ast)}+{\delta m}_{2}^{}\,{\delta m}_{2\ast}^{\prime\,{\rm UV}}\,B_{2}^{\,{\rm UV}}\,\Pi^{(2\ast)} \\
& -({\delta m}_{2}^{})^{2}\,\Pi^{(4b,2\ast2\ast)}+({\delta m}_{2}^{})^{2}\,{\delta m}_{2\ast\ast}^{\,{\rm UV}}\,\Pi^{(2\ast)}-2\,{\delta m}_{4b(1\ast)}^{\rm UV}\,{\delta m}_{2}^{}\,\Pi^{(2\ast)}-{\delta m}_{4b(3^{\prime})}^{\rm UV}\,B_{2}^{\,{\rm UV}}\,\Pi^{(2\ast)} \\
& -{\delta m}_{4b(1^{\prime})}^{\rm UV}\,B_{2}^{\,{\rm UV}}\,\Pi^{(2\ast)}+{\delta m}_{6A}^{\rm UV}\,\Pi^{(2\ast)}
\end{align*}
%
%
\begin{align*}
\Pi^{(p13)} &=
\Delta \Pi^{(p13)}+B_{2}^{\,{\rm UV}}\,\Pi^{(6H)}-L_{4a,2(4^{\prime})}^{\rm UV}\,B_{2}^{\,{\rm UV}}\,\Pi^{(2)}-L_{4a,2(1^{\prime})}^{\rm UV}\,B_{2}^{\,{\rm UV}}\,\Pi^{(2)} \\
& +2\,L_{6D,4}^{\rm UV}\,\Pi^{(2)}+{\delta m}_{2}^{}\,\Pi^{(6H,2\ast)}
\end{align*}
%
%
\begin{align*}
\Pi^{(p14)} &=
\Delta \Pi^{(p14)}+B_{2}^{\,{\rm UV}}\,\Pi^{(6G)}+2\,L_{2}^{\,{\rm UV}}\,\Pi^{(6E)}-2\,L_{2}^{\,{\rm UV}}\,B_{2}^{\,{\rm UV}}\,\Pi^{(4a)}+L_{2}^{\,{\rm UV}}\,L_{2}^{\prime\,{\rm UV}}\,B_{2}^{\,{\rm UV}}\,\Pi^{(2)} \\
& +L_{2}^{\,{\rm UV}}\,L_{2}^{\prime\,{\rm UV}}\,B_{2}^{\,{\rm UV}}\,\Pi^{(2)}-(L_{2}^{\,{\rm UV}})^{2}\,\Pi^{(4b)}+(L_{2}^{\,{\rm UV}})^{2}\,B_{2}^{\,{\rm UV}}\,\Pi^{(2)} \\
& -L_{4b,2(4^{\prime})}^{\rm UV}\,B_{2}^{\,{\rm UV}}\,\Pi^{(2)}-L_{4b,2(1^{\prime})}^{\rm UV}\,B_{2}^{\,{\rm UV}}\,\Pi^{(2)}-2\,L_{4b,1}^{\rm UV}\,L_{2}^{\,{\rm UV}}\,\Pi^{(2)} \\
& +2\,L_{6A,2}^{\rm UV}\,\Pi^{(2)}+{\delta m}_{2}^{}\,\Pi^{(6G,2\ast)}-2\,{\delta m}_{2}^{}\,L_{2}^{\,{\rm UV}}\,\Pi^{(4a,1\ast)}+{\delta m}_{2}^{}\,(L_{2}^{\,{\rm UV}})^{2}\,\Pi^{(2\ast)}
\end{align*}
%
%
\begin{align*}
\Pi^{(p15)} &=
\Delta \Pi^{(p15)}+B_{6H}^{\rm UV}\,\Pi^{(2)}+2\,L_{4a,2}^{\rm UV}\,\Pi^{(4b)}-2\,L_{4a,2}^{\rm UV}\,B_{2}^{\,{\rm UV}}\,\Pi^{(2)}-2\,{\delta m}_{2}^{}\,L_{4a,2}^{\rm UV}\,\Pi^{(2\ast)} \\
& +{\delta m}_{6H}^{\rm UV}\,\Pi^{(2\ast)}
\end{align*}
%
%
\begin{align*}
\Pi^{(p16)} &=
\Delta \Pi^{(p16)}+L_{4a,2}^{\rm UV}\,\Pi^{(4a)}-2\,L_{4a,2}^{\rm UV}\,L_{2}^{\,{\rm UV}}\,\Pi^{(2)}+2\,L_{6H,1}^{\rm UV}\,\Pi^{(2)}
\end{align*}
%
%
\begin{align*}
\Pi^{(p17)} &=
\Delta \Pi^{(p17)}+2\,L_{6H,2}^{\rm UV}\,\Pi^{(2)}
\end{align*}
%
%
\begin{align*}
\Pi^{(p18)} &=
\Delta \Pi^{(p18)}+2\,L_{6G,2}^{\rm UV}\,\Pi^{(2)}
\end{align*}
%
%
\begin{align*}
\Pi^{(p19)} &=
\Delta \Pi^{(p19)}+L_{2}^{\,{\rm UV}}\,\Pi^{(6F)}-(L_{2}^{\,{\rm UV}})^{2}\,\Pi^{(4a)}+2\,(L_{2}^{\,{\rm UV}})^{3}\,\Pi^{(2)}+L_{4b,2}^{\rm UV}\,\Pi^{(4a)} \\
& -2\,L_{4b,2}^{\rm UV}\,L_{2}^{\,{\rm UV}}\,\Pi^{(2)}-2\,L_{4a,1}^{\rm UV}\,L_{2}^{\,{\rm UV}}\,\Pi^{(2)}+2\,L_{6G,5}^{\rm UV}\,\Pi^{(2)}
\end{align*}
%
%
\begin{align*}
\Pi^{(p20)} &=
\Delta \Pi^{(p20)}+B_{4a}^{\rm UV}\,\Pi^{(4b)}-B_{4a}^{\rm UV}\,B_{2}^{\prime\,{\rm UV}}\,\Pi^{(2)}+B_{6C}^{\rm UV}\,\Pi^{(2)}+2\,L_{2}^{\,{\rm UV}}\,\Pi^{(6C)} \\
& -2\,L_{2}^{\,{\rm UV}}\,B_{2}^{\,{\rm UV}}\,\Pi^{(4b)}+2\,L_{2}^{\,{\rm UV}}\,B_{2}^{\,{\rm UV}}\,B_{2}^{\prime\,{\rm UV}}\,\Pi^{(2)}-2\,L_{2}^{\,{\rm UV}}\,B_{4b}^{\rm UV}\,\Pi^{(2)} \\
& -{\delta m}_{2}^{\prime\,{\rm UV}}\,B_{4a}^{\rm UV}\,\Pi^{(2\ast)}+2\,{\delta m}_{2}^{\prime\,{\rm UV}}\,L_{2}^{\,{\rm UV}}\,B_{2}^{\,{\rm UV}}\,\Pi^{(2\ast)}-2\,{\delta m}_{2}^{}\,L_{2}^{\,{\rm UV}}\,\Pi^{(4b,2\ast)} \\
& +2\,{\delta m}_{2}^{}\,{\delta m}_{2\ast}^{\,{\rm UV}}\,L_{2}^{\,{\rm UV}}\,\Pi^{(2\ast)}-2\,{\delta m}_{4b}^{\rm UV}\,L_{2}^{\,{\rm UV}}\,\Pi^{(2\ast)}+{\delta m}_{4a}^{\rm UV}\,\Pi^{(4b,2\ast)} \\
& -{\delta m}_{4a}^{\rm UV}\,{\delta m}_{2\ast}^{\,{\rm UV}}\,\Pi^{(2\ast)}+{\delta m}_{6C}^{\rm UV}\,\Pi^{(2\ast)}
\end{align*}
%
%
\begin{align*}
\Pi^{(p21)} &=
\Delta \Pi^{(p21)}+B_{2}^{\,{\rm UV}}\,\Pi^{(6D)}-B_{4a(2^{\prime})}^{\rm UV}\,B_{2}^{\,{\rm UV}}\,\Pi^{(2)}+B_{6E}^{\rm UV}\,\Pi^{(2)}-L_{2}^{\prime\,{\rm UV}}\,B_{2}^{\,{\rm UV}}\,\Pi^{(4b)} \\
& +L_{2}^{\prime\,{\rm UV}}\,(B_{2}^{\,{\rm UV}})^{2}\,\Pi^{(2)}-L_{2}^{\prime\,{\rm UV}}\,B_{2}^{\,{\rm UV}}\,\Pi^{(4b)}+L_{2}^{\prime\,{\rm UV}}\,(B_{2}^{\,{\rm UV}})^{2}\,\Pi^{(2)} \\
& +2\,L_{4b,1}^{\rm UV}\,\Pi^{(4b)}-2\,L_{4b,1}^{\rm UV}\,B_{2}^{\,{\rm UV}}\,\Pi^{(2)}+{\delta m}_{2}^{}\,\Pi^{(6D,3\ast)}+{\delta m}_{2}^{}\,L_{2}^{\prime\,{\rm UV}}\,B_{2}^{\,{\rm UV}}\,\Pi^{(2\ast)} \\
& +{\delta m}_{2}^{}\,L_{2}^{\prime\,{\rm UV}}\,B_{2}^{\,{\rm UV}}\,\Pi^{(2\ast)}-2\,{\delta m}_{2}^{}\,L_{4b,1}^{\rm UV}\,\Pi^{(2\ast)}-{\delta m}_{4a(2\ast)}^{\rm UV}\,{\delta m}_{2}^{}\,\Pi^{(2\ast)} \\
& -{\delta m}_{4a(2^{\prime})}^{\rm UV}\,B_{2}^{\,{\rm UV}}\,\Pi^{(2\ast)}+{\delta m}_{6E}^{\rm UV}\,\Pi^{(2\ast)}
\end{align*}
%
%
\begin{align*}
\Pi^{(p22)} &=
\Delta \Pi^{(p22)}+B_{2}^{\,{\rm UV}}\,\Pi^{(6C)}-B_{2}^{\,{\rm UV}}\,B_{2}^{\prime\,{\rm UV}}\,\Pi^{(4b)}+B_{2}^{\,{\rm UV}}\,(B_{2}^{\prime\,{\rm UV}})^{2}\,\Pi^{(2)} \\
& -B_{4b(2^{\prime})}^{\rm UV}\,B_{2}^{\,{\rm UV}}\,\Pi^{(2)}+B_{4b}^{\rm UV}\,\Pi^{(4b)}-B_{4b}^{\rm UV}\,B_{2}^{\prime\,{\rm UV}}\,\Pi^{(2)}+B_{6B}^{\rm UV}\,\Pi^{(2)} \\
& -{\delta m}_{2}^{\prime\,{\rm UV}}\,B_{2}^{\,{\rm UV}}\,\Pi^{(4b,2\ast)}+{\delta m}_{2}^{\prime\,{\rm UV}}\,B_{2}^{\,{\rm UV}}\,B_{2}^{\prime\,{\rm UV}}\,\Pi^{(2\ast)}-{\delta m}_{2}^{\prime\,{\rm UV}}\,B_{4b}^{\rm UV}\,\Pi^{(2\ast)} \\
& +{\delta m}_{2}^{\prime\,{\rm UV}}\,{\delta m}_{2\ast}^{\,{\rm UV}}\,B_{2}^{\,{\rm UV}}\,\Pi^{(2\ast)}+{\delta m}_{2}^{}\,\Pi^{(6C,3\ast)}-{\delta m}_{2}^{}\,{\delta m}_{2\ast}^{\,{\rm UV}}\,\Pi^{(4b,2\ast)} \\
& +{\delta m}_{2}^{}\,({\delta m}_{2\ast}^{\,{\rm UV}})^{2}\,\Pi^{(2\ast)}-{\delta m}_{4b(2\ast)}^{\rm UV}\,{\delta m}_{2}^{}\,\Pi^{(2\ast)}-{\delta m}_{4b(2^{\prime})}^{\rm UV}\,B_{2}^{\,{\rm UV}}\,\Pi^{(2\ast)} \\
& +{\delta m}_{4b}^{\rm UV}\,\Pi^{(4b,2\ast)}-{\delta m}_{4b}^{\rm UV}\,{\delta m}_{2\ast}^{\,{\rm UV}}\,\Pi^{(2\ast)}+{\delta m}_{6B}^{\rm UV}\,\Pi^{(2\ast)}
\end{align*}
%
%
\begin{align*}
\Pi^{(p23)} &=
\Delta \Pi^{(p23)}+2\,L_{6G,3}^{\rm UV}\,\Pi^{(2)}
\end{align*}
%
%
\begin{align*}
\Pi^{(p24)} &=
\Delta \Pi^{(p24)}+L_{2}^{\,{\rm UV}}\,\Pi^{(6H)}+L_{4a,2}^{\rm UV}\,\Pi^{(4a)}-3\,L_{4a,2}^{\rm UV}\,L_{2}^{\,{\rm UV}}\,\Pi^{(2)}+L_{6E,3}^{\rm UV}\,\Pi^{(2)} \\
& +L_{6C,3}^{\rm UV}\,\Pi^{(2)}
\end{align*}
%
%
\begin{align*}
\Pi^{(p25)} &=
\Delta \Pi^{(p25)}+2\,B_{2}^{\,{\rm UV}}\,\Pi^{(6E)}-(B_{2}^{\,{\rm UV}})^{2}\,\Pi^{(4a)}+L_{2}^{\prime\prime\,{\rm UV}}\,(B_{2}^{\,{\rm UV}})^{2}\,\Pi^{(2)}+L_{2}^{\,{\rm UV}}\,\Pi^{(6A)} \\
& -2\,L_{2}^{\,{\rm UV}}\,B_{2}^{\,{\rm UV}}\,\Pi^{(4b)}+L_{2}^{\,{\rm UV}}\,(B_{2}^{\,{\rm UV}})^{2}\,\Pi^{(2)}-L_{4b,1(4^{\prime})}^{\rm UV}\,B_{2}^{\,{\rm UV}}\,\Pi^{(2)} \\
& -L_{4b,1(2^{\prime})}^{\rm UV}\,B_{2}^{\,{\rm UV}}\,\Pi^{(2)}+L_{6A,1}^{\rm UV}\,\Pi^{(2)}+{\delta m}_{2}^{}\,\Pi^{(6E,4\ast)}+{\delta m}_{2}^{}\,\Pi^{(6E,2\ast)} \\
& -2\,{\delta m}_{2}^{}\,B_{2}^{\,{\rm UV}}\,\Pi^{(4a,1\ast)}-2\,{\delta m}_{2}^{}\,L_{2}^{\,{\rm UV}}\,\Pi^{(4b,1\ast)}+2\,{\delta m}_{2}^{}\,L_{2}^{\,{\rm UV}}\,B_{2}^{\,{\rm UV}}\,\Pi^{(2\ast)} \\
& -({\delta m}_{2}^{})^{2}\,\Pi^{(4a,1\ast1\ast)}+({\delta m}_{2}^{})^{2}\,L_{2}^{\,{\rm UV}}\,\Pi^{(2\ast\ast)}
\end{align*}
%
%
\begin{align*}
\Pi^{(p26)} &=
\Delta \Pi^{(p26)}+B_{2}^{\,{\rm UV}}\,\Pi^{(6D)}+B_{4a}^{\rm UV}\,\Pi^{(4b)}-B_{4a}^{\rm UV}\,B_{2}^{\,{\rm UV}}\,\Pi^{(2)}+2\,L_{2}^{\,{\rm UV}}\,\Pi^{(6A)} \\
& -4\,L_{2}^{\,{\rm UV}}\,B_{2}^{\,{\rm UV}}\,\Pi^{(4b)}+2\,L_{2}^{\,{\rm UV}}\,(B_{2}^{\,{\rm UV}})^{2}\,\Pi^{(2)}+{\delta m}_{2}^{}\,\Pi^{(6D,1\ast)}-{\delta m}_{2}^{}\,B_{4a}^{\rm UV}\,\Pi^{(2\ast)} \\
& -4\,{\delta m}_{2}^{}\,L_{2}^{\,{\rm UV}}\,\Pi^{(4b,1\ast)}+4\,{\delta m}_{2}^{}\,L_{2}^{\,{\rm UV}}\,B_{2}^{\,{\rm UV}}\,\Pi^{(2\ast)}+2\,({\delta m}_{2}^{})^{2}\,L_{2}^{\,{\rm UV}}\,\Pi^{(2\ast\ast)} \\
& +{\delta m}_{4a}^{\rm UV}\,\Pi^{(4b,1\ast)}-{\delta m}_{4a}^{\rm UV}\,B_{2}^{\,{\rm UV}}\,\Pi^{(2\ast)}-{\delta m}_{4a}^{\rm UV}\,{\delta m}_{2}^{}\,\Pi^{(2\ast\ast)}
\end{align*}
%
%
\begin{align*}
\Pi^{(p27)} &=
\Delta \Pi^{(p27)}+B_{2}^{\,{\rm UV}}\,\Pi^{(6F)}+L_{2}^{\,{\rm UV}}\,\Pi^{(6E)}-L_{2}^{\,{\rm UV}}\,B_{2}^{\,{\rm UV}}\,\Pi^{(4a)}+L_{2}^{\,{\rm UV}}\,L_{2}^{\prime\,{\rm UV}}\,B_{2}^{\,{\rm UV}}\,\Pi^{(2)} \\
& -(L_{2}^{\,{\rm UV}})^{2}\,\Pi^{(4b)}+(L_{2}^{\,{\rm UV}})^{2}\,B_{2}^{\,{\rm UV}}\,\Pi^{(2)}-L_{4b,1}^{\rm UV}\,L_{2}^{\,{\rm UV}}\,\Pi^{(2)}-L_{4a,1(2^{\prime})}^{\rm UV}\,B_{2}^{\,{\rm UV}}\,\Pi^{(2)} \\
& +L_{4a,1}^{\rm UV}\,\Pi^{(4b)}-L_{4a,1}^{\rm UV}\,B_{2}^{\,{\rm UV}}\,\Pi^{(2)}+L_{6D,1}^{\rm UV}\,\Pi^{(2)}+{\delta m}_{2}^{}\,\Pi^{(6F,1\ast)} \\
& -{\delta m}_{2}^{}\,L_{2}^{\,{\rm UV}}\,\Pi^{(4a,1\ast)}+{\delta m}_{2}^{}\,(L_{2}^{\,{\rm UV}})^{2}\,\Pi^{(2\ast)}-{\delta m}_{2}^{}\,L_{4a,1}^{\rm UV}\,\Pi^{(2\ast)}
\end{align*}
%
%
\begin{align*}
\Pi^{(p28)} &=
\Delta \Pi^{(p28)}+B_{2}^{\,{\rm UV}}\,\Pi^{(6C)}+B_{2}^{\,{\rm UV}}\,\Pi^{(6A)}-B_{2}^{\,{\rm UV}}\,B_{2}^{\prime\,{\rm UV}}\,\Pi^{(4b)}-(B_{2}^{\,{\rm UV}})^{2}\,\Pi^{(4b)} \\
& +(B_{2}^{\,{\rm UV}})^{2}\,B_{2}^{\prime\,{\rm UV}}\,\Pi^{(2)}+B_{4b}^{\rm UV}\,\Pi^{(4b)}-B_{4b}^{\rm UV}\,B_{2}^{\,{\rm UV}}\,\Pi^{(2)}-{\delta m}_{2}^{\prime\,{\rm UV}}\,B_{2}^{\,{\rm UV}}\,\Pi^{(4b,1\ast)} \\
& +{\delta m}_{2}^{\prime\,{\rm UV}}\,(B_{2}^{\,{\rm UV}})^{2}\,\Pi^{(2\ast)}+{\delta m}_{2}^{}\,\Pi^{(6C,1\ast)}+{\delta m}_{2}^{}\,\Pi^{(6A,2\ast)}-{\delta m}_{2}^{}\,B_{2}^{\,{\rm UV}}\,\Pi^{(4b,2\ast)} \\
& -{\delta m}_{2}^{}\,B_{2}^{\,{\rm UV}}\,\Pi^{(4b,1\ast)}+{\delta m}_{2}^{}\,B_{2}^{\,{\rm UV}}\,B_{2}^{\prime\,{\rm UV}}\,\Pi^{(2\ast)}-{\delta m}_{2}^{}\,B_{4b}^{\rm UV}\,\Pi^{(2\ast)} \\
& -{\delta m}_{2}^{}\,{\delta m}_{2\ast}^{\,{\rm UV}}\,\Pi^{(4b,1\ast)}+{\delta m}_{2}^{}\,{\delta m}_{2\ast}^{\,{\rm UV}}\,B_{2}^{\,{\rm UV}}\,\Pi^{(2\ast)}+{\delta m}_{2}^{}\,{\delta m}_{2}^{\prime\,{\rm UV}}\,B_{2}^{\,{\rm UV}}\,\Pi^{(2\ast\ast)} \\
& -({\delta m}_{2}^{})^{2}\,\Pi^{(4b,1\ast2\ast)}+({\delta m}_{2}^{})^{2}\,{\delta m}_{2\ast}^{\,{\rm UV}}\,\Pi^{(2\ast\ast)}+{\delta m}_{4b}^{\rm UV}\,\Pi^{(4b,1\ast)}-{\delta m}_{4b}^{\rm UV}\,B_{2}^{\,{\rm UV}}\,\Pi^{(2\ast)} \\
& -{\delta m}_{4b}^{\rm UV}\,{\delta m}_{2}^{}\,\Pi^{(2\ast\ast)}
\end{align*}
%
%
\begin{align*}
\Pi^{(p29)} &=
\Delta \Pi^{(p29)}+B_{2}^{\,{\rm UV}}\,\Pi^{(6H)}-L_{4a,2(3^{\prime})}^{\rm UV}\,B_{2}^{\,{\rm UV}}\,\Pi^{(2)}+L_{4a,2}^{\rm UV}\,\Pi^{(4b)}-L_{4a,2}^{\rm UV}\,B_{2}^{\,{\rm UV}}\,\Pi^{(2)} \\
& +L_{6E,2}^{\rm UV}\,\Pi^{(2)}+{\delta m}_{2}^{}\,\Pi^{(6H,1\ast)}-{\delta m}_{2}^{}\,L_{4a,2}^{\rm UV}\,\Pi^{(2\ast)}
\end{align*}
%
%
\begin{align*}
\Pi^{(p30)} &=
\Delta \Pi^{(p30)}+B_{2}^{\,{\rm UV}}\,\Pi^{(6G)}-L_{2}^{\prime\,{\rm UV}}\,B_{2}^{\,{\rm UV}}\,\Pi^{(4a)}+L_{2}^{\,{\rm UV}}\,\Pi^{(6E)}-L_{2}^{\,{\rm UV}}\,B_{2}^{\,{\rm UV}}\,\Pi^{(4a)} \\
& +2\,L_{2}^{\,{\rm UV}}\,L_{2}^{\prime\,{\rm UV}}\,B_{2}^{\,{\rm UV}}\,\Pi^{(2)}-(L_{2}^{\,{\rm UV}})^{2}\,\Pi^{(4b)}+(L_{2}^{\,{\rm UV}})^{2}\,B_{2}^{\,{\rm UV}}\,\Pi^{(2)} \\
& -L_{4b,2(3^{\prime})}^{\rm UV}\,B_{2}^{\,{\rm UV}}\,\Pi^{(2)}+L_{4b,2}^{\rm UV}\,\Pi^{(4b)}-L_{4b,2}^{\rm UV}\,B_{2}^{\,{\rm UV}}\,\Pi^{(2)}+L_{4b,1}^{\rm UV}\,\Pi^{(4a)} \\
& -2\,L_{4b,1}^{\rm UV}\,L_{2}^{\,{\rm UV}}\,\Pi^{(2)}+L_{6B,2}^{\rm UV}\,\Pi^{(2)}+{\delta m}_{2}^{}\,\Pi^{(6G,1\ast)}-{\delta m}_{2}^{}\,L_{2}^{\,{\rm UV}}\,\Pi^{(4a,1\ast)} \\
& +{\delta m}_{2}^{}\,(L_{2}^{\,{\rm UV}})^{2}\,\Pi^{(2\ast)}-{\delta m}_{2}^{}\,L_{4b,2}^{\rm UV}\,\Pi^{(2\ast)}
\end{align*}
%
%
\begin{align*}
\Pi^{(p31)} &=
\Delta \Pi^{(p31)}+B_{2}^{\,{\rm UV}}\,\Pi^{(6F)}+L_{2}^{\,{\rm UV}}\,\Pi^{(6E)}-L_{2}^{\,{\rm UV}}\,B_{2}^{\,{\rm UV}}\,\Pi^{(4a)}+L_{2}^{\,{\rm UV}}\,L_{2}^{\prime\,{\rm UV}}\,B_{2}^{\,{\rm UV}}\,\Pi^{(2)} \\
& -(L_{2}^{\,{\rm UV}})^{2}\,\Pi^{(4b)}+(L_{2}^{\,{\rm UV}})^{2}\,B_{2}^{\,{\rm UV}}\,\Pi^{(2)}-L_{4b,1}^{\rm UV}\,L_{2}^{\,{\rm UV}}\,\Pi^{(2)}-L_{4a,1(1^{\prime})}^{\rm UV}\,B_{2}^{\,{\rm UV}}\,\Pi^{(2)} \\
& +L_{4a,1}^{\rm UV}\,\Pi^{(4b)}-L_{4a,1}^{\rm UV}\,B_{2}^{\,{\rm UV}}\,\Pi^{(2)}+L_{6D,3}^{\rm UV}\,\Pi^{(2)}+{\delta m}_{2}^{}\,\Pi^{(6F,5\ast)} \\
& -{\delta m}_{2}^{}\,L_{2}^{\,{\rm UV}}\,\Pi^{(4a,1\ast)}+{\delta m}_{2}^{}\,(L_{2}^{\,{\rm UV}})^{2}\,\Pi^{(2\ast)}-{\delta m}_{2}^{}\,L_{4a,1}^{\rm UV}\,\Pi^{(2\ast)}
\end{align*}
%
%
\begin{align*}
\Pi^{(p32)} &=
\Delta \Pi^{(p32)}+B_{4a}^{\rm UV}\,\Pi^{(4a)}-L_{2}^{\prime\,{\rm UV}}\,B_{4a}^{\rm UV}\,\Pi^{(2)}+2\,L_{2}^{\,{\rm UV}}\,\Pi^{(6E)}+L_{2}^{\,{\rm UV}}\,\Pi^{(6D)} \\
& -2\,L_{2}^{\,{\rm UV}}\,B_{2}^{\,{\rm UV}}\,\Pi^{(4a)}-L_{2}^{\,{\rm UV}}\,B_{4a}^{\rm UV}\,\Pi^{(2)}+2\,L_{2}^{\,{\rm UV}}\,L_{2}^{\prime\,{\rm UV}}\,B_{2}^{\,{\rm UV}}\,\Pi^{(2)} \\
& -2\,(L_{2}^{\,{\rm UV}})^{2}\,\Pi^{(4b)}+2\,(L_{2}^{\,{\rm UV}})^{2}\,B_{2}^{\,{\rm UV}}\,\Pi^{(2)}-2\,L_{4b,1}^{\rm UV}\,L_{2}^{\,{\rm UV}}\,\Pi^{(2)}+L_{6C,1}^{\rm UV}\,\Pi^{(2)} \\
& -2\,{\delta m}_{2}^{}\,L_{2}^{\,{\rm UV}}\,\Pi^{(4a,1\ast)}+2\,{\delta m}_{2}^{}\,(L_{2}^{\,{\rm UV}})^{2}\,\Pi^{(2\ast)}+{\delta m}_{4a}^{\rm UV}\,\Pi^{(4a,1\ast)}-{\delta m}_{4a}^{\rm UV}\,L_{2}^{\,{\rm UV}}\,\Pi^{(2\ast)}
\end{align*}
%
%
\begin{align*}
\Pi^{(p33)} &=
\Delta \Pi^{(p33)}+L_{2}^{\,{\rm UV}}\,\Pi^{(6F)}-(L_{2}^{\,{\rm UV}})^{2}\,\Pi^{(4a)}+2\,(L_{2}^{\,{\rm UV}})^{3}\,\Pi^{(2)}+L_{4a,1}^{\rm UV}\,\Pi^{(4a)} \\
& -4\,L_{4a,1}^{\rm UV}\,L_{2}^{\,{\rm UV}}\,\Pi^{(2)}+L_{6G,1}^{\rm UV}\,\Pi^{(2)}+L_{6F,1}^{\rm UV}\,\Pi^{(2)}
\end{align*}
%
%
\begin{align*}
\Pi^{(p34)} &=
\Delta \Pi^{(p34)}+B_{2}^{\,{\rm UV}}\,\Pi^{(6D)}-B_{4a(3^{\prime})}^{\rm UV}\,B_{2}^{\,{\rm UV}}\,\Pi^{(2)}+B_{6D}^{\rm UV}\,\Pi^{(2)}-L_{2}^{\prime\,{\rm UV}}\,B_{2}^{\,{\rm UV}}\,\Pi^{(4b)} \\
& +L_{2}^{\prime\,{\rm UV}}\,(B_{2}^{\,{\rm UV}})^{2}\,\Pi^{(2)}+L_{2}^{\,{\rm UV}}\,\Pi^{(6C)}-L_{2}^{\,{\rm UV}}\,B_{2}^{\,{\rm UV}}\,\Pi^{(4b)}+L_{2}^{\,{\rm UV}}\,B_{2}^{\,{\rm UV}}\,B_{2}^{\prime\,{\rm UV}}\,\Pi^{(2)} \\
& -L_{2}^{\,{\rm UV}}\,B_{4b}^{\rm UV}\,\Pi^{(2)}+L_{4b,1}^{\rm UV}\,\Pi^{(4b)}-L_{4b,1}^{\rm UV}\,B_{2}^{\,{\rm UV}}\,\Pi^{(2)}+{\delta m}_{2}^{\prime\,{\rm UV}}\,L_{2}^{\,{\rm UV}}\,B_{2}^{\,{\rm UV}}\,\Pi^{(2\ast)} \\
& +{\delta m}_{2}^{}\,\Pi^{(6D,2\ast)}+{\delta m}_{2}^{}\,L_{2}^{\prime\,{\rm UV}}\,B_{2}^{\,{\rm UV}}\,\Pi^{(2\ast)}-{\delta m}_{2}^{}\,L_{2}^{\,{\rm UV}}\,\Pi^{(4b,2\ast)} \\
& -{\delta m}_{2}^{}\,L_{4b,1}^{\rm UV}\,\Pi^{(2\ast)}+{\delta m}_{2}^{}\,{\delta m}_{2\ast}^{\,{\rm UV}}\,L_{2}^{\,{\rm UV}}\,\Pi^{(2\ast)}-{\delta m}_{4b}^{\rm UV}\,L_{2}^{\,{\rm UV}}\,\Pi^{(2\ast)} \\
& -{\delta m}_{4a(1\ast)}^{\rm UV}\,{\delta m}_{2}^{}\,\Pi^{(2\ast)}-{\delta m}_{4a(3^{\prime})}^{\rm UV}\,B_{2}^{\,{\rm UV}}\,\Pi^{(2\ast)}+{\delta m}_{6D}^{\rm UV}\,\Pi^{(2\ast)}
\end{align*}
%
%
\begin{align*}
\Pi^{(p35)} &=
\Delta \Pi^{(p35)}+L_{2}^{\,{\rm UV}}\,\Pi^{(6H)}-2\,L_{4a,2}^{\rm UV}\,L_{2}^{\,{\rm UV}}\,\Pi^{(2)}+L_{6G,4}^{\rm UV}\,\Pi^{(2)}+L_{6F,2}^{\rm UV}\,\Pi^{(2)}
\end{align*}
%
%
\begin{align*}
\Pi^{(p36)} &=
\Delta \Pi^{(p36)}+L_{2}^{\,{\rm UV}}\,\Pi^{(6G)}+L_{2}^{\,{\rm UV}}\,\Pi^{(6F)}-2\,(L_{2}^{\,{\rm UV}})^{2}\,\Pi^{(4a)}+3\,(L_{2}^{\,{\rm UV}})^{3}\,\Pi^{(2)} \\
& -2\,L_{4b,2}^{\rm UV}\,L_{2}^{\,{\rm UV}}\,\Pi^{(2)}+L_{4a,1}^{\rm UV}\,\Pi^{(4a)}-3\,L_{4a,1}^{\rm UV}\,L_{2}^{\,{\rm UV}}\,\Pi^{(2)}+L_{6D,2}^{\rm UV}\,\Pi^{(2)} \\
& +L_{6C,2}^{\rm UV}\,\Pi^{(2)}
\end{align*}
%
%
\begin{align*}
\Pi^{(p37)} &=
\Delta \Pi^{(p37)}+B_{2}^{\,{\rm UV}}\,\Pi^{(6E)}-B_{2}^{\,{\rm UV}}\,B_{2}^{\prime\,{\rm UV}}\,\Pi^{(4a)}+B_{4b}^{\rm UV}\,\Pi^{(4a)}+L_{2}^{\prime\,{\rm UV}}\,B_{2}^{\,{\rm UV}}\,B_{2}^{\prime\,{\rm UV}}\,\Pi^{(2)} \\
& -L_{2}^{\prime\,{\rm UV}}\,B_{4b}^{\rm UV}\,\Pi^{(2)}+L_{2}^{\,{\rm UV}}\,\Pi^{(6C)}-L_{2}^{\,{\rm UV}}\,B_{2}^{\,{\rm UV}}\,\Pi^{(4b)}+L_{2}^{\,{\rm UV}}\,B_{2}^{\,{\rm UV}}\,B_{2}^{\prime\,{\rm UV}}\,\Pi^{(2)} \\
& -L_{2}^{\,{\rm UV}}\,B_{4b}^{\rm UV}\,\Pi^{(2)}-L_{4b,1(3^{\prime})}^{\rm UV}\,B_{2}^{\,{\rm UV}}\,\Pi^{(2)}+L_{6B,1}^{\rm UV}\,\Pi^{(2)}-{\delta m}_{2}^{\prime\,{\rm UV}}\,B_{2}^{\,{\rm UV}}\,\Pi^{(4a,1\ast)} \\
& +{\delta m}_{2}^{\prime\,{\rm UV}}\,L_{2}^{\,{\rm UV}}\,B_{2}^{\,{\rm UV}}\,\Pi^{(2\ast)}+{\delta m}_{2}^{}\,\Pi^{(6E,3\ast)}-{\delta m}_{2}^{}\,L_{2}^{\,{\rm UV}}\,\Pi^{(4b,2\ast)} \\
& -{\delta m}_{2}^{}\,{\delta m}_{2\ast}^{\,{\rm UV}}\,\Pi^{(4a,1\ast)}+{\delta m}_{2}^{}\,{\delta m}_{2\ast}^{\,{\rm UV}}\,L_{2}^{\,{\rm UV}}\,\Pi^{(2\ast)}+{\delta m}_{4b}^{\rm UV}\,\Pi^{(4a,1\ast)} \\
& -{\delta m}_{4b}^{\rm UV}\,L_{2}^{\,{\rm UV}}\,\Pi^{(2\ast)}
\end{align*}
%
%
\begin{align*}
\Pi^{(p38)} &=
\Delta \Pi^{(p38)}+B_{2}^{\,{\rm UV}}\,\Pi^{(6F)}-L_{2}^{\prime\,{\rm UV}}\,B_{2}^{\,{\rm UV}}\,\Pi^{(4a)}+2\,L_{2}^{\,{\rm UV}}\,L_{2}^{\prime\,{\rm UV}}\,B_{2}^{\,{\rm UV}}\,\Pi^{(2)} \\
& +L_{4b,1}^{\rm UV}\,\Pi^{(4a)}-2\,L_{4b,1}^{\rm UV}\,L_{2}^{\,{\rm UV}}\,\Pi^{(2)}-L_{4a,1(4^{\prime})}^{\rm UV}\,B_{2}^{\,{\rm UV}}\,\Pi^{(2)}-L_{4a,1(3^{\prime})}^{\rm UV}\,B_{2}^{\,{\rm UV}}\,\Pi^{(2)} \\
& +L_{6E,1}^{\rm UV}\,\Pi^{(2)}+L_{6D,5}^{\rm UV}\,\Pi^{(2)}+{\delta m}_{2}^{}\,\Pi^{(6F,2\ast)}
\end{align*}
%
%
\begin{align*}
\Pi^{(p39)} &=
\Delta \Pi^{(p39)}+B_{6G}^{\rm UV}\,\Pi^{(2)}+L_{2}^{\,{\rm UV}}\,\Pi^{(6D)}-L_{2}^{\,{\rm UV}}\,B_{4a}^{\rm UV}\,\Pi^{(2)}-2\,(L_{2}^{\,{\rm UV}})^{2}\,\Pi^{(4b)} \\
& +2\,(L_{2}^{\,{\rm UV}})^{2}\,B_{2}^{\,{\rm UV}}\,\Pi^{(2)}+L_{4b,2}^{\rm UV}\,\Pi^{(4b)}-L_{4b,2}^{\rm UV}\,B_{2}^{\,{\rm UV}}\,\Pi^{(2)}+L_{4a,1}^{\rm UV}\,\Pi^{(4b)} \\
& -L_{4a,1}^{\rm UV}\,B_{2}^{\,{\rm UV}}\,\Pi^{(2)}+2\,{\delta m}_{2}^{}\,(L_{2}^{\,{\rm UV}})^{2}\,\Pi^{(2\ast)}-{\delta m}_{2}^{}\,L_{4b,2}^{\rm UV}\,\Pi^{(2\ast)} \\
& -{\delta m}_{2}^{}\,L_{4a,1}^{\rm UV}\,\Pi^{(2\ast)}-{\delta m}_{4a}^{\rm UV}\,L_{2}^{\,{\rm UV}}\,\Pi^{(2\ast)}+{\delta m}_{6G}^{\rm UV}\,\Pi^{(2\ast)}
\end{align*}


\section{Divergence structure of renormalization constants of 
sixth and lower orders}
\label{sec:app:div.str.}

Throughout this Appendix 
$\widetilde{L}_n$,
$\widetilde{B}_n$, and
$\delta \widetilde{m}_n$
denote quantities obtained by removing the 
overall UV divergences of
$L_n$, $B_n$, and $\delta m_n$ by the \textit{K}-operation.
They may still have subdiagram UV divergences which are 
subtracted by subdiagram \textit{K}-operations.
The resulting UV-finite quantities are denoted 
as $L_n^{\rm R}$, $B_n^{\rm R}$, and $\delta m_n^{\rm R}$.
These quantities may have IR divergences,
which are subtracted by \textit{R}-subtraction and \textit{I}-subtraction.
These operations create UV- and IR-finite quantities
which are denoted as $\Delta L_n$, $\Delta B_n$, and $\Delta \delta m_n$.

\subsection{Second-order renormalization constants}

\begin{gather*}
  L_2 = L_2^{\rm UV} + \widetilde{L}_2,
  \qquad
  L_2^{\rm R} = \widetilde{L}_2 = I_2, \\
  B_2 = B_2^{\rm UV} + \widetilde{B}_2,
  \qquad
  B_2^{\rm R} = \widetilde{B}_2 = - I_2 + \Delta B_2, \\
  \Delta \LB_2 = L_2^{\rm R} + B_2^{\rm R} = \Delta B_2, \\
  B_{2^*} = - 2 L_{2^*}, 
  \qquad
  L_{2^*} = I_{2^*} + \Delta L_{2^*}, \\
  B_{2^{**}} = - 2 ( 2 L_{2(1*1*)} + L_{2(1*2*)} ), \\
  {\delta m}_{2^*} = {\delta m}_{2^*}^{\rm UV} + I_2 + \Delta {\delta m}_{2^*}.
\end{gather*}

\subsection{Fourth-order renormalization constants}

%
%
\begin{align*}
{\delta m}_{4a}^{{\rm R}} &=
{\delta\widetilde{m}}_{4a}
\end{align*}
%
%
\begin{align*}
{\delta m}_{4b}^{{\rm R}} &=
{\delta\widetilde{m}}_{4b}-{\delta\widetilde{m}}_{2\prime}\,{B}_{2}^{{\rm UV}}-{\delta\widetilde{m}}_{2\ast}\,{\delta m}_{2}
\end{align*}
%
%
\begin{align*}
\Delta {\delta m}_{4a} &=
{\delta m}_{4a}^{{\rm R}}
\end{align*}
%
%
\begin{align*}
\Delta {\delta m}_{4b} &=
{\delta m}_{4b}^{{\rm R}}
\end{align*}
%
%
\begin{align*}
\Delta {\delta m}_{4} &=
\Delta {\delta m}_{4a}+\Delta {\delta m}_{4b}
\end{align*}
%
%
%
%
\begin{align*}
{B}_{4a}^{{\rm R}} &=
\widetilde{B}_{4a}-2\,{L}_{2}^{{\rm UV}}\,\widetilde{B}_{2}
\end{align*}
%
%
\begin{align*}
{B}_{4b}^{{\rm R}} &=
\widetilde{B}_{4b}-\widetilde{B}_{2\prime}\,{B}_{2}^{{\rm UV}}-{\delta m}_{2}\,{B}_{2\ast}
\end{align*}
%
%
\begin{align*}
{L}_{4a,1}^{{\rm R}} &=
\widetilde{L}_{4a,1}-\widetilde{L}_{2}\,{L}_{2}^{{\rm UV}}
\end{align*}
%
%
\begin{align*}
{L}_{4a,2}^{{\rm R}} &=
\widetilde{L}_{4a,2}
\end{align*}
%
%
\begin{align*}
{L}_{4b,1}^{{\rm R}} &=
\widetilde{L}_{4b,1}-\widetilde{L}_{2\prime}\,{B}_{2}^{{\rm UV}}-{\delta m}_{2}\,{L}_{2\ast}
\end{align*}
%
%
\begin{align*}
{L}_{4b,2}^{{\rm R}} &=
\widetilde{L}_{4b,2}-\widetilde{L}_{2}\,{L}_{2}^{{\rm UV}}
\end{align*}
%
%
\begin{align*}
\Delta {\LB}_{4a} &=
{B}_{4a}^{{\rm R}}+2\,{L}_{4a,1}^{{\rm R}}+{L}_{4a,2}^{{\rm R}}
\end{align*}
%
%
\begin{align*}
\Delta {\LB}_{4b} &=
{B}_{4b}^{{\rm R}}+2\,{L}_{4b,1}^{{\rm R}}+{L}_{4b,2}^{{\rm R}}-{L}_{2}^{{\rm R}}\,{B}_{2}^{{\rm R}}-({L}_{2}^{{\rm R}})^{2}
\end{align*}
%
%
\begin{align*}
\Delta {\LB}_{4} &=
\Delta {\LB}_{4a}+\Delta {\LB}_{4b}
\end{align*}
%
%
 
\subsection{Sixth-order renormalization constants}

%
%
\begin{align*}
{\delta m}_{6A}^{{\rm R}} &=
{\delta\widetilde{m}}_{6A}-{B}_{2}^{{\rm UV}}\,{\delta\widetilde{m}}_{4b(1^{\prime})}-{B}_{2}^{{\rm UV}}\,{\delta\widetilde{m}}_{4b(3^{\prime})}-2\,{\delta m}_{2}\,{\delta\widetilde{m}}_{4b(1\ast)}+{\delta\widetilde{m}}_{2\prime\prime}\,({B}_{2}^{{\rm UV}})^{2}+{\delta\widetilde{m}}_{2\ast\prime}\,{\delta m}_{2}\,{B}_{2}^{{\rm UV}} \\
& +{\delta\widetilde{m}}_{2\ast\prime}\,{\delta m}_{2}\,{B}_{2}^{{\rm UV}}+{\delta\widetilde{m}}_{2\ast\ast}\,({\delta m}_{2})^{2}
\end{align*}
%
%
\begin{align*}
{\delta m}_{6B}^{{\rm R}} &=
{\delta\widetilde{m}}_{6B}-{B}_{2}^{{\rm UV}}\,{\delta\widetilde{m}}_{4b(2^{\prime})}-{\delta m}_{2}\,{\delta\widetilde{m}}_{4b(2\ast)}-{\delta\widetilde{m}}_{2\prime}\,{B}_{4b}^{{\rm UV}}+{\delta\widetilde{m}}_{2\prime}\,{B}_{2\prime}^{{\rm UV}}\,{B}_{2}^{{\rm UV}} \\
& -{\delta\widetilde{m}}_{2\ast}\,{\delta m}_{4b}^{{\rm UV}}+{\delta\widetilde{m}}_{2\ast}\,{\delta m}_{2\prime}^{{\rm UV}}\,{B}_{2}^{{\rm UV}}+{\delta\widetilde{m}}_{2\ast}\,{\delta m}_{2\ast}^{{\rm UV}}\,{\delta m}_{2}
\end{align*}
%
%
\begin{align*}
{\delta m}_{6C}^{{\rm R}} &=
{\delta\widetilde{m}}_{6C}-2\,{L}_{2}^{{\rm UV}}\,{\delta\widetilde{m}}_{4b}-{\delta\widetilde{m}}_{2\prime}\,{B}_{4a}^{{\rm UV}}+2\,{\delta\widetilde{m}}_{2\prime}\,{L}_{2}^{{\rm UV}}\,{B}_{2}^{{\rm UV}}-{\delta\widetilde{m}}_{2\ast}\,{\delta m}_{4a}^{{\rm UV}} \\
& +2\,{\delta\widetilde{m}}_{2\ast}\,{\delta m}_{2}\,{L}_{2}^{{\rm UV}}
\end{align*}
%
%
\begin{align*}
{\delta m}_{6D}^{{\rm R}} &=
{\delta\widetilde{m}}_{6D}-{B}_{2}^{{\rm UV}}\,{\delta\widetilde{m}}_{4a(1^{\prime})}-{L}_{2}^{{\rm UV}}\,{\delta\widetilde{m}}_{4b}-{\delta m}_{2}\,{\delta\widetilde{m}}_{4a(1\ast)}+{\delta\widetilde{m}}_{2\prime}\,{L}_{2}^{{\rm UV}}\,{B}_{2}^{{\rm UV}} \\
& +{\delta\widetilde{m}}_{2\ast}\,{\delta m}_{2}\,{L}_{2}^{{\rm UV}}
\end{align*}
%
%
\begin{align*}
{\delta m}_{6E}^{{\rm R}} &=
{\delta\widetilde{m}}_{6E}-{B}_{2}^{{\rm UV}}\,{\delta\widetilde{m}}_{4a(2^{\prime})}-{\delta m}_{2}\,{\delta\widetilde{m}}_{4a(2\ast)}
\end{align*}
%
%
\begin{align*}
{\delta m}_{6F}^{{\rm R}} &=
{\delta\widetilde{m}}_{6F}-2\,{L}_{2}^{{\rm UV}}\,{\delta\widetilde{m}}_{4a}
\end{align*}
%
%
\begin{align*}
{\delta m}_{6G}^{{\rm R}} &=
{\delta\widetilde{m}}_{6G}-{L}_{2}^{{\rm UV}}\,{\delta\widetilde{m}}_{4a}
\end{align*}
%
%
\begin{align*}
{\delta m}_{6H}^{{\rm R}} &=
{\delta\widetilde{m}}_{6H}
\end{align*}
%
%
\begin{align*}
\Delta {\delta m}_{6A} &=
{\delta m}_{6A}^{{\rm R}}
\end{align*}
%
%
\begin{align*}
\Delta {\delta m}_{6B} &=
{\delta m}_{6B}^{{\rm R}}-\widetilde{L}_{2}\,{\delta m}_{4b}^{{\rm R}}
\end{align*}
%
%
\begin{align*}
\Delta {\delta m}_{6C} &=
{\delta m}_{6C}^{{\rm R}}-\widetilde{L}_{2}\,{\delta m}_{4a}^{{\rm R}}
\end{align*}
%
%
\begin{align*}
\Delta {\delta m}_{6D} &=
{\delta m}_{6D}^{{\rm R}}
\end{align*}
%
%
\begin{align*}
\Delta {\delta m}_{6E} &=
{\delta m}_{6E}^{{\rm R}}
\end{align*}
%
%
\begin{align*}
\Delta {\delta m}_{6F} &=
{\delta m}_{6F}^{{\rm R}}
\end{align*}
%
%
\begin{align*}
\Delta {\delta m}_{6G} &=
{\delta m}_{6G}^{{\rm R}}
\end{align*}
%
%
\begin{align*}
\Delta {\delta m}_{6H} &=
{\delta m}_{6H}^{{\rm R}}
\end{align*}
%
%
\begin{align*}
\Delta {\delta m}_{6} &=
{\delta m}_{6A}^{{\rm R}}+{\delta m}_{6B}^{{\rm R}}+{\delta m}_{6C}^{{\rm R}}+2\,{\delta m}_{6D}^{{\rm R}}+{\delta m}_{6E}^{{\rm R}}+{\delta m}_{6F}^{{\rm R}}+2\,{\delta m}_{6G}^{{\rm R}}+{\delta m}_{6H}^{{\rm R}}-\widetilde{L}_{2}\,{\delta m}_{4a}^{{\rm R}} \\
& -\widetilde{L}_{2}\,{\delta m}_{4b}^{{\rm R}}
\end{align*}


%
%
\begin{align*}
{B}_{6A}^{{\rm R}} &=
\widetilde{B}_{6A}-{B}_{2}^{{\rm UV}}\,\widetilde{B}_{4b(1^{\prime})}-{B}_{2}^{{\rm UV}}\,\widetilde{B}_{4b(3^{\prime})}+\widetilde{B}_{2\prime\prime}\,({B}_{2}^{{\rm UV}})^{2}-2\,{\delta m}_{2}\,{B}_{4b(1\ast)}+{\delta m}_{2}\,{B}_{2}^{{\rm UV}}\,{B}_{2\ast\prime} \\
& +{\delta m}_{2}\,{B}_{2}^{{\rm UV}}\,{B}_{2\ast\prime}+({\delta m}_{2})^{2}\,{B}_{2\ast\ast}
\end{align*}
%
%
\begin{align*}
{B}_{6B}^{{\rm R}} &=
\widetilde{B}_{6B}-{B}_{2\ast}\,{\delta m}_{4b}^{{\rm UV}}-{B}_{2}^{{\rm UV}}\,\widetilde{B}_{4b(2^{\prime})}-\widetilde{B}_{2\prime}\,{B}_{4b}^{{\rm UV}}+\widetilde{B}_{2\prime}\,{B}_{2\prime}^{{\rm UV}}\,{B}_{2}^{{\rm UV}} \\
& -{\delta m}_{2}\,{B}_{4b(2\ast)}+{\delta m}_{2\prime}^{{\rm UV}}\,{B}_{2}^{{\rm UV}}\,{B}_{2\ast}+{\delta m}_{2\ast}^{{\rm UV}}\,{\delta m}_{2}\,{B}_{2\ast}
\end{align*}
%
%
\begin{align*}
{B}_{6C}^{{\rm R}} &=
\widetilde{B}_{6C}-{B}_{2\ast}\,{\delta m}_{4a}^{{\rm UV}}-\widetilde{B}_{2\prime}\,{B}_{4a}^{{\rm UV}}-2\,{L}_{2}^{{\rm UV}}\,\widetilde{B}_{4b}+2\,{L}_{2}^{{\rm UV}}\,\widetilde{B}_{2\prime}\,{B}_{2}^{{\rm UV}} \\
& +2\,{\delta m}_{2}\,{L}_{2}^{{\rm UV}}\,{B}_{2\ast}
\end{align*}
%
%
\begin{align*}
{B}_{6D}^{{\rm R}} &=
\widetilde{B}_{6D}-{B}_{2}^{{\rm UV}}\,\widetilde{B}_{4a(1^{\prime})}-\widetilde{B}_{2}\,{L}_{4b,1}^{{\rm UV}}-{L}_{2}^{{\rm UV}}\,\widetilde{B}_{4b}+{L}_{2}^{{\rm UV}}\,\widetilde{B}_{2\prime}\,{B}_{2}^{{\rm UV}} \\
& +{L}_{2\prime}^{{\rm UV}}\,\widetilde{B}_{2}\,{B}_{2}^{{\rm UV}}-{\delta m}_{2}\,{B}_{4a(1\ast)}+{\delta m}_{2}\,{L}_{2}^{{\rm UV}}\,{B}_{2\ast}
\end{align*}
%
%
\begin{align*}
{B}_{6E}^{{\rm R}} &=
\widetilde{B}_{6E}-{B}_{2}^{{\rm UV}}\,\widetilde{B}_{4a(2^{\prime})}-2\,\widetilde{B}_{2}\,{L}_{4b,1}^{{\rm UV}}+{L}_{2\prime}^{{\rm UV}}\,\widetilde{B}_{2}\,{B}_{2}^{{\rm UV}}+{L}_{2\prime}^{{\rm UV}}\,\widetilde{B}_{2}\,{B}_{2}^{{\rm UV}} \\
& -{\delta m}_{2}\,{B}_{4a(2\ast)}
\end{align*}
%
%
\begin{align*}
{B}_{6F}^{{\rm R}} &=
\widetilde{B}_{6F}-2\,\widetilde{B}_{2}\,{L}_{4a,1}^{{\rm UV}}-2\,{L}_{2}^{{\rm UV}}\,\widetilde{B}_{4a}+3\,({L}_{2}^{{\rm UV}})^{2}\,\widetilde{B}_{2}
\end{align*}
%
%
\begin{align*}
{B}_{6G}^{{\rm R}} &=
\widetilde{B}_{6G}-\widetilde{B}_{2}\,{L}_{4a,1}^{{\rm UV}}-\widetilde{B}_{2}\,{L}_{4b,2}^{{\rm UV}}-{L}_{2}^{{\rm UV}}\,\widetilde{B}_{4a}+2\,({L}_{2}^{{\rm UV}})^{2}\,\widetilde{B}_{2}
\end{align*}
%
%
\begin{align*}
{B}_{6H}^{{\rm R}} &=
\widetilde{B}_{6H}-2\,\widetilde{B}_{2}\,{L}_{4a,2}^{{\rm UV}}
\end{align*}
%
%
\begin{align*}
{L}_{6A,1}^{{\rm R}} &=
\widetilde{L}_{6A,1}-{B}_{2}^{{\rm UV}}\,\widetilde{L}_{4b,1((1^{\prime})^{\prime})}-{B}_{2}^{{\rm UV}}\,\widetilde{L}_{4b,1(3^{\prime})}+\widetilde{L}_{2\prime\prime}\,({B}_{2}^{{\rm UV}})^{2}-{\delta m}_{2}\,{L}_{4b,1((1^{\prime})\ast)}-{\delta m}_{2}\,{L}_{4b,1(3\ast)} \\
& +{\delta m}_{2}\,{L}_{2\ast\prime}\,{B}_{2}^{{\rm UV}}+{\delta m}_{2}\,{L}_{2\ast\prime}\,{B}_{2}^{{\rm UV}}+({\delta m}_{2})^{2}\,{L}_{2(1\ast1\ast)}
\end{align*}
%
%
\begin{align*}
{L}_{6A,2}^{{\rm R}} &=
\widetilde{L}_{6A,2}-{B}_{2}^{{\rm UV}}\,\widetilde{L}_{4b,2(3^{\prime})}-{L}_{2}^{{\rm UV}}\,\widetilde{L}_{4b,1}+\widetilde{L}_{2\prime}\,{L}_{2}^{{\rm UV}}\,{B}_{2}^{{\rm UV}}-{\delta m}_{2}\,{L}_{4b,2(1\ast)} \\
& +{\delta m}_{2}\,{L}_{2}^{{\rm UV}}\,{L}_{2\ast}
\end{align*}
%
%
\begin{align*}
{L}_{6A,3}^{{\rm R}} &=
\widetilde{L}_{6A,3}-2\,{B}_{2}^{{\rm UV}}\,\widetilde{L}_{4b,1(1^{\prime})}+\widetilde{L}_{2\prime\prime}\,({B}_{2}^{{\rm UV}})^{2}-2\,{\delta m}_{2}\,{L}_{4b,1(1\ast)}+2\,{\delta m}_{2}\,{L}_{2\ast\prime}\,{B}_{2}^{{\rm UV}} \\
& +({\delta m}_{2})^{2}\,{L}_{2(1\ast(1^{\prime})\ast)}
\end{align*}
%
%
\begin{align*}
{L}_{6B,1}^{{\rm R}} &=
\widetilde{L}_{6B,1}-{B}_{2}^{{\rm UV}}\,\widetilde{L}_{4b,1(2^{\prime})}-{L}_{2\ast}\,{\delta m}_{4b}^{{\rm UV}}-\widetilde{L}_{2\prime}\,{B}_{4b}^{{\rm UV}}+\widetilde{L}_{2\prime}\,{B}_{2\prime}^{{\rm UV}}\,{B}_{2}^{{\rm UV}} \\
& -{\delta m}_{2}\,{L}_{4b,1(2\ast)}+{\delta m}_{2\prime}^{{\rm UV}}\,{L}_{2\ast}\,{B}_{2}^{{\rm UV}}+{\delta m}_{2\ast}^{{\rm UV}}\,{\delta m}_{2}\,{L}_{2\ast}
\end{align*}
%
%
\begin{align*}
{L}_{6B,2}^{{\rm R}} &=
\widetilde{L}_{6B,2}-{B}_{2}^{{\rm UV}}\,\widetilde{L}_{4b,2((2^{\prime})^{\prime})}-\widetilde{L}_{2}\,{L}_{4b,1}^{{\rm UV}}+\widetilde{L}_{2}\,{L}_{2\prime}^{{\rm UV}}\,{B}_{2}^{{\rm UV}}-{\delta m}_{2}\,{L}_{4b,2(2\ast)}
\end{align*}
%
%
\begin{align*}
{L}_{6B,3}^{{\rm R}} &=
\widetilde{L}_{6B,3}-{L}_{2}^{{\rm UV}}\,\widetilde{L}_{4b,2}-\widetilde{L}_{2}\,{L}_{4b,2}^{{\rm UV}}+\widetilde{L}_{2}\,({L}_{2}^{{\rm UV}})^{2}
\end{align*}
%
%
\begin{align*}
{L}_{6C,1}^{{\rm R}} &=
\widetilde{L}_{6C,1}-{L}_{2\ast}\,{\delta m}_{4a}^{{\rm UV}}-2\,{L}_{2}^{{\rm UV}}\,\widetilde{L}_{4b,1}-\widetilde{L}_{2\prime}\,{B}_{4a}^{{\rm UV}}+2\,\widetilde{L}_{2\prime}\,{L}_{2}^{{\rm UV}}\,{B}_{2}^{{\rm UV}} \\
& +2\,{\delta m}_{2}\,{L}_{2}^{{\rm UV}}\,{L}_{2\ast}
\end{align*}
%
%
\begin{align*}
{L}_{6C,2}^{{\rm R}} &=
\widetilde{L}_{6C,2}-{L}_{2}^{{\rm UV}}\,\widetilde{L}_{4b,2}-\widetilde{L}_{2}\,{L}_{4a,1}^{{\rm UV}}+\widetilde{L}_{2}\,({L}_{2}^{{\rm UV}})^{2}
\end{align*}
%
%
\begin{align*}
{L}_{6C,3}^{{\rm R}} &=
\widetilde{L}_{6C,3}-\widetilde{L}_{2}\,{L}_{4a,2}^{{\rm UV}}
\end{align*}
%
%
\begin{align*}
{L}_{6D,1}^{{\rm R}} &=
\widetilde{L}_{6D,1}-{B}_{2}^{{\rm UV}}\,\widetilde{L}_{4a,1((1^{\prime})^{\prime})}-{L}_{2}^{{\rm UV}}\,\widetilde{L}_{4b,1}+\widetilde{L}_{2\prime}\,{L}_{2}^{{\rm UV}}\,{B}_{2}^{{\rm UV}}-{\delta m}_{2}\,{L}_{4a,1((1^{\prime})\ast)} \\
& +{\delta m}_{2}\,{L}_{2}^{{\rm UV}}\,{L}_{2\ast}
\end{align*}
%
%
\begin{align*}
{L}_{6D,2}^{{\rm R}} &=
\widetilde{L}_{6D,2}-{L}_{2}^{{\rm UV}}\,\widetilde{L}_{4a,1}-{L}_{2}^{{\rm UV}}\,\widetilde{L}_{4b,2}+\widetilde{L}_{2}\,({L}_{2}^{{\rm UV}})^{2}
\end{align*}
%
%
\begin{align*}
{L}_{6D,3}^{{\rm R}} &=
\widetilde{L}_{6D,3}-{B}_{2}^{{\rm UV}}\,\widetilde{L}_{4a,1(1^{\prime})}-{L}_{2}^{{\rm UV}}\,\widetilde{L}_{4b,1}+\widetilde{L}_{2\prime}\,{L}_{2}^{{\rm UV}}\,{B}_{2}^{{\rm UV}}-{\delta m}_{2}\,{L}_{4a,1(1\ast)} \\
& +{\delta m}_{2}\,{L}_{2}^{{\rm UV}}\,{L}_{2\ast}
\end{align*}
%
%
\begin{align*}
{L}_{6D,4}^{{\rm R}} &=
\widetilde{L}_{6D,4}-{B}_{2}^{{\rm UV}}\,\widetilde{L}_{4a,2(1^{\prime})}-{\delta m}_{2}\,{L}_{4a,2(1\ast)}
\end{align*}
%
%
\begin{align*}
{L}_{6D,5}^{{\rm R}} &=
\widetilde{L}_{6D,5}-{B}_{2}^{{\rm UV}}\,\widetilde{L}_{4a,1(3^{\prime})}-\widetilde{L}_{2}\,{L}_{4b,1}^{{\rm UV}}+\widetilde{L}_{2}\,{L}_{2\prime}^{{\rm UV}}\,{B}_{2}^{{\rm UV}}-{\delta m}_{2}\,{L}_{4a,1(3\ast)}
\end{align*}
%
%
\begin{align*}
{L}_{6E,1}^{{\rm R}} &=
\widetilde{L}_{6E,1}-{B}_{2}^{{\rm UV}}\,\widetilde{L}_{4a,1(2^{\prime})}-\widetilde{L}_{2}\,{L}_{4b,1}^{{\rm UV}}+\widetilde{L}_{2}\,{L}_{2\prime}^{{\rm UV}}\,{B}_{2}^{{\rm UV}}-{\delta m}_{2}\,{L}_{4a,1(2\ast)}
\end{align*}
%
%
\begin{align*}
{L}_{6E,2}^{{\rm R}} &=
\widetilde{L}_{6E,2}-{B}_{2}^{{\rm UV}}\,\widetilde{L}_{4a,2((2^{\prime})^{\prime})}-{\delta m}_{2}\,{L}_{4a,2(2\ast)}
\end{align*}
%
%
\begin{align*}
{L}_{6E,3}^{{\rm R}} &=
\widetilde{L}_{6E,3}-{L}_{2}^{{\rm UV}}\,\widetilde{L}_{4a,2}
\end{align*}
%
%
\begin{align*}
{L}_{6F,1}^{{\rm R}} &=
\widetilde{L}_{6F,1}-{L}_{2}^{{\rm UV}}\,\widetilde{L}_{4a,1}-\widetilde{L}_{2}\,{L}_{4a,1}^{{\rm UV}}+\widetilde{L}_{2}\,({L}_{2}^{{\rm UV}})^{2}
\end{align*}
%
%
\begin{align*}
{L}_{6F,2}^{{\rm R}} &=
\widetilde{L}_{6F,2}-{L}_{2}^{{\rm UV}}\,\widetilde{L}_{4a,2}
\end{align*}
%
%
\begin{align*}
{L}_{6F,3}^{{\rm R}} &=
\widetilde{L}_{6F,3}-2\,{L}_{2}^{{\rm UV}}\,\widetilde{L}_{4a,1}+\widetilde{L}_{2}\,({L}_{2}^{{\rm UV}})^{2}
\end{align*}
%
%
\begin{align*}
{L}_{6G,1}^{{\rm R}} &=
\widetilde{L}_{6G,1}-{L}_{2}^{{\rm UV}}\,\widetilde{L}_{4a,1}-\widetilde{L}_{2}\,{L}_{4a,1}^{{\rm UV}}+\widetilde{L}_{2}\,({L}_{2}^{{\rm UV}})^{2}
\end{align*}
%
%
\begin{align*}
{L}_{6G,2}^{{\rm R}} &=
\widetilde{L}_{6G,2}
\end{align*}
%
%
\begin{align*}
{L}_{6G,3}^{{\rm R}} &=
\widetilde{L}_{6G,3}
\end{align*}
%
%
\begin{align*}
{L}_{6G,4}^{{\rm R}} &=
\widetilde{L}_{6G,4}-{L}_{2}^{{\rm UV}}\,\widetilde{L}_{4a,2}
\end{align*}
%
%
\begin{align*}
{L}_{6G,5}^{{\rm R}} &=
\widetilde{L}_{6G,5}-{L}_{2}^{{\rm UV}}\,\widetilde{L}_{4a,1}-\widetilde{L}_{2}\,{L}_{4b,2}^{{\rm UV}}+\widetilde{L}_{2}\,({L}_{2}^{{\rm UV}})^{2}
\end{align*}
%
%
\begin{align*}
{L}_{6H,1}^{{\rm R}} &=
\widetilde{L}_{6H,1}-\widetilde{L}_{2}\,{L}_{4a,2}^{{\rm UV}}
\end{align*}
%
%
\begin{align*}
{L}_{6H,2}^{{\rm R}} &=
\widetilde{L}_{6H,2}
\end{align*}
%
%
\begin{align*}
{L}_{6H,3}^{{\rm R}} &=
\widetilde{L}_{6H,3}
\end{align*}
%
%
\begin{align*}
\Delta {\LB}_{6A} &=
{B}_{6A}^{{\rm R}}+2\,{L}_{6A,1}^{{\rm R}}+2\,{L}_{6A,2}^{{\rm R}}+{L}_{6A,3}^{{\rm R}} - 2 \Delta {\LB}_{2}\,{L}_{4b,1}^{{\rm R}}
\end{align*}
%
%
\begin{align*}
\Delta {\LB}_{6B} &=
{B}_{6B}^{{\rm R}}+2\,{L}_{6B,1}^{{\rm R}}+2\,{L}_{6B,2}^{{\rm R}}+{L}_{6B,3}^{{\rm R}}-\Delta {\LB}_{4b}\,{L}_{2}^{{\rm R}}-\Delta {\LB}_{2}\,{L}_{4b,2}^{{\rm R}}
\end{align*}
%
%
\begin{align*}
\Delta {\LB}_{6C} &=
{B}_{6C}^{{\rm R}}+2\,{L}_{6C,1}^{{\rm R}}+2\,{L}_{6C,2}^{{\rm R}}+{L}_{6C,3}^{{\rm R}}-\Delta {\LB}_{4a}\,{L}_{2}^{{\rm R}}
\end{align*}
%
%
\begin{align*}
\Delta {\LB}_{6D} &=
{B}_{6D}^{{\rm R}}+{L}_{6D,1}^{{\rm R}}+{L}_{6D,2}^{{\rm R}}+{L}_{6D,3}^{{\rm R}}+{L}_{6D,4}^{{\rm R}}+{L}_{6D,5}^{{\rm R}}-\Delta {\LB}_{2}\,{L}_{4a,1}^{{\rm R}}
\end{align*}
%
%
\begin{align*}
\Delta {\LB}_{6E} &=
{B}_{6E}^{{\rm R}}+2\,{L}_{6E,1}^{{\rm R}}+2\,{L}_{6E,2}^{{\rm R}}+{L}_{6E,3}^{{\rm R}}-\Delta {\LB}_{2}\,{L}_{4a,2}^{{\rm R}}
\end{align*}
%
%
\begin{align*}
\Delta {\LB}_{6F} &=
{B}_{6F}^{{\rm R}}+2\,{L}_{6F,1}^{{\rm R}}+2\,{L}_{6F,2}^{{\rm R}}+{L}_{6F,3}^{{\rm R}}
\end{align*}
%
%
\begin{align*}
\Delta {\LB}_{6G} &=
{B}_{6G}^{{\rm R}}+{L}_{6G,1}^{{\rm R}}+{L}_{6G,2}^{{\rm R}}+{L}_{6G,3}^{{\rm R}}+{L}_{6G,4}^{{\rm R}}+{L}_{6G,5}^{{\rm R}}
\end{align*}
%
%
\begin{align*}
\Delta {\LB}_{6H} &=
{B}_{6H}^{{\rm R}}+2\,{L}_{6H,1}^{{\rm R}}+2\,{L}_{6H,2}^{{\rm R}}+{L}_{6H,3}^{{\rm R}}
\end{align*}
%
%
%
\begin{equation}
\Delta\LB_6 = \sum_{\beta=A}^H \lambda_\beta \Delta\LB_{6\beta},
\end{equation}
where $\lambda_A=\lambda_B=\lambda_C=\lambda_E=\lambda_F=\lambda_H=1$,
and $\lambda_D=\lambda_G=2$. 

$\Delta L_6$ and $\Delta B_6$ defined in Ref. \cite{Kinoshita:2005sm} are related to
$\Delta\LB_6$ through
\begin{equation}
\Delta\LB_6 = \Delta L_6 + \Delta B_6 + \Delta L_4 \Delta B_2 + \Delta \delta 
m_4 B_{2^*} [I].
\end{equation}
%



\bibliographystyle{apsrev}
\bibliography{b}

\end{document}